\documentclass[final,5p,times,twoside,twocolumn]{elsarticle}

\usepackage{graphicx,amssymb,amsmath}
\graphicspath{{./figures/}}

\usepackage[colorlinks=true]{hyperref}

\newcommand{\draft}{
  \setlength{\unitlength}{1mm}{\hbox{\begin{picture}(0,0)
        \put(0,10){\mbox{\footnotesize%
            Draft, \today}}\end{picture}}}}

\biboptions{sort&compress}
\journal{Physics Letters B}

\begin{document}

\begin{frontmatter}
  \title{\draft
    Spectral functions and critical dynamics of the $O(4)$ model\\
    from classical-statistical lattice simulations} 

\author[BU]{S\"oren Schlichting}
\author[JLU]{Dominik Smith}
\author[JLU]{Lorenz von Smekal}

\address[BU]{Fakult\"{a}t f\"{u}r Physik, Universit\"{a}t Bielefeld, 33615 Bielefeld, Germany}
\address[JLU]{Institut f\"ur Theoretische Physik,
  Justus-Liebig-Universit\"at, Heinrich-Buff-Ring 16, 35392 Gie\ss en, Germany}

\begin{abstract}
We calculate spectral functions of the relativistic $O(4)$ model from real-time lattice simulations in classical-statistical field theory. While in the low and high temperature phase of the model, the spectral functions of longitudinal $(\sigma)$ and transverse $(\pi)$ modes are well described by relativistic quasi-particle peaks, we find a highly non-trivial behavior of the spectral functions in the cross over region, where additional structures appear. Similarly, we observe a significant broadening of the quasi-particle peaks, when the amount explicit $O(4)$ symmetry breaking is reduced. We further demonstrate that in the vicinity of the $O(4)$ critical point, the spectral functions develop an infrared power law associated with the critical dynamics, and comment on the extraction of the dynamical critical exponent $z$ from our simulations.
\end{abstract}


\end{frontmatter}

\section{Introduction}
\label{sec:intro}

Besides static equilibrium properties, real-time correlation functions are of great interest in a wide range of physical settings, ranging from
heavy-ion collisions to condensed-matter physics, as they carry important information
about the dynamical properties of classical and quantum systems. Specifically, for equilibrium systems, the corresponding spectral functions contain information about the quasi-particle spectrum of a theory and can be used to reconstruct all real-time and Euclidean correlation functions in thermal equilibrium, via the 
fluctuation-dissipation relation. In addition,
one may also extract transport properties such as the bulk viscosity \cite{Moore2008}, the life times of resonances or particle production rates
from appropriate spectral functions. In the vicinity of second-order phase transitions, one can even use spectral functions to identify the dynamic 
universality class of a system  \cite{Berges:2009jz}. 

Non-perturbative calculations of spectral functions are tremendously difficult. Lattice field-theory simulations 
offer a first-principles approach, but these are typically carried out in Euclidean space-time.
Subsequently an analytic continuation to real time must be performed, which is an ill-posed numerical
problem as it involves computing an inverse Laplace transform from a finite set of data points of finite accuracy. 
Different reconstruction schemes exist, such as Maximum Entropy Methods \cite{JARRELL1996133,ASAKAWA2001459,PhysRevLett.111.182003}, 
the Backus-Gilbert method \cite{PhysRevD.92.094510} the Schlessinger Point or Resonances-via-Pad\'e method 
\cite{PhysRev.167.1411,TRIPOLT2017411} or Tikhonov regularization  \cite{PhysRevD.89.014010} but each
of these comes with its own set of limitations (see Ref. \cite{Tripolt:2018xeo} for a comparison). One interesting alternative to such reconstructions of
spectral functions from lattice data, as done for the present model in Ref.~\cite{Engels:2009tv}, 
is given by functional approaches such as $n-$PI \cite{Roder:2005vt} and Functional Renormalization Group (FRG) methods \cite{Kamikado:2013sia,Tripolt:2013jra,Tripolt:2014wra,Mesterhazy:2015uja,Pawlowski:2015mia,Strodthoff:2016pxx,Pawlowski:2017gxj} or Dyson-Schwinger equations (DSE) \cite{Mueller:2010ah,Fischer:2017kbq}, which can be analytically continued or formulated directly in the real-frequency domain.
However, such approaches necessarily require truncations of an infinite set of evolution equations or equations of motion for $n$-point correlation functions, and thus greatly 
benefit from additional insights into the structure and dynamics of excitations.

In this work, we use classical-statistical lattice simulations in real time to compute the single-particle spectral 
function of a scalar field theory. 
Since critical phenomena in quantum field theories are governed by classical dynamics, universal properties
can be computed in a corresponding classical theory \cite{Aarts:1997kp}. Likewise, spectral functions can be approximated by products
of classical fields close to a second-order phase transition. This approach is based on the fluctuation-dissipation 
relation or Kubo-Martin-Schwinger periodicity condition \cite{Kubo57,PhysRev.115.1342} and becomes exact as one
approaches the critical point. 
In the past, this method has been successfully applied to a single-component scalar
field theory in 2+1 dimensions \cite{Aarts:2001yx} and used to verify that this theory belongs to the dynamic universality 
class of relaxational models with conserved density (Model C) \cite{Berges:2009jz} according to the classification scheme
of Hohenberg and Halperin \cite{RevModPhys.49.435}.

Here we focus on the relativistic isovector Lorentz-scalar field theory with internal $O(4)$ symmetry
(``$O(4)$ model'') in $3\!+\!1$ space-time dimensions, which also exhibits a second-order phase transition. Clearly, this model is of
particular relevance as an effective theory for low energy QCD; in particular the chiral phase transition of QCD for two degenerate light-quark 
flavours is believed to be in the same $O(4)$ universality class \cite{PhysRevD.29.338,Wilczek92,RAJAGOPAL1993395,Engels:2009tv}. Other $O(N)$ models are
of interest in a QCD context as well, such as e.g. the $O(3)$ model in 1+1 dimensions, which exhibits instanton solutions, 
asymptotic freedom and a trace anomaly \cite{POLYAKOV197579,NOVIKOV1984103,Andersen:2003va,Seel:2012vj}.
Central objective of our study is to calculate and analyze the features of real-time spectral functions in the $O(4)$ model within the classical-statistical approach.
 Even though strictly speaking the classical-statistical approximation is only justifiable at very high temperatures or in the vicinity of the critical point, we will also explore the behavior away from criticality, where our results can still provide qualitative insights which may serve as a valuable input to the non-perturbative functional methods mentioned above. 

Starting with a brief outline of the methodology and simulation setup in Secs.~\ref{sec:background} and \ref{sec:setup}, we proceed to the extraction of the phase diagram and analysis of the static critical behavior of the $O(4)$ scalar-field model in Sec.~\ref{sec:static_u}. Simulation results for real-time spectral functions and dynamic critical behavior are presented in Sec. \ref{sec:spectral_f}, where we discuss the behavior of the spectral functions across a crossover transition and in the vicinity of the critical point. Our conclusions are provided in Sec.~\ref{sec:conclusion}.

\section{Spectral functions, fluctuation-dissipation theorem and classical-statistical approximation}
\label{sec:background}
Consider an arbitrary bosonic Heisenberg operator $\hat{O}(t,{\mathbf x})$ 
in a quantum field theory described by the Hamiltonian $\hat{H}$. The spectral function of this
operator is defined via the commutator
\begin{equation}
\rho(t-t', {\mathbf x}-{\mathbf y}, T) \, = \, i \langle [ \hat{O}(t,{\mathbf x}),\hat{O}^\dagger(t',{\mathbf y}) ]_- \rangle \, ,
\label{eq:rhodef}
\end{equation}
where the expectation value in thermal equilibrium is  
\begin{equation}
\langle \hat{O}(t,{\mathbf x}) \rangle \, = \, 
\frac{1}{Z}\, \mathrm{Tr}\left( e^{-\hat{H}/T} \hat{O}(t,{\mathbf x}) \right)~, \quad Z \, = \, \mathrm{Tr}\, e^{-\hat{H}/T} \, .
\label{eq:expectation}
\end{equation}
Besides the spectral function, which characterizes the structure of possible excitations, we can also consider the statistical two-point function, which characterizes statistical fluctuations of the fields, and is defined in the quantum theory from the anti-commutator:
\begin{equation}
F(t-t', {\mathbf x}-{\mathbf y}, T) = \frac{1}{2} \langle [ \hat{O}(t,{\mathbf x}),\hat{O}^\dagger(t',{\mathbf y}) ]_+ \rangle 
- \langle \hat{O}(t, {\mathbf x}) \rangle \langle \hat{O}^\dagger(t', {\mathbf y}) \rangle \, .
\label{eq:Fdef}
\end{equation}
In thermal equilibrium the statistical fluctuations $F(\cdot)$ are connected to the spectral function $\rho(\cdot)$ by the 
fluctuation-dissipation relation or Kubo-Martin-Schwinger (KMS) condition \cite{Kubo57,PhysRev.115.1342,Parisi:1988nd}, which follows from the
imaginary-time periodicity of the Euclidean propagator and is stated in 
Fourier space as
\begin{equation}
F(\omega,p,T) \, = \, \left( n_T(\omega) + \frac{1}{2} \right) \rho(\omega,p,T) \, .
\label{eq:flucdiss}
\end{equation}
Here, $n_T(\omega)$ is the Bose-Einstein distribution. We furthermore denote the magnitude of the spatial momentum as $p \equiv |{\mathbf p}|$ and define the Fourier transformations by  
\begin{eqnarray}
F(\omega, p, T) &=&~~~~\int {\mathrm d}t\, {\mathrm d}^3 x \, e^{i(\omega t - {\mathbf p}{\mathbf x})} F(t,\mathbf{x},T) \,, \\
\rho(\omega, p, T) &=&-i \int {\mathrm d}t\, {\mathrm d}^3 x \, e^{i(\omega t - {\mathbf p}{\mathbf x})} \rho(t,\mathbf{x},T) \, . \nonumber
\label{eq:FFourier}
\end{eqnarray}

In the limit of small frequencies $\omega \ll T$ (or high temperatures) the Bose Einstein distribution $n_T(\omega) = 1/(\exp(\omega/T) - 1)$ is well approximated by $n_T(\omega) \approx T/\omega$, which is precisely the Rayleigh-Jeans
distribution of the occupation-number  in a classical-statistical bosonic field theory. Since the universal properties in the vicinity of a finite-temperature phase transition are governed by infrared field modes with $\omega \ll T$, it is exactly this limit which is relevant to the study of critical dynamics.
In the absence of quantum anomalies critical phenomena at a finite temperature phase transition are therefore rigorously characterized by classical dynamics and we will argue in the following that (\ref{eq:rhodef}) is approximated with increasing precision by a product of classical field variables
computed in a corresponding classical-statistical theory as one approaches a critical point. 

We further note that a classical-statistical description of the dynamics also becomes applicable when statistical fluctuations $\sim F$ dominate over quantum fluctuations $\sim \rho$, as the classical-statistical approximation (CSA) can formally be seen as a leading order expansion in $F \gg \rho$, as discussed in detail in \cite{Berges:2004yj}. Based on this idea, the classical-statistical description has also been applied to the study of equilibrium spectral functions in the high-temperature regime of scalar field theories  \cite{Aarts:1997kp}.

In the classical limit there are no commutators, so 
the spectral function is given by
\begin{equation}
\rho_\mathrm{cl}(t-t', {\mathbf x}-{\mathbf y}, T) \, = \, -\langle \{ O(t,{\mathbf x}),O^{*}(t',{\mathbf y}) \} \rangle_\mathrm{cl}\, ,
\label{eq:clrhodef}
\end{equation}
where $\{\cdot,\cdot\}$ denotes the Poisson bracket, the expectation value is now computed with respect to a
classical-statistical ensemble and $O(t,{\mathbf x})$ becomes a functional of classical fields $\phi(t,{\mathbf x})$ and
their conjugate momenta $\pi(t,{\mathbf x})$. Even though one could in principle compute the spectral function directly using
Eq. (\ref{eq:clrhodef}) (see e.g.~\cite{PineiroOrioli:2018hst}), it turns out that handling the Poisson bracket is impractical and there is a more elegant way to calculate equilibrium spectral functions in classical-statistical field theory  \cite{Aarts:1997kp}.

Exploiting the fluctuation dissipation relation for  $\omega \ll T$ in the classical-statistical theory, Eq. (\ref{eq:flucdiss}) is approximated by
\begin{equation}
F(\omega, p, T) \, \approx \, \frac{T}{\omega}\, \rho(\omega, p, T) \, , 
\label{eq:clflucdiss}
\end{equation}
which in the time domain can be expressed as
\begin{equation}
  \rho(t,p, T) \, \approx \, - \frac{1}{T} \, \partial_{t} F(t,p, T) \, .
\label{eq:KMStime}
\end{equation}
By using Eqs.~(\ref{eq:clflucdiss}), (\ref{eq:KMStime}),  and the fact that in the classical limit the statistical two-point function becomes
\begin{equation}
F_\mathrm{cl}(t-t', {\mathbf x}-{\mathbf y}, T) =\langle O(t,{\mathbf x}) O^{*}(t',{\mathbf y}) \rangle_\mathrm{cl} \, 
- \left\langle O(t, {\mathbf x}) \right\rangle_\mathrm{cl} \left\langle O^{*}(t', \mathbf{y}) \right\rangle_\mathrm{cl} \, ,
\label{eq:Fcldef}
\end{equation}
we can construct simple expressions for different spectral functions which make use only of products
of field variables, are exact for classical-statistical theories but also
describe the universal critical behavior of quantum field theories in the same universality class  \cite{Berges:2009jz}. 

In this work, we numerically obtain the single particle spectral function (i.e. $O(t,{\mathbf x})\equiv\phi(t,{\mathbf x})$) in
the momentum domain.
We consider a real scalar field theory, so
$O^\dagger(t,{\mathbf x})=O(t,{\mathbf x})$. Using Eq.~(\ref{eq:KMStime}), we can write for the spectral function in real time
\begin{align}
\label{eq:clrhotime}
&\rho_\mathrm{cl}(t,p, T) \,  \\
&= \, - \frac{1}{T}  \int  {\mathrm d}^3 x  \, e^{-i \mathbf{p} \mathbf{x}} \,
 \partial_{t} F_\mathrm{cl}(t, {\mathbf x}, T)
\, \nonumber \\
&= \, - \frac{1}{2T} \int {\mathrm d}^3 x \, e^{-i \mathbf{p} \mathbf{x}} \, \langle \, \pi(t,\mathbf{x})\, \phi(0,\mathbf{0}) - \phi(t,\mathbf{x}) \, \pi(0,\mathbf{0})\,  \rangle_\mathrm{cl}  \nonumber
\, ,
\end{align}
where $ \pi(t,{\mathbf x})=\partial_t \phi(t,{\mathbf x})$, and we used the fact that the disconnected part vanishes due to $\langle\pi(t,{\mathbf x})\rangle=0$. We will focus for simplicity on the $p=0$ component, for which the spectral function is explicitly given by
\begin{equation}
\rho_\mathrm{cl}(t-t',p=0, T) \, = \, - \frac{V}{2T} \,\left\langle \Pi(t) \Phi(t') - \Phi(t)\Pi(t') \right\rangle_\mathrm{cl} 
\, ,\label{eq:clrhozerop}
\end{equation}
with $V=\int  {\mathrm d}^3 x$ and
\begin{equation}
\Pi(t)=\frac{1}{V}\int {\mathrm d}^3 x \,\pi(t,\mathbf{x}) ~,\quad\Phi(t)=\frac{1}{V}\int {\mathrm d}^3 x \,\phi(t,\mathbf{x})~. \label{eq:fieldint}
\end{equation}
Since in practice the spectral functions $\rho_\mathrm{cl}(t-t',0, T)$ are obtained directly in the time domain, it is then straightforward to obtain the corresponding spectral functions in the frequency domain by a Fourier transform.

In the above discussion, we have used a single-component scalar field for illustration. We note that for the $O(4)$ model, the spectral function $\rho_{ab}(t,\mathbf{x})$ is
computed individually for the different field components $\phi^a(t,{\mathbf x})$. By introducing an explicit symmetry breaking, the $O(4)$ symmetry is broken down to $O(3)$, and we can distinguish between the directions parallel and perpendicular to the vacuum alignment of the order parameter $\langle \phi^{a}(t,{\mathbf x}) \rangle$, which we will refer to as the $\sigma$ (parallel) and $\pi$ (perpendicular) components.

\section{Simulation setup}
\label{sec:setup}

We study the classical equilibrium properties of the
 $3\!+\!1$ dimensional $O(4)$ model defined by the lattice Hamiltonian 
\begin{align}
\label{eq:hamiltonian}
{H} &= \sum_{i} a_s^d \left\{ \frac{1}{2} \pi^{a}_i\pi^{a}_i - \frac{1}{2a_s^2} \sum_{j\sim i} \phi^{a}_i \phi^{a}_j + 
\left(\frac{m^2}{2}+\frac{d}{a_s^2}\right) \phi^{a}_i\phi^{a}_i \right. 
\nonumber\\
  &\hskip 4cm \left. + \frac{\lambda}{4! N} \Big(\phi^{a}_i\phi^{a}_i\Big)^2 + J^{a} \phi^{a}_{i} \right\}
  \, , 
\end{align}
where $\phi^{a}_i$ are real valued field variables associated with the sites of a cubic lattice with periodic boundary conditions, 
$\pi^{a}_i$ are conjugate momenta, $a=\{1,\dots \,4\}$ label the components of the fields, $J^{a}=\delta^{a1} J$ denotes
an explicit symmetry breaking term, $d$ is the number of spatial dimensions (we consider the case $d=3$)
and $a_s$ denotes the spatial lattice spacing. We set $a_s=1$ in the following, 
which implies that all dimensionful quantities are understood to be expressed in units of $a_s$ from here on.
The sum $\sum_{j \sim i}$ runs over all nearest neighbors  $j$ of site $i$. 

In classical thermal equilibrium with inverse temperature $\beta=1/T$, the expectation value of a static observable $O[\phi,\pi]$ is defined as
\begin{equation}
\langle O \rangle = \int D\phi D\pi~O[\phi,\pi]~e^{-\beta {H}}\;,
\end{equation}
and can be computed in a straightforward way by generating an ensemble of classical field configurations with distribution $e^{-\beta {H}}$ and subsequently evaluating the observable $O[\phi,\pi]$ as a function of the fundamental fields. In practice we generate our configurations using a Langevin prescription 
\begin{equation}
\partial_{t_L} \phi^{a}_i=\frac{\partial H}{\partial \pi^{a}_i}\;, \quad \partial_{t_L}  \pi^{a}_i= -\frac{\partial H}{\partial \phi^{a}_i} - \gamma\, \pi^{a}_i + \sqrt{2\gamma T}\,  \xi^{a}_i\; ,
\end{equation}
where $t_{L}$ denotes the Langevin time, $\xi^{a}_{i}$ corresponds to a Gaussian white noise $\langle \xi^{a}_i\xi^{b}_j \rangle=\delta^{ab} \delta_{ij}$  and
\begin{equation}
\frac{\partial H}{\partial \pi^{a}_i}=\pi^{a}_{i}\;, \quad  \frac{\partial H}{\partial \phi^{a}_i}=-(\Delta \phi^a)_i+\Big( m^2  + \frac{\lambda}{6 N} \phi^{b}_i\phi^{b}_i \Big) \phi^{a}_i \, + \, \delta^{a1} J\; ,
 \end{equation}
where the Laplacian is discretized as
\begin{equation}
 (\Delta \phi^a)_i = \sum_{j \sim i} (\phi^a_j - \phi^a_i)~.
 \end{equation}
The stochastic differential equation is solved numerically using the Euler-Maruyama scheme with an update step $a_t/a_s=0.01$ and -- if not stated otherwise -- we employ the set of parameters $m^2=-1$, $\lambda=1$ and $\gamma=0.3$. Note that in order to assess the universal critical behavior of the model, the coupling constant $\lambda$ can be tuned to an optimal value to reduce scaling corrections~\cite{Hasenbusch:1999cc}. However, we did not pursue this in our study.

Besides the static observables it is also straightforward to compute unequal time correlation functions in the classical-statistical field theory. This allows for a simple prescription to extract the classical-statistical spectral function in real time through Eqs. (\ref{eq:clrhozerop}) and (\ref{eq:fieldint}). Since the classical fields $\phi,\pi$ obey Hamilton's equations of motion
\begin{equation}
\partial_{t} \phi^{a}_i=\frac{\partial H}{\partial \pi^{a}_i}\;, \qquad \partial_{t}  \pi^{a}_i= -\frac{\partial H}{\partial \phi^{a}_i}\; , 
\end{equation}
it is straightforward to compute the unequal time correlation function $\langle \sum_i\phi^{a}_i(t)  \sum_j\pi^{a}_j(t') \rangle$ entering Eqs. (\ref{eq:clrhozerop}) and (\ref{eq:fieldint}). We first generate an ensemble of initial field configurations, and then independently evolve the classical field configurations up to a time $max(t,t')$ based on a leap-frog scheme with $a_t/a_s=0.05$ unless stated otherwise. By saving the evolution of the order-parameter field along the classical trajectories, we subsequently extract the correlation functions between different time slices.

\section{Results: Static universality} 
\label{sec:static_u}
Critical exponents and scaling functions of the three-dimensional $O(4)$ spin model 
have been studied extensively using lattice simulations \cite{PhysRevD.51.2404,Engels:1999wf,Pelissetto:2000ek,ENGELS2001299,Engels:2003nq,Engels:2014bra}. 
Before we discuss our results for real-time spectral functions, we verify that we reproduce the expected static critical properties 
and extract the phase diagram of our field theoretical model (\ref{eq:hamiltonian}) in the $J-T$ plane. 
Our basic observables for this purpose are cumulants of the 
ferromagnetic order parameter, which in the presence of an explicit symmetry breaking is defined as
\begin{equation}
\phi_{J}= \frac{1}{|J|V}\Big(\sum_{i} J_{a} \phi^{a}_i\Big)\;.
\end{equation}
Conversely, in the absence of an explicit symmetry breaking, we employ 
\begin{equation}
|\phi|=\frac{1}{V}\sqrt{ \Big(\sum_i \phi^{a}_i\Big) \Big(\sum_j \phi^{a}_j\Big)}\;, 
\end{equation}
as a proxy for the order parameter. In the following we will also use the symbol $\phi$ to generically refer to either $|\phi|$ or $\phi_J$, when relations of the same form apply to both equally.


\begin{figure*}
\begin{minipage}{0.45\textwidth}
\includegraphics[width=\textwidth]{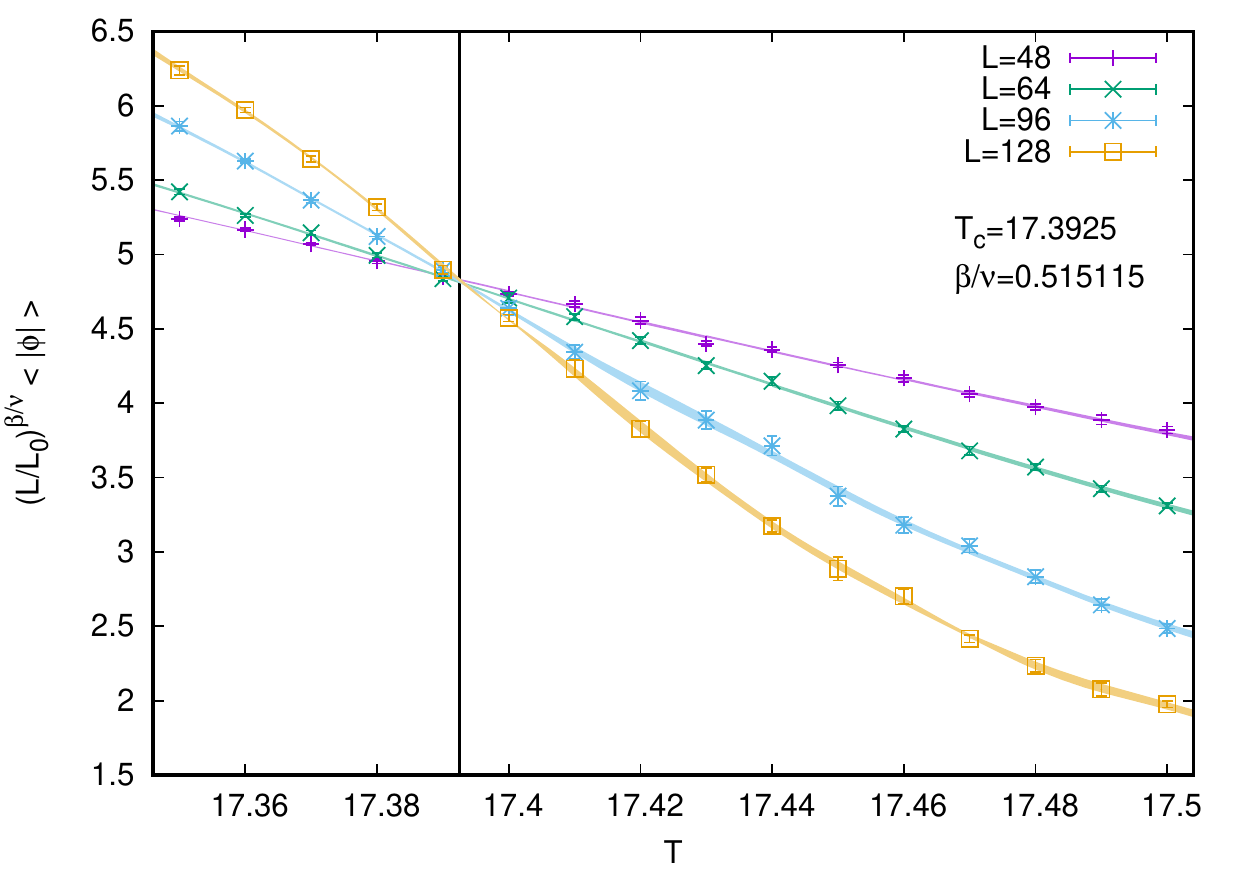}
\end{minipage}
\hspace{1cm}
\begin{minipage}{0.45\textwidth}
\includegraphics[width=\textwidth]{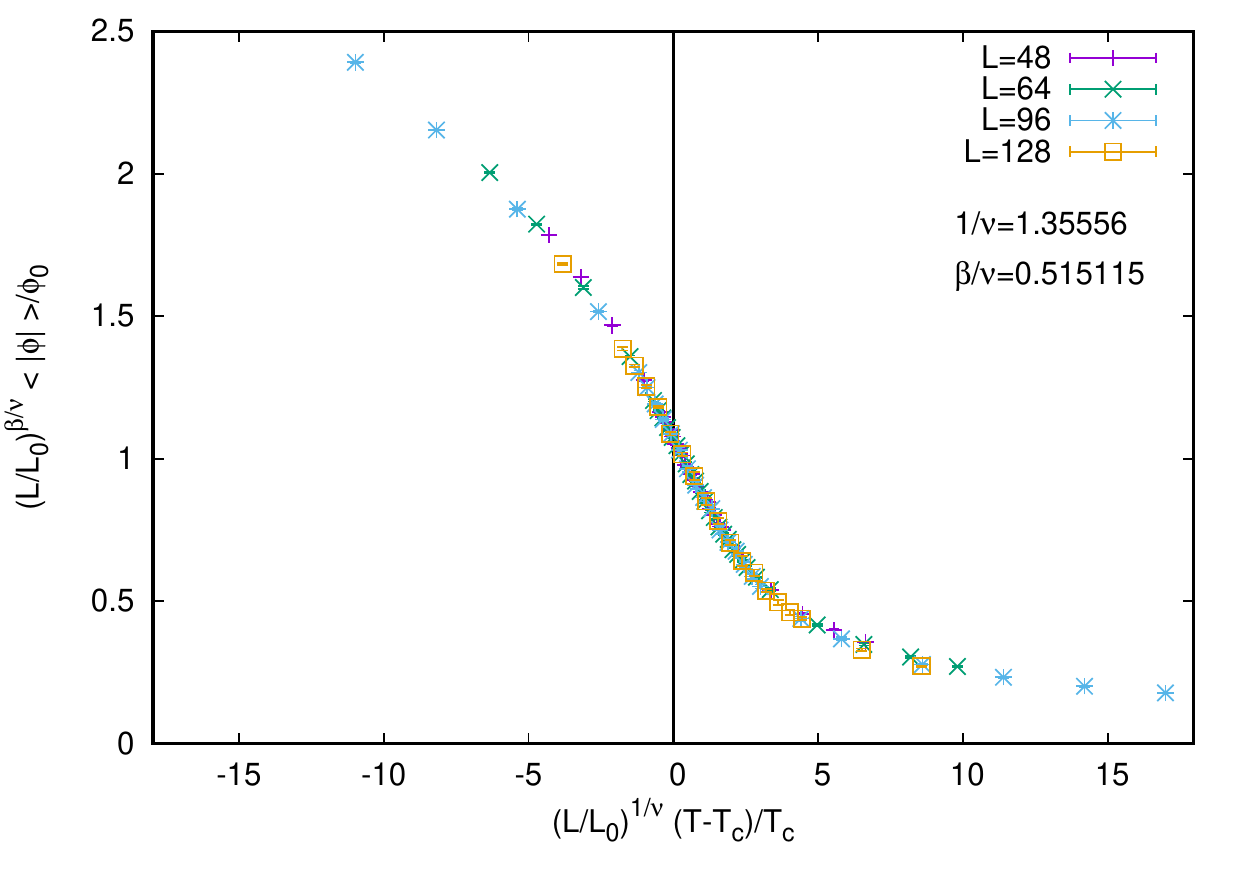}
\end{minipage}
\caption{Temperature dependence and critical scaling of ferromagnetic order parameter 
at $J=0$ for $L=48,64,96,128$ ($\beta$ and $\nu$ taken from \cite{Engels:2014bra}). Left: $T_c=17.3925(10)$ is determined by the intersection of $L^{\beta/\nu}\langle|\phi|\rangle$
for different $L$. Right: Collapse 
of data points of different $L$ onto a universal scaling function.
\label{fig:ordertempscaling}}
\end{figure*}

\begin{figure*}
\begin{minipage}{0.45\textwidth}
\includegraphics[width=\textwidth]{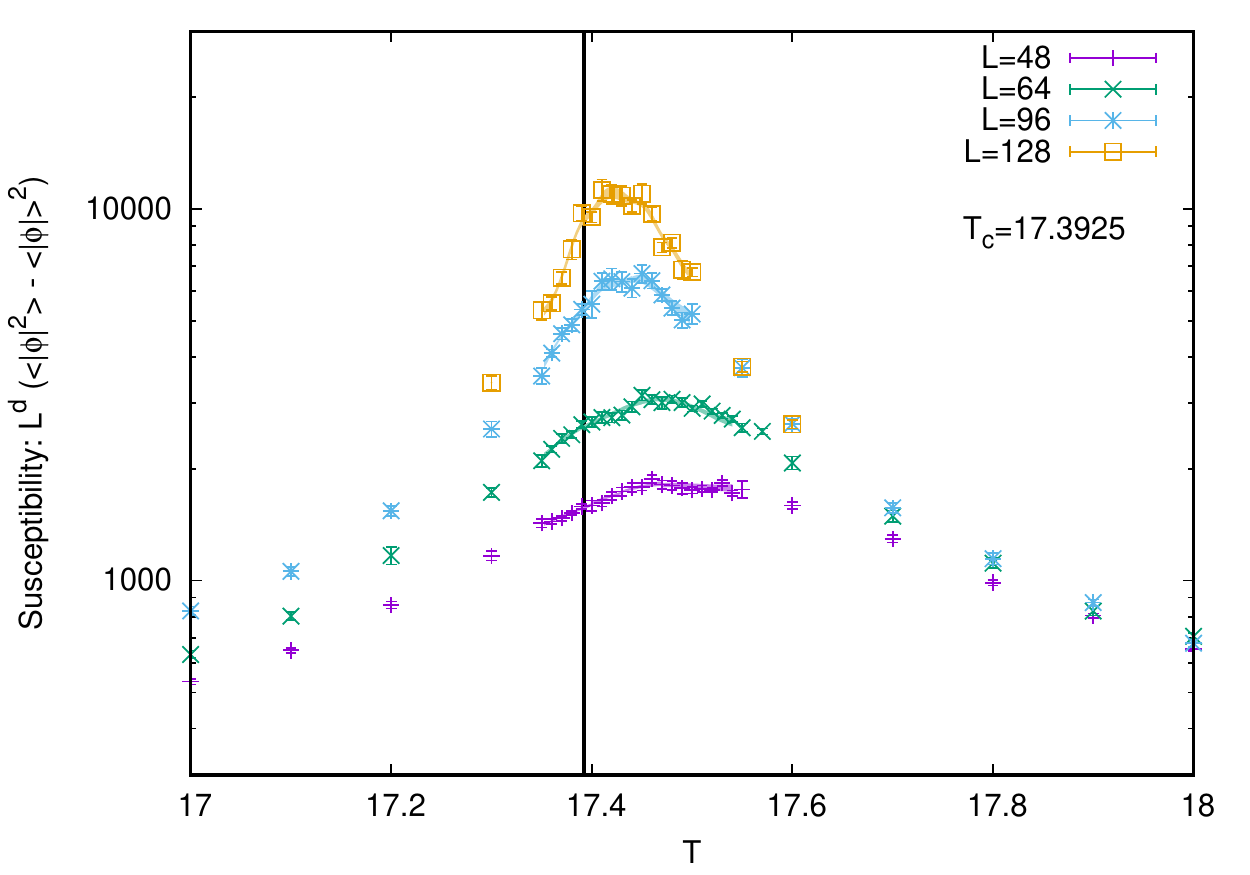}
\end{minipage}
\hspace{1cm}
\begin{minipage}{0.45\textwidth}
\includegraphics[width=\textwidth]{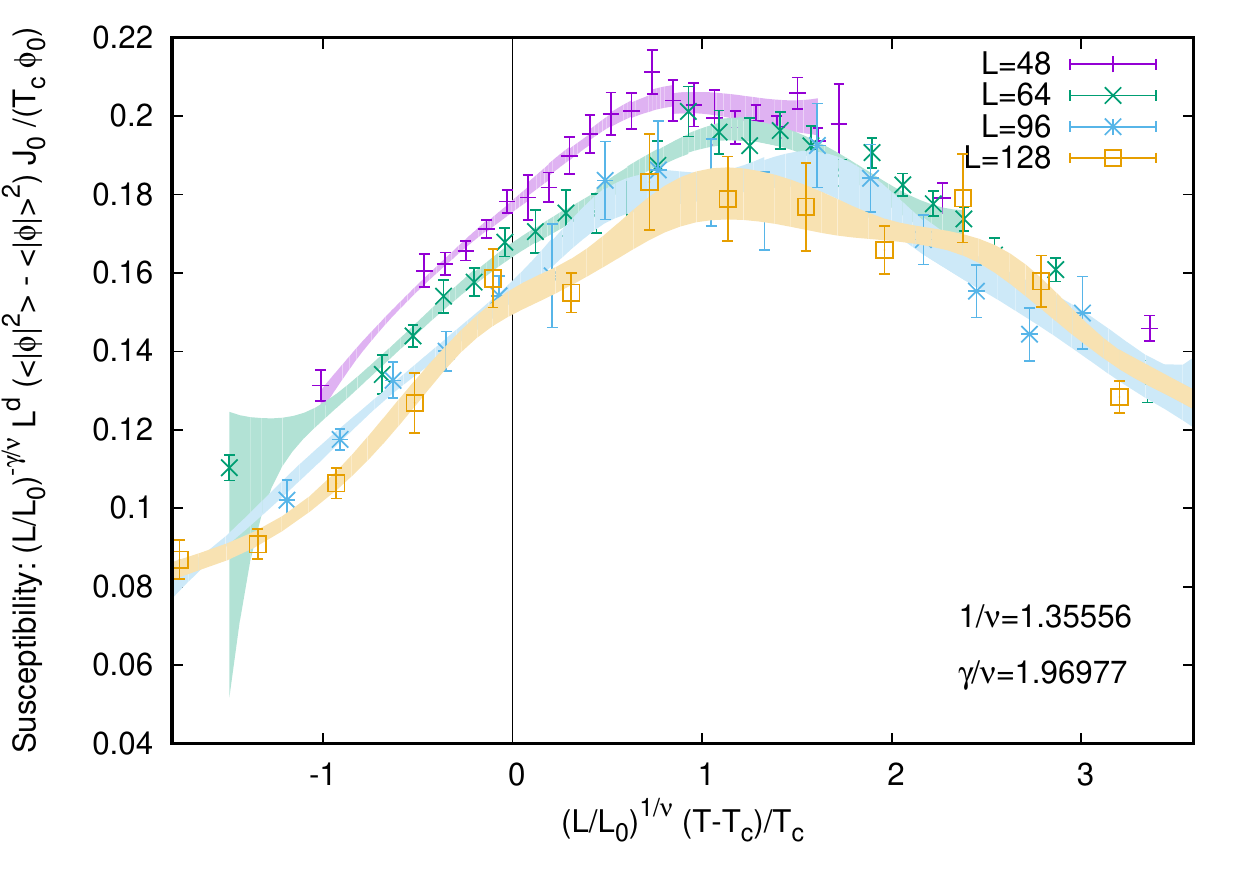}
\end{minipage}
\caption{Temperature dependence (left) and critical scaling (right) of ferromagnetic susceptibility at $J=0$ for $L=48,64,96,128$ ($\gamma$ and $\nu$ taken from \cite{Engels:2014bra}). $T_c$ as obtained from finite-size scaling of $\langle |\phi|\rangle$ is marked by a vertical line (deviations of the peak of $\chi_{|\phi|}$ from this line are due to finite-size effects). 
\label{fig:susctempscaling}} 
\end{figure*}

\begin{figure*}
\begin{minipage}{0.45\textwidth}
\includegraphics[width=\textwidth]{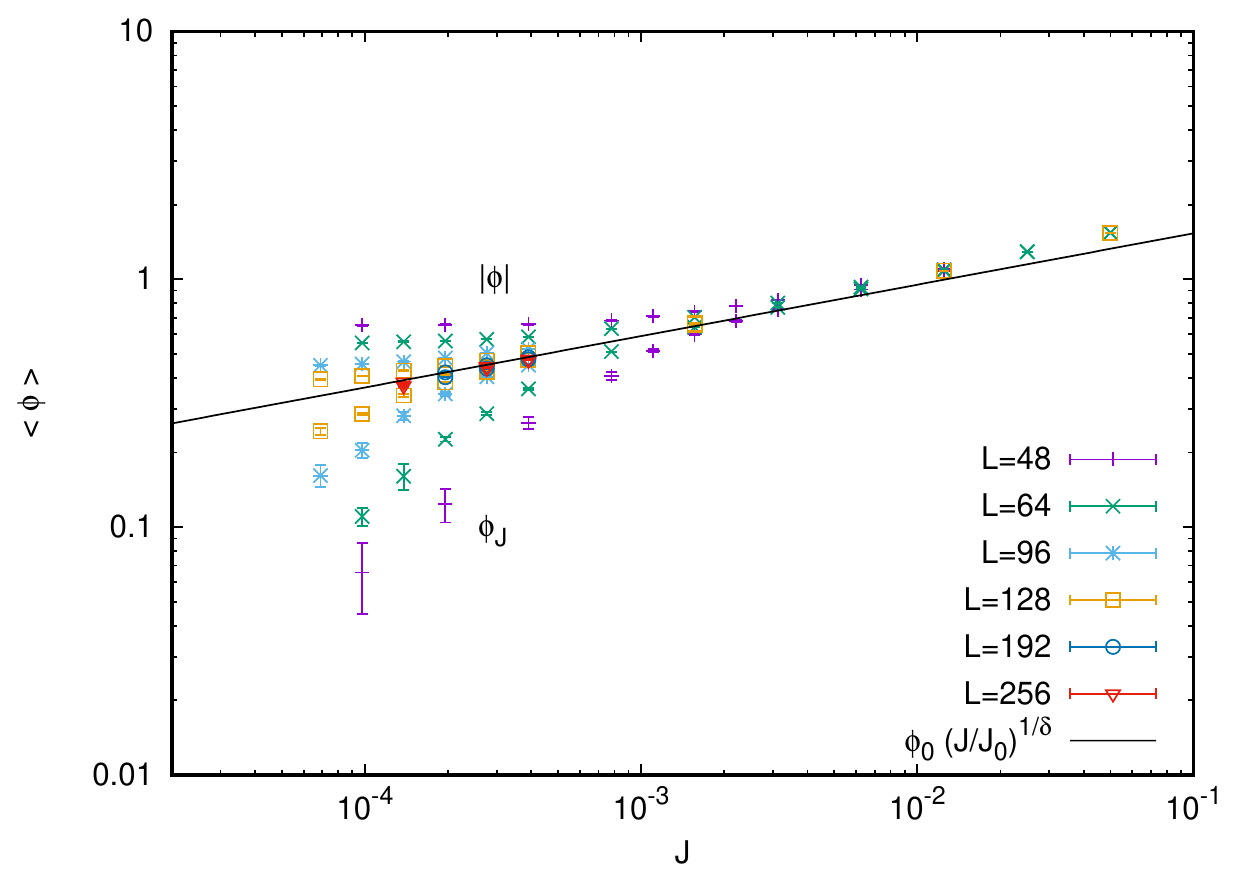}
\end{minipage}
\hspace{1cm}
\begin{minipage}{0.45\textwidth}
\includegraphics[width=\textwidth]{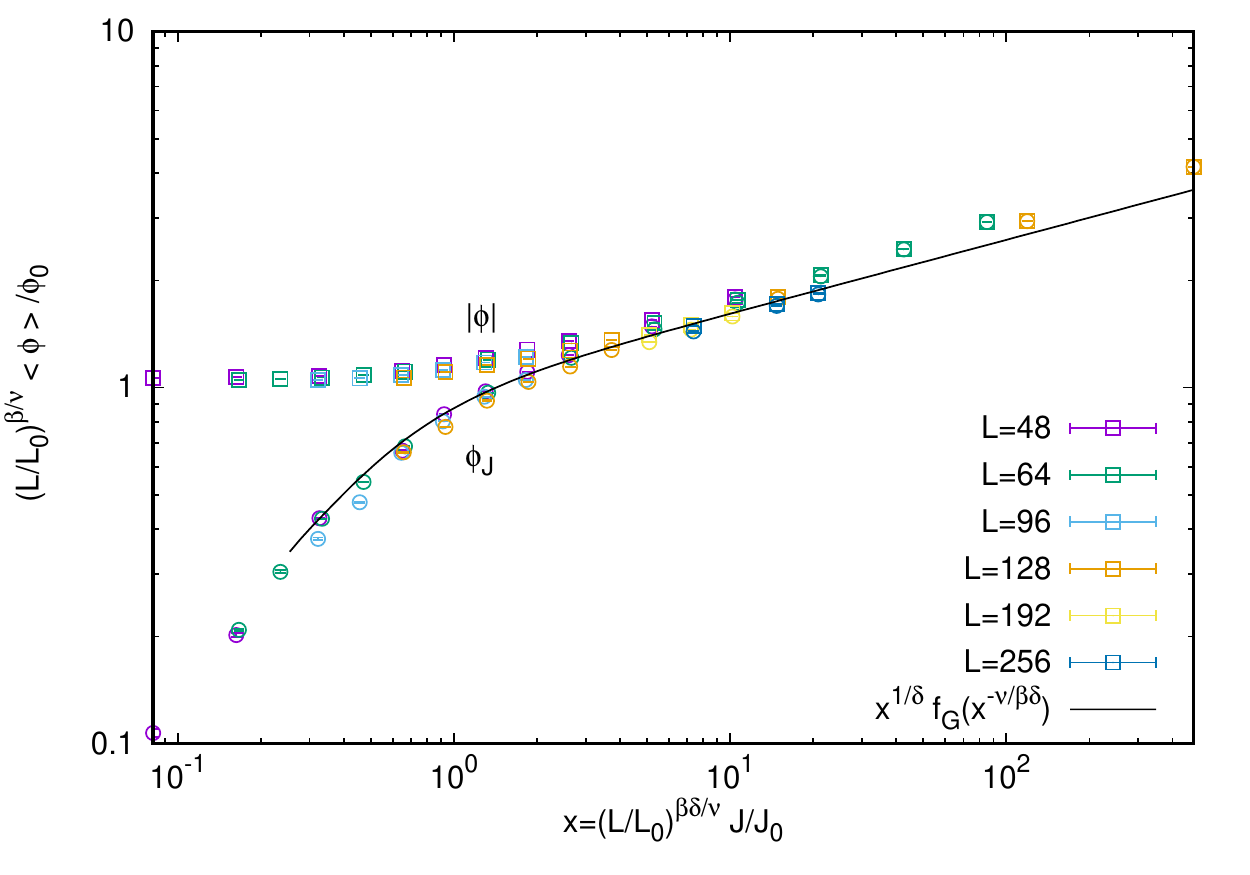}
\end{minipage}
\caption{$J$ dependence (left) and magnetic scaling (right) of order parameter $\phi$ at $T=T_c$. 
Absolute value $\langle|\phi|\rangle$ and component $\phi_J$ in direction of the external field $\langle \phi_J \rangle$ 
are shown. For small $L^{\beta\delta/\nu}J$ the magnetic scaling of $|\phi|$ and $\phi_J$ differs. As $J$ decreases, $L$ must be increased for $\langle|\phi|\rangle=\langle \phi_J \rangle$ to hold. Solid line in left panel shows $\phi=c J^{1/\delta}$ with $\delta=4.824$ taken from \cite{Engels:2014bra} and $c$ chosen such that the
points for the largest lattices at small $J$ are traversed. Solid line in right panel shows universal finite size scaling function for $\phi_J$ as obtained in \cite{Engels:2014bra} from $O(4)$ spin model simulations. 
\label{fig:ordermagneticscaling}} 
\end{figure*}


In the vicinity of the  critical point, the leading dependence of $\phi$ on the reduced temperature $T_{r}=\frac{T-T_c}{T_c}$, the explicit symmetry breaking $J$ and the linear system size $L$ (we always consider 3d-lattices with $L_x=L_y=L_z$), follows from the scaling relation 
\begin{equation}
\label{eq:scalingrel}
\phi(T,J,L^{-1})=s^{\beta}\phi_0~\Phi\left(s^{-1} T_{r},s^{-\beta \delta} \frac{J}{J_0}, s^{-\nu} \left(\frac{L}{L_0}\right)^{-1}\right)\;,
\end{equation}
with a universal scaling function $\Phi$ and non-universal amplitudes $\phi_0$, $J_0$, $L_0$. By adapting the normalization conditions $\Phi(1,0,0)=\Phi(0,1,0)=\Phi(0,0,1)=1$, the critical behavior of the order parameter is determined by\footnote{We have obtained the following values $\phi_0=4.5\pm 0.5$, $\phi_0 J_0^{-1/\delta}=2.4\pm0.2$, $\phi_0 L_0^{\beta/\nu}=4.5\pm 0.3$ as rough estimates for the non-universal amplitudes in our model.}
\begin{eqnarray}
\phi(T_{r},0,0)&=&|T_{r}|^{\beta}~\phi_0\;,\nonumber \\
\phi(0,J,0)&=&|J|^{1/\delta}~\phi_0 J_0^{-1/\delta}\;, \nonumber \\
\phi(0,0,L)&=&|L|^{-\beta/\nu}~\phi_0 L_0^{\beta/\nu}\;,  \label{eq:critscaling}
\end{eqnarray}
and finite-size scaling relations take the form
\begin{align}
\phi(T_{r},0,L)&=\left( \frac{L}{L_0}\right)^{-\beta/\nu} f_T\Big( \left( \frac{L}{L_0}\right)^{1/\nu}T_{r}\Big)\;, \label{eq:fss_order_T}\\
\phi(0,J,L)&=\left( \frac{L}{L_0}\right)^{-\beta/\nu} f_J\Big( \left( \frac{L}{L_0}\right)^{\frac{\beta}{\nu \delta}}\frac{J}{J_0}\Big)\;. \label{eq:fss_order_J}
\end{align}
We will also consider the static susceptibilities
\begin{equation}
\chi_{|\phi|} = V(\langle |\phi|^2 \rangle  - \langle  |\phi| \rangle^2) \;, \quad \chi_{J} = V(\langle \phi_{J}^2 \rangle - \langle \phi_{J} \rangle^2) \;, \label{eq:chi_J_and_abs}
\end{equation}
and the Binder cumulant of the order parameter, which is given by
\begin{equation}
\chi_4=1-  \frac{\langle |\phi|^4 \rangle}{3\langle |\phi|^2 \rangle^2 }~.
\end{equation}

\begin{figure*}[t!]
\begin{minipage}{0.45\textwidth}
\includegraphics[width=\textwidth]{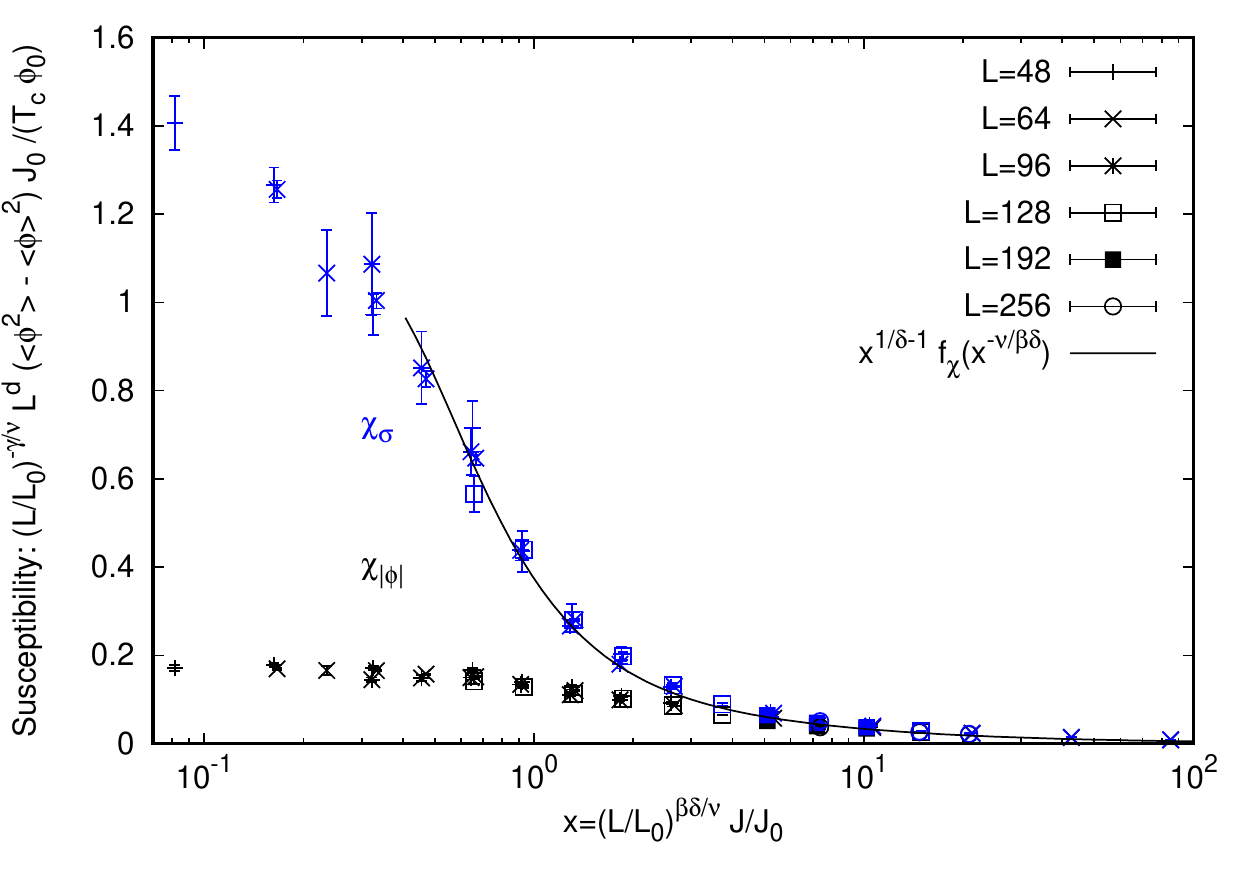}
\end{minipage}
\hspace{1cm}
\begin{minipage}{0.45\textwidth}
\includegraphics[width=\textwidth]{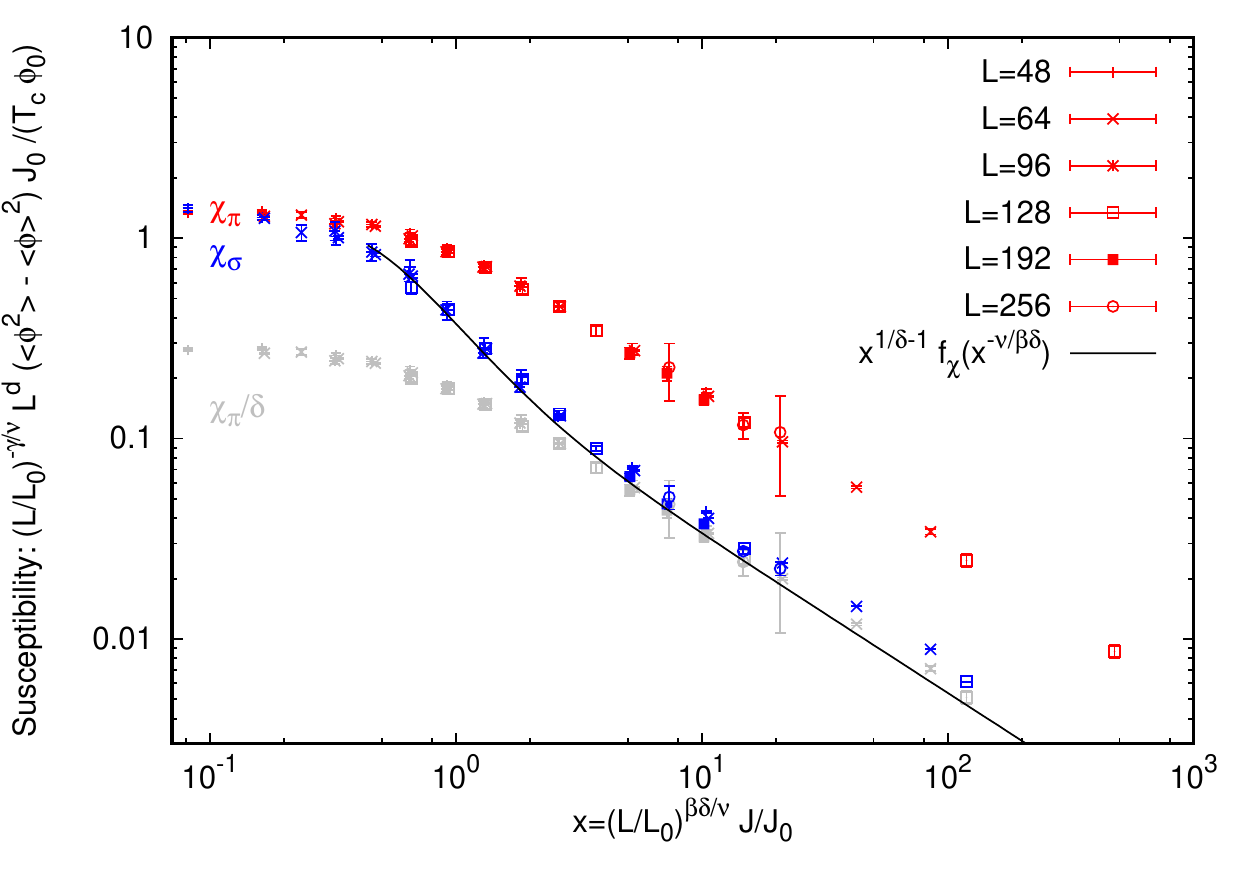}
\end{minipage}
\caption{Magnetic scaling of susceptiblity $\chi$. (left) Susceptibilities of 
absolute value $|\phi|$ and component $\phi_J$ in direction of external field differ unless
$L^{\delta\beta/\nu}J$ is sufficiently large.  (right) Longitudinal and transverse susceptibilities $\chi_{\pi/\sigma}$ exhibit universal scaling at large $L^{\delta\beta/\nu}J$, but become indistinguishable at small $L^{\delta\beta/\nu}J$.
Solid lines in both panels show universal finite size scaling function for $\chi_\sigma$ as obtained in \cite{Engels:2014bra} from $O(4)$ spin model simulations. 
\label{fig:suscmagneticscaling}} 
\end{figure*}

\subsection{Static universality -- $T$ dependence}

We begin with studying the temperature dependence at $J=0$, which is summarized in Figs.~\ref{fig:ordertempscaling}, \ref{fig:susctempscaling}. Individual points in each figure correspond to the data obtained from simulations at the corresponding temperature values, while solid bands are obtained by performing a multi-histogram re-weighting analysis~\cite{Ferrenberg:1988yz}, using the data from the closest six temperature points. Errorbars are obtained from a jacknife analysis. 

We first estimate the the critical temperature $T_c$ at $J=0$. Since the critical exponents $\beta=0.380(2)$ and $\nu=0.7377(41)$ have been determined very precisely from spin-model simulations~\cite{Engels:2014bra}, we use these results and exploit the third identity in Eq. (\ref{eq:critscaling}) along with the finite-size scaling relation (\ref{eq:fss_order_T}) to infer the critical temperature from the order parameter. 
We find by plotting $L^{\beta/\nu}\langle|\phi|\rangle$ for different lattice sizes $L=48,64,96,128$ as a function of $T$ that all curves intersect in a single point with good accuracy (Fig.~\ref{fig:ordertempscaling}, left), which then determines the critical temperature $T_c = 17.3925(10)$. 
Subsequently, we explicitly verify the universal finite-size scaling of our data, by plotting the same observable $L^{\beta/\nu}\langle|\phi|\rangle$ as a function of the rescaled reduced temperature $L^{1/\nu}T_r$ (Fig.~\ref{fig:ordertempscaling}, right).  All data points collapse onto a single universal scaling curve, indicating that for typical ranges of $T_r$ and lattice sizes $L$ our simulations are well within the scaling window. 

Even though the $O(4)$ universality class is strongly favored by these consistency checks, we find that our data do not constrain the critical exponents 
at the same level of accuracy as in the spin models. By optimizing the scaling collapse across different data sets we can, for instance, obtain the estimates $T_c\approx 17.395 \pm 0.02$, $\beta/\nu \approx 0.53 \pm 0.015$ and $1/\nu \approx 1.38 \pm 0.06$, which are consistent with the determination of $T_c$ above and the critical-exponent values from \cite{Engels:2014bra}. Here we have estimated the errors, which are always dominated by the systematic uncertainties, by sequentially excluding different lattice sizes from our analysis. We have also checked that the value of the critical Binder cumulant $\chi_{4}(T_c)=0.63 \pm 0.01$ agrees well with the values reported in the literature \cite{Pelissetto:2000ek,Springer2013}.

Next, we turn to the susceptibility $\chi_{|\phi|}$ for which the critical behavior at $J=0$
is determined by 
\begin{equation}
\chi_{|\phi|}(T,L=\infty)\sim|T_{r}|^{-\gamma},\quad~
\chi_{|\phi|}(T,L)=\left( \frac{L}{L_0}\right)^{\gamma/\nu} g_T\Big( \left( \frac{L}{L_0}\right)^{1/\nu}T_{r}\Big)\;. \label{eq:fss_susc_T}\\
\end{equation}
This relation is obtained from Eq. (\ref{eq:scalingrel}) by differentiating with respect to $J$ and using the hyperscaling relation $\gamma=\beta(\delta-1)$. We first study the $T$ dependence of $\chi_{|\phi|}$
for $L=48,64,96,128$ (Fig.~\ref{fig:susctempscaling}, left) and verify that the pseudo-critical transition temperature $T_{pc}(L)$, corresponding to the position of the peak, moves towards our estimate
of $T_c$ with increasing system size $L\to\infty$. We also confirm the finite-size scaling law (\ref{eq:fss_susc_T}), by plotting $L^{-\gamma/\nu}\chi_{|\phi|}$ as a function of $L^{1/\nu}T_r$ (with $\gamma=1.4531(104)$ taken from \cite{Engels:2014bra}) and verifying that the results collapse onto a single curve (Fig.~\ref{fig:susctempscaling}, right). While for temperatures $T>T_{pc}(L)$ (above the pseudo-critical transition temperature) we find good agreement between different data sets, such scaling breaks down below the pseudo-critical temperature. Since for $T<T_{pc}(L)$ the susceptibility receives additional contributions of massless Goldstone modes, one expects to find a linear scaling of the susceptibility with the volume, which has been discussed in detail in~\cite{Engels:1999wf} and is confirmed by our data.

\subsection{Static universality -- $J$ dependence}

So far we have verified the static critical behavior in the absence of explicit symmetry breaking, using the absolute value $|\phi|$ as an approximate order parameter. We now proceed by setting the temperature $T$ to $T_c\approx 17.3925$ and study the dependence on the external 
field $J$. Again, we first consider the order parameter itself and verify the power-law behavior 
$\phi\sim|J|^{1/\delta}$ and the magnetic scaling (\ref{eq:fss_order_J}), taking $\delta= 4.824(9)$ from \cite{Engels:2014bra}.
Fig.~\ref{fig:ordermagneticscaling} summarizes these results. In the presence of an explicit symmetry breaking we can distinguish between the absolute 
value $\langle|\phi|\rangle$ and the component $\langle\phi_J\rangle$ in direction of
the external field. In principle both exhibit identical universal properties, but as $J$ becomes smaller the quantities
differ unless the system size $L$ is simultaneously increased by a sufficient amount. What is striking is that 
both quantities independently show magnetic scaling, whereby the data points for $L^{\beta/\nu}\phi$ 
collapse onto single but distinct curves (Fig.~\ref{fig:ordermagneticscaling}, right).

\begin{figure*}
\begin{minipage}{0.45\textwidth}
\includegraphics[width=\textwidth]{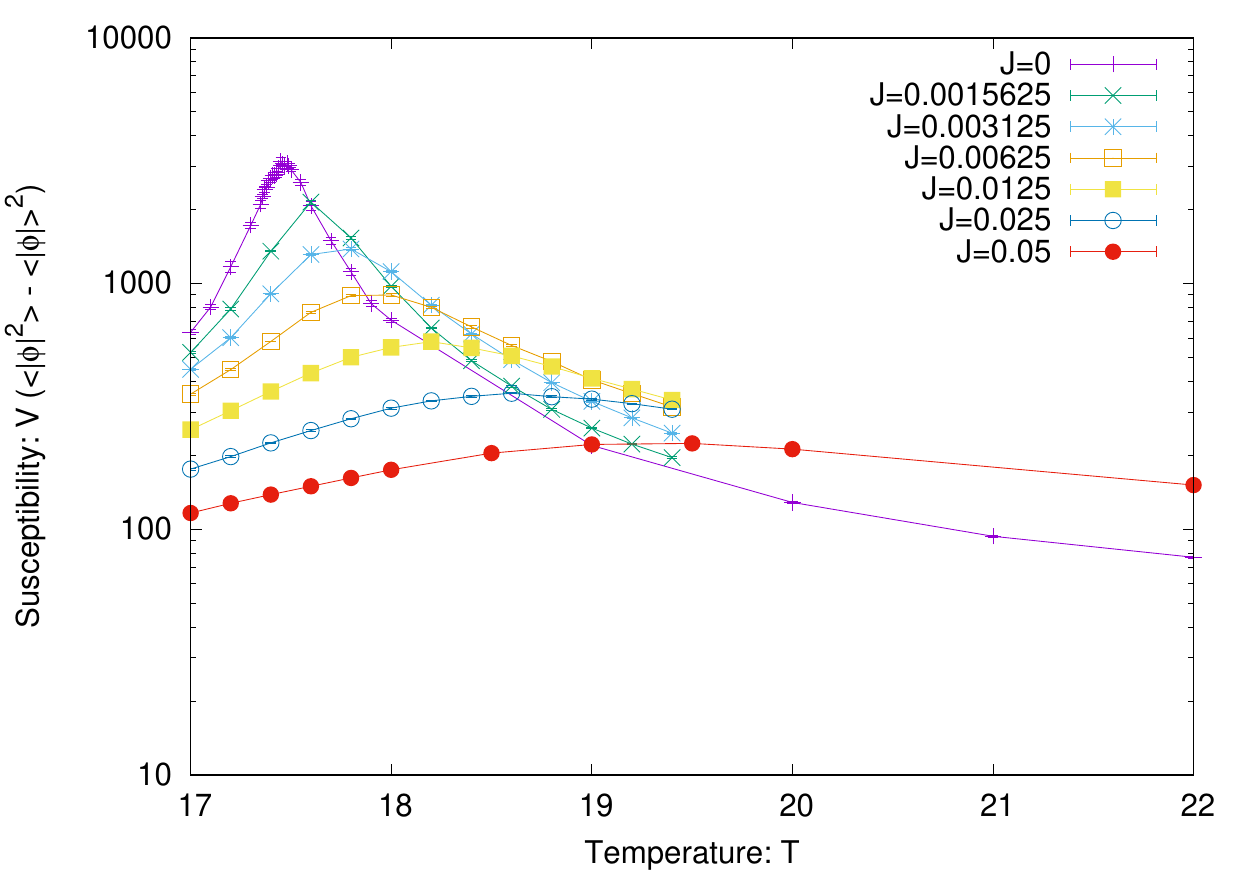}
\end{minipage}
\hspace{1cm}
\begin{minipage}{0.45\textwidth}
  \includegraphics[width=\textwidth]{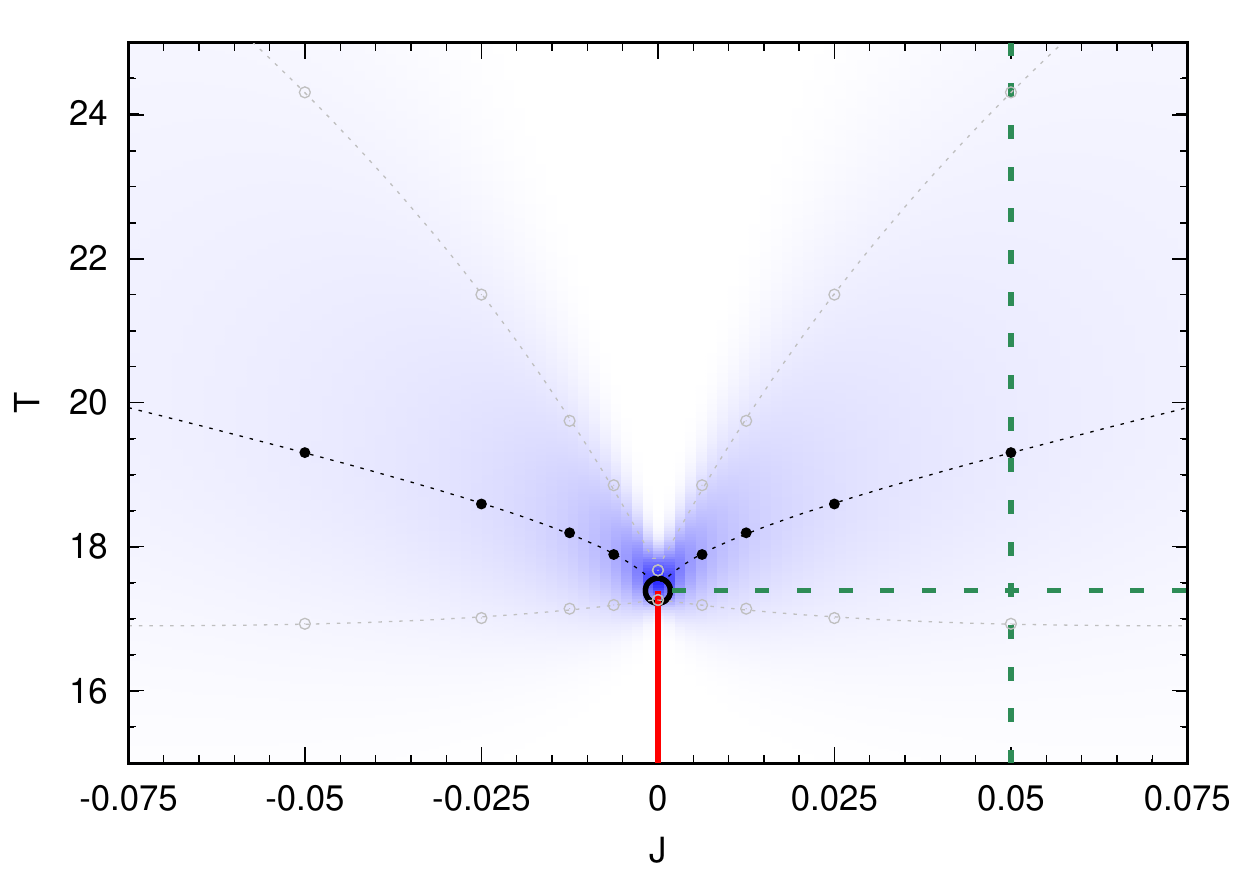}
\end{minipage}
\caption{Left: Temperature dependence of magnetic susceptibility $\chi_{|\phi|}$ for different values of $J$ on a $L=64$ lattice. Pseudocritical temperatures are estimated from location of maximum. Right: Tentative phase diagram of $3+1$ dimensional $O(4)$ model in $J-T$ plane. Blue shading denotes the interpolated magnitude of the susceptibility $\chi_{|\phi|}$ at each point while dashed black and grey lines mark its maximum and inflection points respectively. Computation of spectral functions were carried out along dashed green lines. 
\label{fig:phasediag}} 
\end{figure*}

Similarly, we also verify the $J$ dependent critical properties of the susceptibility, which are given by
\begin{equation}
\chi(J,L=\infty)\sim|J|^{-\frac{\gamma}{\beta\delta}},\quad~
\chi(J,L)=\left( \frac{L}{L_0}\right)^{\gamma/\nu} g_J\Big( \left( \frac{L}{L_0}\right)^{\frac{\beta\delta}{\nu}}\frac{J}{J_0}\Big)\;, \label{eq:fss_susc_J}\\
\end{equation}
where the symbol $\chi$ is a generic placeholder for $\chi_{|\phi|}$ and $\chi_{J}$. 
Just as the order parameters $|\phi|$ and $\phi_J$, the corresponding susceptibilities  
$\chi_{|\phi|}$ and $\chi_{J}$ can be distinguished and independently collapse onto distinct critical scaling functions, which 
are different  in the finite size scaling regime (for small $L^{\beta\delta/\nu}J$) but merge for sufficiently large $L^{\beta\delta/\nu}J$ 
(see Fig.~\ref{fig:suscmagneticscaling}, left). 

In the presence of a non-zero explicit symmetry breaking term $J$, we can also distinguish between the longitudinal ($\sigma)$ and 
transverse ($\pi$) components $\chi_{\sigma/\pi}$ of the susceptibility, where $\chi_{\sigma}=\chi_{J}$ (cf.~Eq.~(\ref{eq:chi_J_and_abs})) 
and  $\chi_{\pi}$ is given by
\begin{equation}
\chi_{\pi} = \frac{V}{N-1} \left[ \langle |\phi|^2 \rangle  - \langle  \phi_{J}^2 \rangle\right]\;.
\end{equation}
Independent finite size scaling of longitudinal and transverse susceptibilities is again observed (Fig.~\ref{fig:suscmagneticscaling}, right). 
For sufficiently large values of $L^{\beta\delta/\nu}J$, i.e. close to the infinite volume limit, both curves are expected to approach the scaling behavior in Eq.~(\ref{eq:fss_susc_J}), with a universal amplitude ratio $\chi_{\sigma}/\chi_{\pi} = 1/\delta$ (see \cite{Engels:2003nq}). 
This is nicely confirmed by our data. Even though most of our data points are outside the infinite volume scaling regime $(L^{\beta\delta/\nu}J \to \infty)$ where $\chi$ exhibits a power law dependence (cf.~Eq.~(\ref{eq:fss_susc_J})), we also observe that the finite size scaling regime extends to much smaller values of $L^{\beta\delta/\nu}J$. In particular, for very small values of $L^{\beta\delta/\nu}J$ the two scaling curves become almost indistinguishable, as the distinction between longitudinal and transverse components becomes less and less meaningful. 

When comparing our results for $\phi_J$ and $\chi_\sigma$ to the universal finite size scaling functions determined in \cite{Engels:2014bra} for the $O(4)$ spin model (displayed as solid lines in Fig.~\ref{fig:suscmagneticscaling} and the right panel of Fig.~\ref{fig:ordermagneticscaling}), good agreement is found across the entire range where the parametrization is available.

\subsection{Static universality -- Conclusion}
We conclude from all of the above that both, the $T$ and $J$ dependent static critical properties are indeed correctly reproduced in our classical statistical simulations, and we can now safely proceed to study real-time properties. In order to set the stage for our study of real-time correlation functions, we finally sketch the phase diagram in the
$J$-$T$ plane. For this purpose we compute the $T$ dependence of the susceptibility $\chi_{|\phi|}$ on a $L=64$ lattice for several different
values of $J$ and carry out an interpolation for regions in between data points. For each line of constant $J$ we then
determine the maximum of $\chi_{|\phi|}$, which serves as an estimate for the pseudo-critical
temperature along this line, and the inflection points. Our results for the phase-diagram are shown in Fig.~\ref{fig:phasediag}, where the color coding indicates the magnitude of the susceptibility.  Most importantly, the horizontal and vertical dashed green lines correspond to the values of $T,J$ considered in our study of spectral functions.

\section{Results: Spectral functions}
\label{sec:spectral_f}
We now study spectral functions $\rho(t,t')$ which we can directly extract as a function of the real-time variables $t,t'$ according to the procedure 
discussed in Secs.~\ref{sec:background}, \ref{sec:setup}. We prepare $N_{conf}$ independent initial configurations for each choice of parameters and 
evolve each of them up to maximum time $t_{Max}$ (typically $\sim10^{4} - 10^{5} a_s$), recording the evolution of $\sum_i \phi^{a}_i$ and $\sum_{i} \pi^{a}_i$. Based on 
this data, we then construct the un-equal time correlation function as a function of $t-t'$ according to the right-hand side of 
Eq.~(\ref{eq:clrhozerop}), evaluated separately for each configuration, while immediately averaging over different positions $t+t'$ in the real-time evolution. 
Statistical averages and errors of the spectral function $\rho(t-t')$ and its Fourier transform $\rho(\omega)$ are computed from averaging over the 
ensemble of typically $N_{conf}=32$ independent configurations. If not stated otherwise, all results have been obtained on $L=64$ lattices, and we have 
checked at the example of a few data points that except for the immediate vicinity of the critical point our results remain unchanged when going to larger 
lattices. We further emphasize that in order to properly distinguish the longitudinal ($\pi$) and transverse ($\sigma$) field components in our simulations, 
we always have to introduce a non-zero explicit symmetry breaking $J$. 

\subsection{Spectral functions in the crossover regime -- $T$ dependence}
\begin{figure*}
\begin{minipage}{0.33\textwidth}
\includegraphics[width=\textwidth]{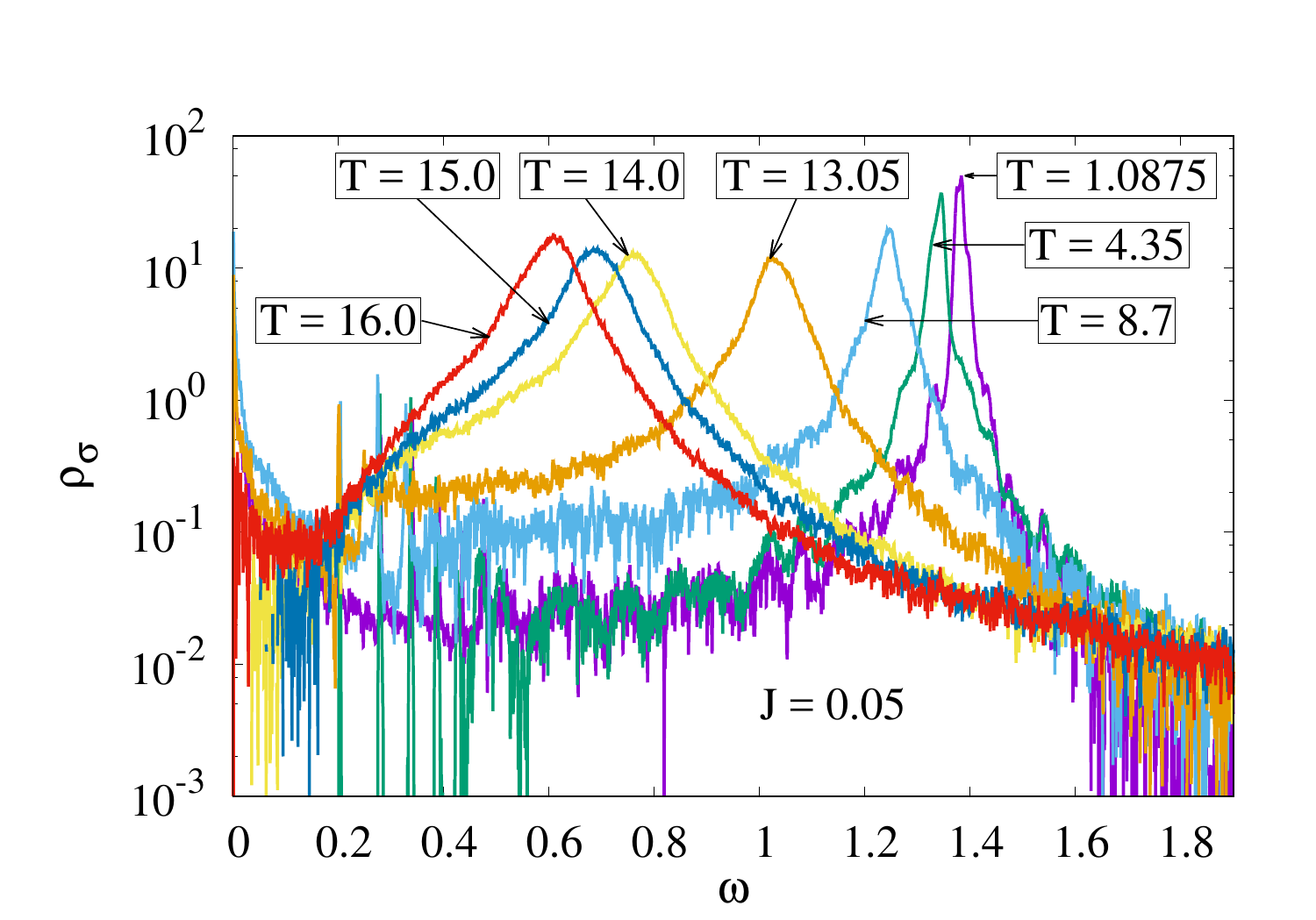}
\end{minipage}
\begin{minipage}{0.33\textwidth}
\includegraphics[width=\textwidth]{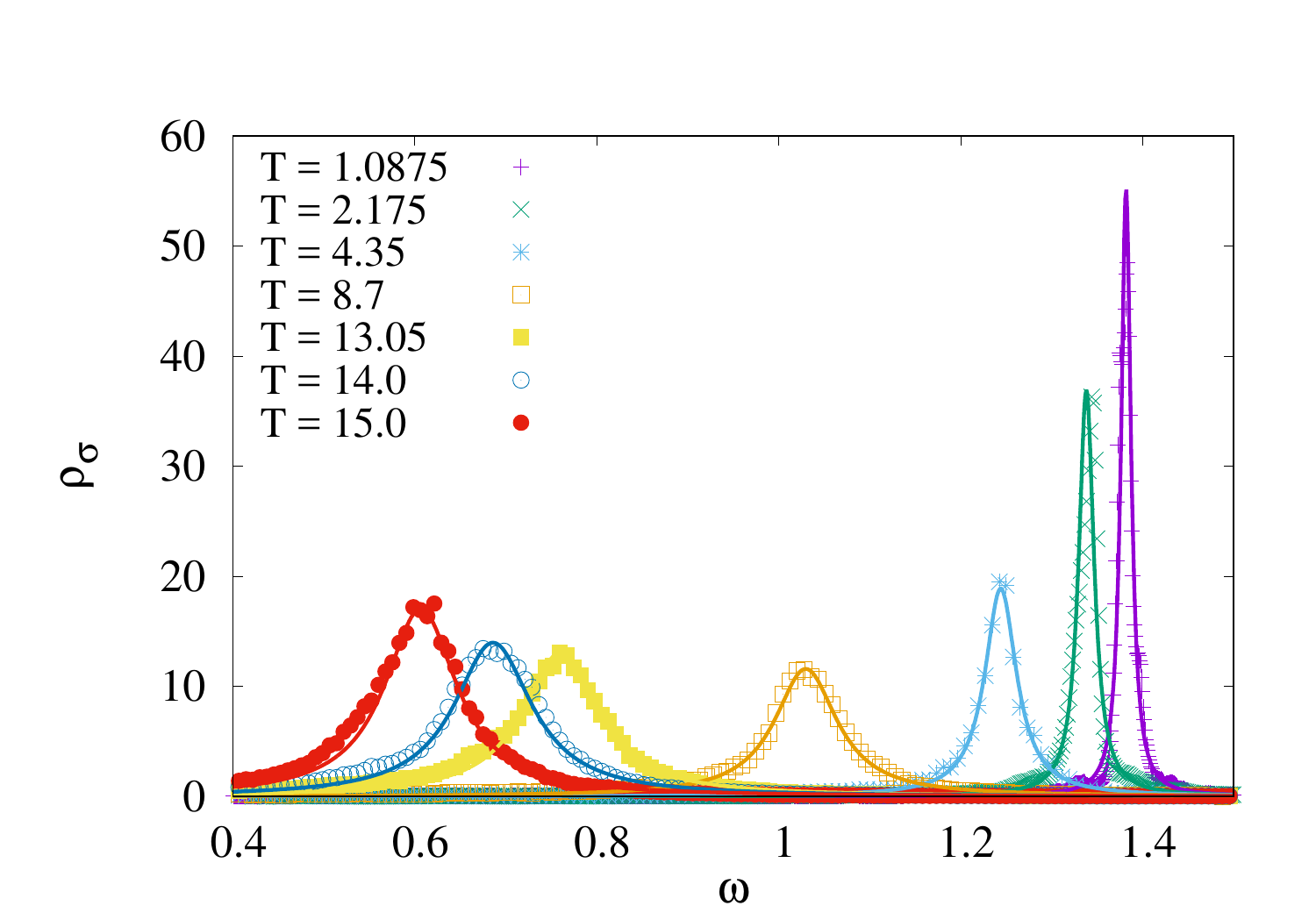}
\end{minipage}
\begin{minipage}{0.33\textwidth}
\includegraphics[width=\textwidth]{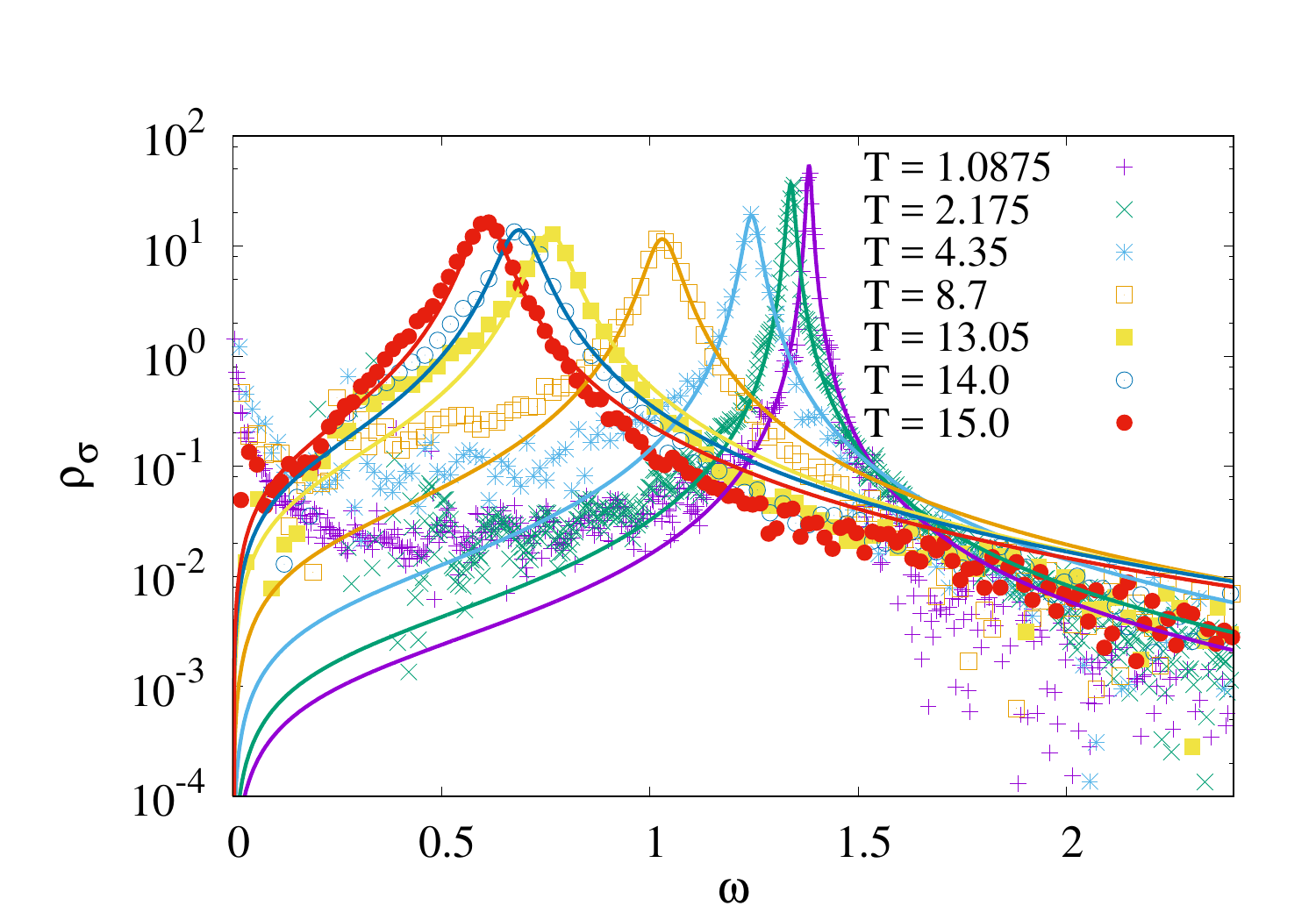}
\end{minipage}
\begin{minipage}{0.33\textwidth}
\includegraphics[width=\textwidth]{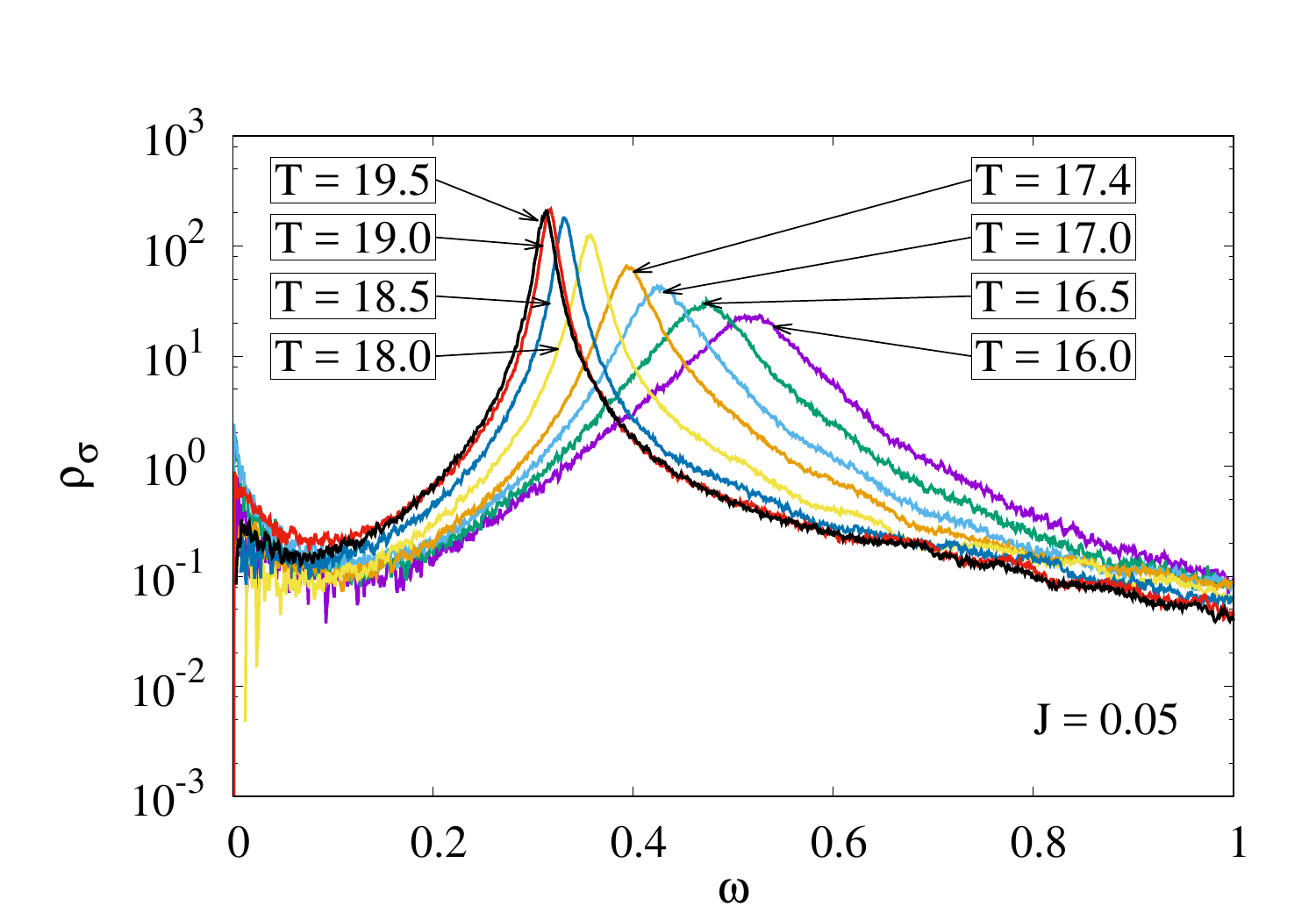}
\end{minipage}
\begin{minipage}{0.33\textwidth}
\includegraphics[width=\textwidth]{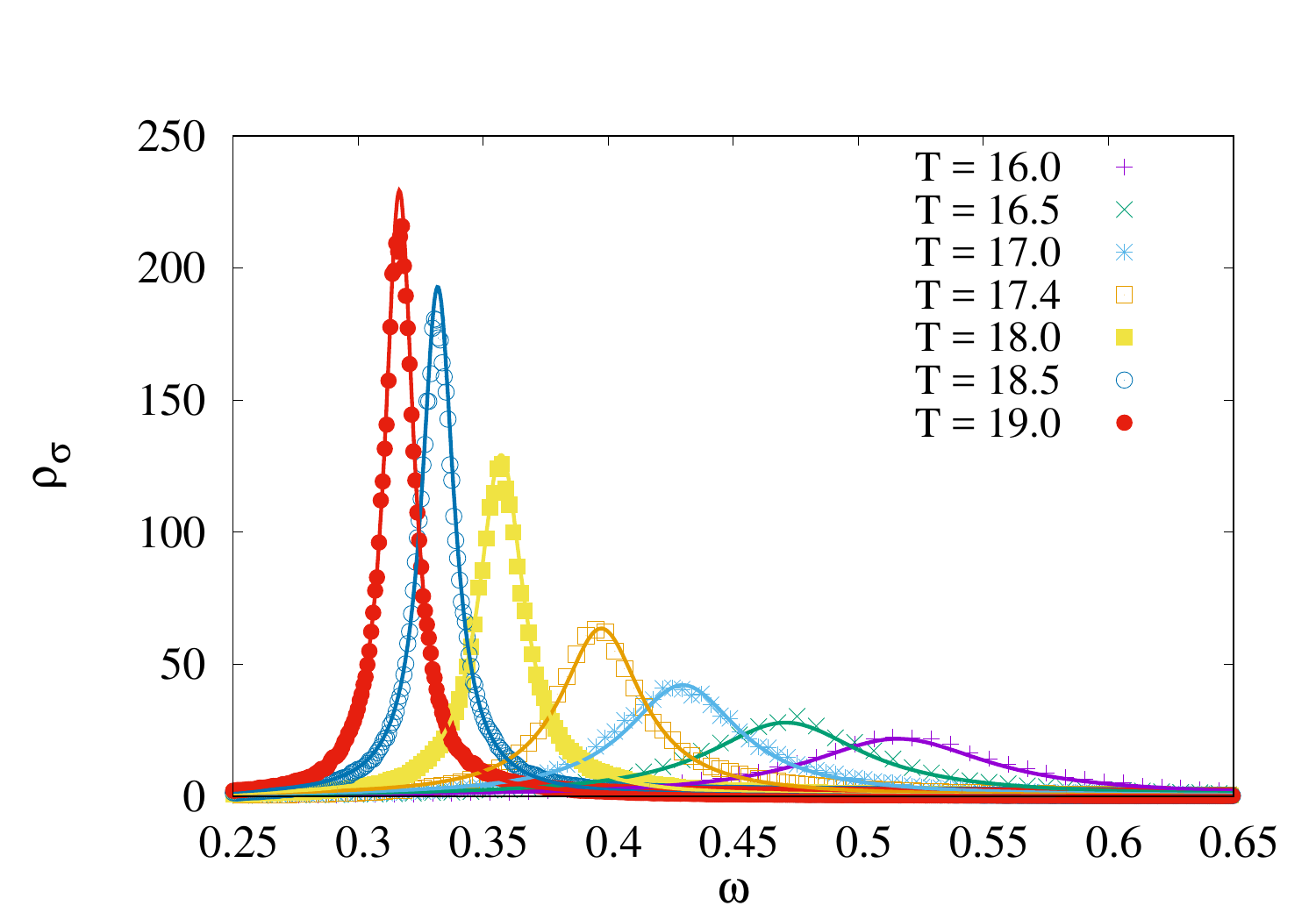}
\end{minipage}
\begin{minipage}{0.33\textwidth}
\includegraphics[width=\textwidth]{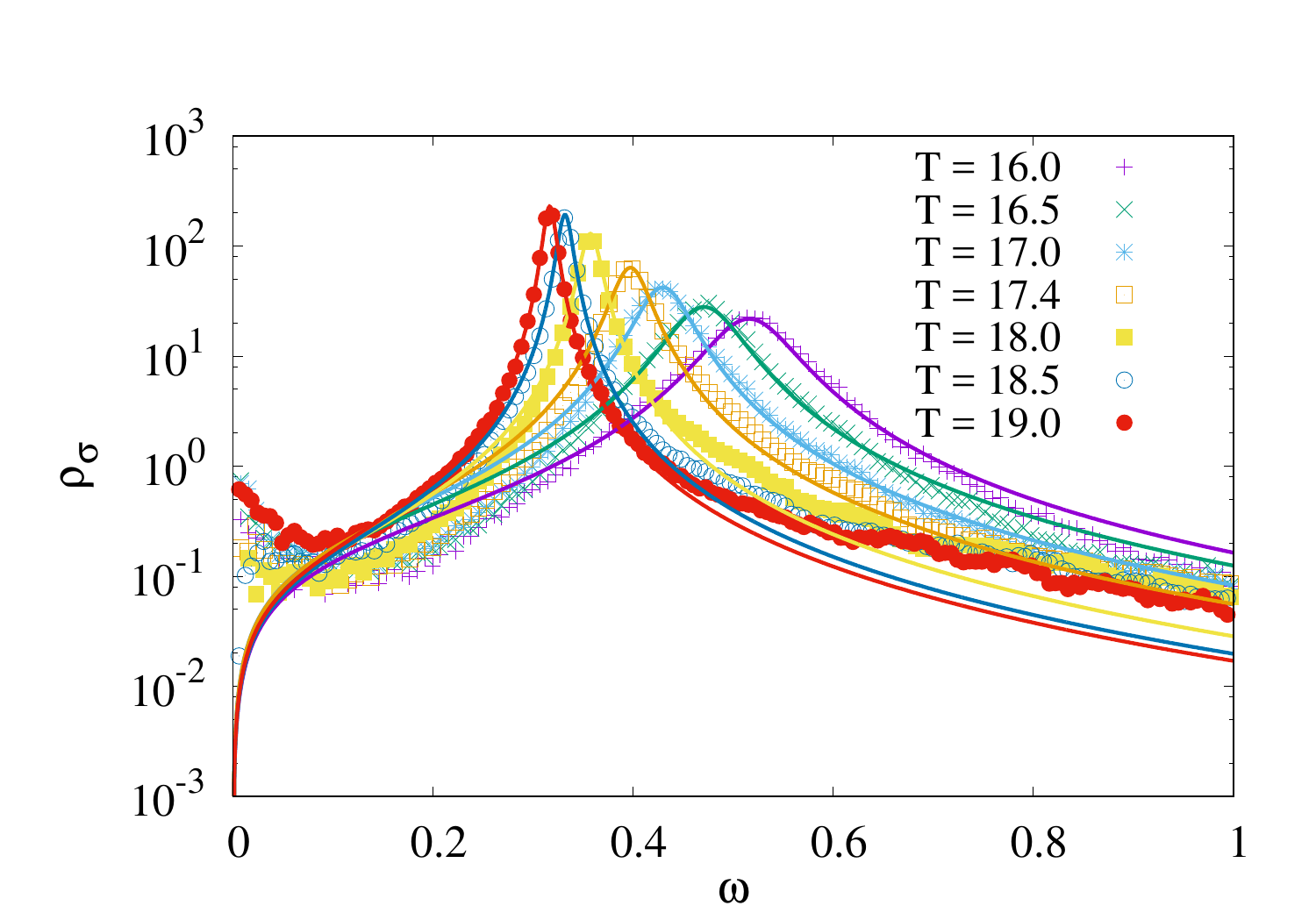}
\end{minipage}
\begin{minipage}{0.33\textwidth}
\includegraphics[width=\textwidth]{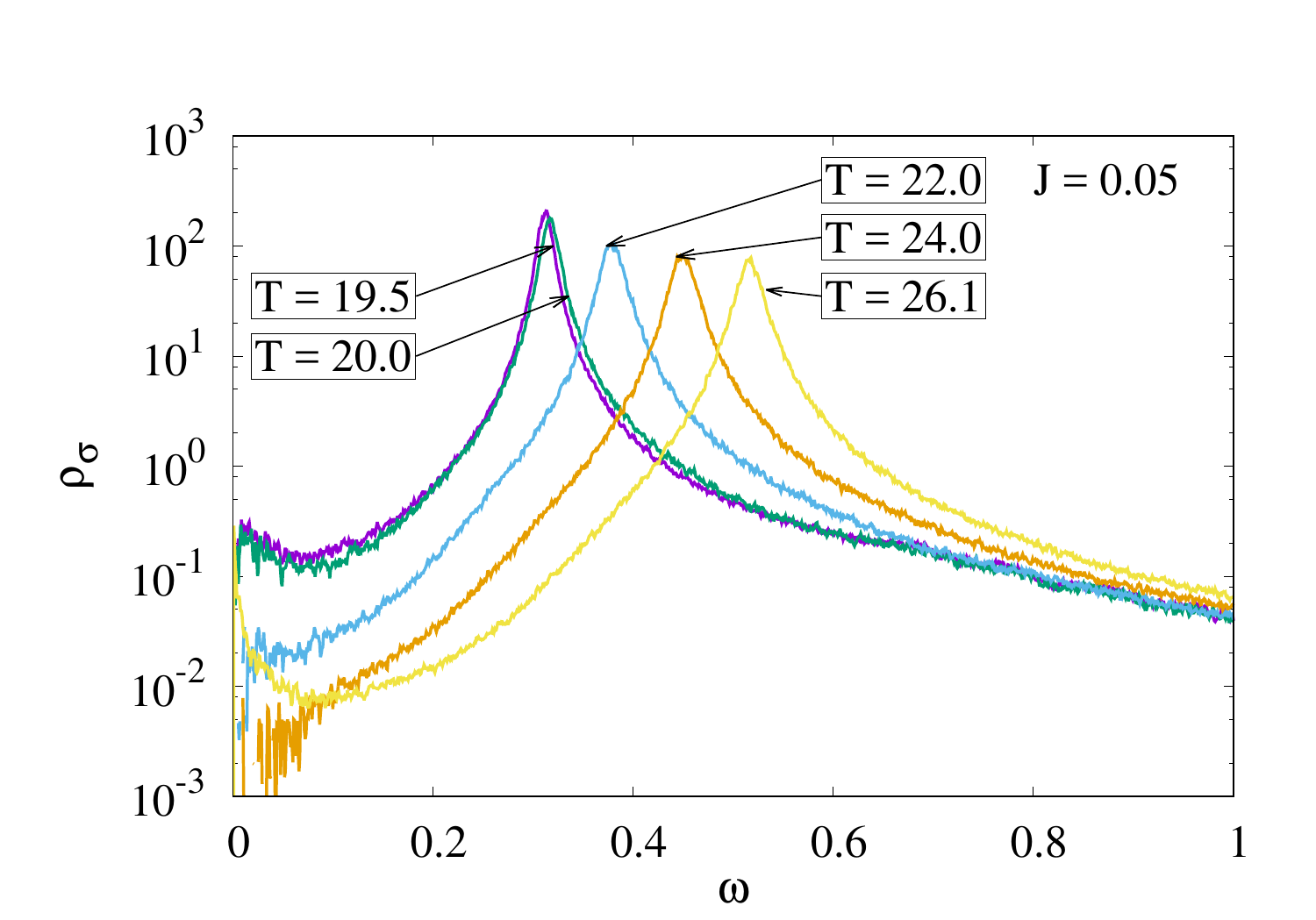}
\end{minipage}
\begin{minipage}{0.33\textwidth}
\includegraphics[width=\textwidth]{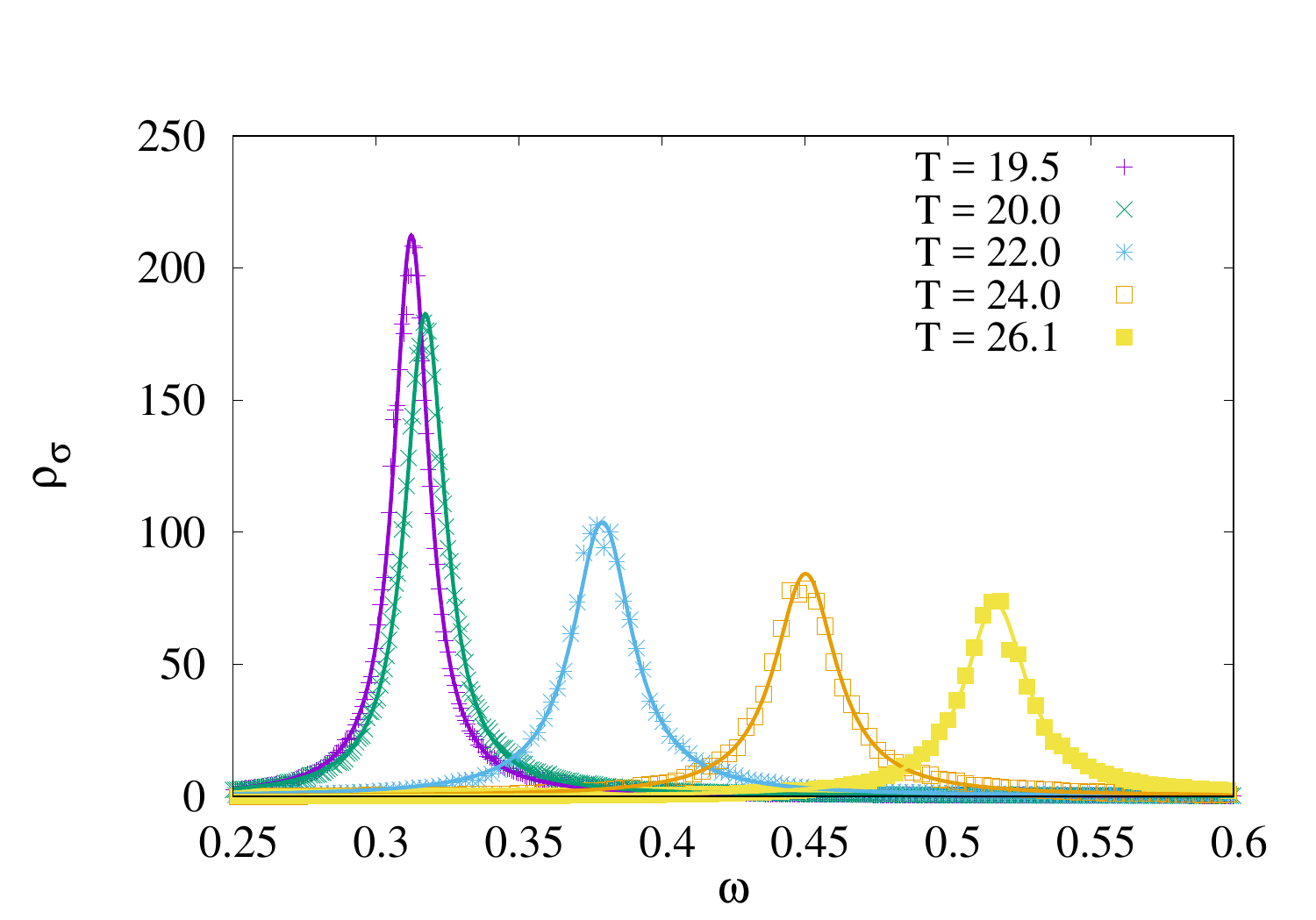}
\end{minipage}
\begin{minipage}{0.33\textwidth}
\includegraphics[width=\textwidth]{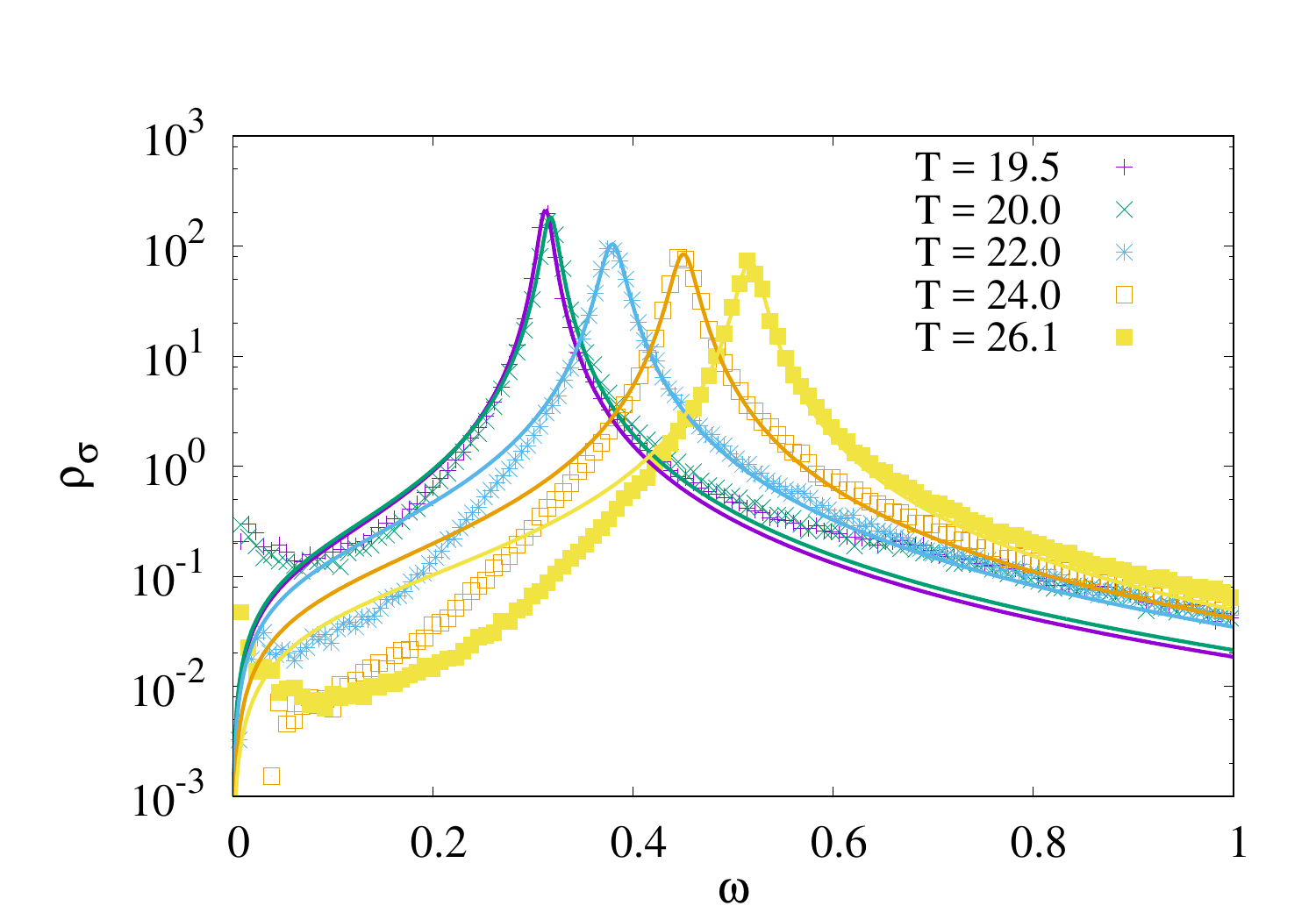}
\end{minipage}
\caption{\label{fig:sigmaJsf} $T$ dependence of $\sigma$ spectral functions for $J=0.05$ below (top row), around
(middle row) and above (bottom row) the pseudocritical temperature $T_{pc}\approx19.5$.
Left column: Raw data only. Middle column: Data with Breit-Wigner fits (\ref{eq:breit_wigner}).
Error bars are of the order of the pointsize. Several intermediate points are not displayed
for better visibility but are considered for the fits. Right column: Same as middle column
but with log-scale.} 
\end{figure*}

\begin{figure*}
\begin{minipage}{0.33\textwidth}
\includegraphics[width=\textwidth]{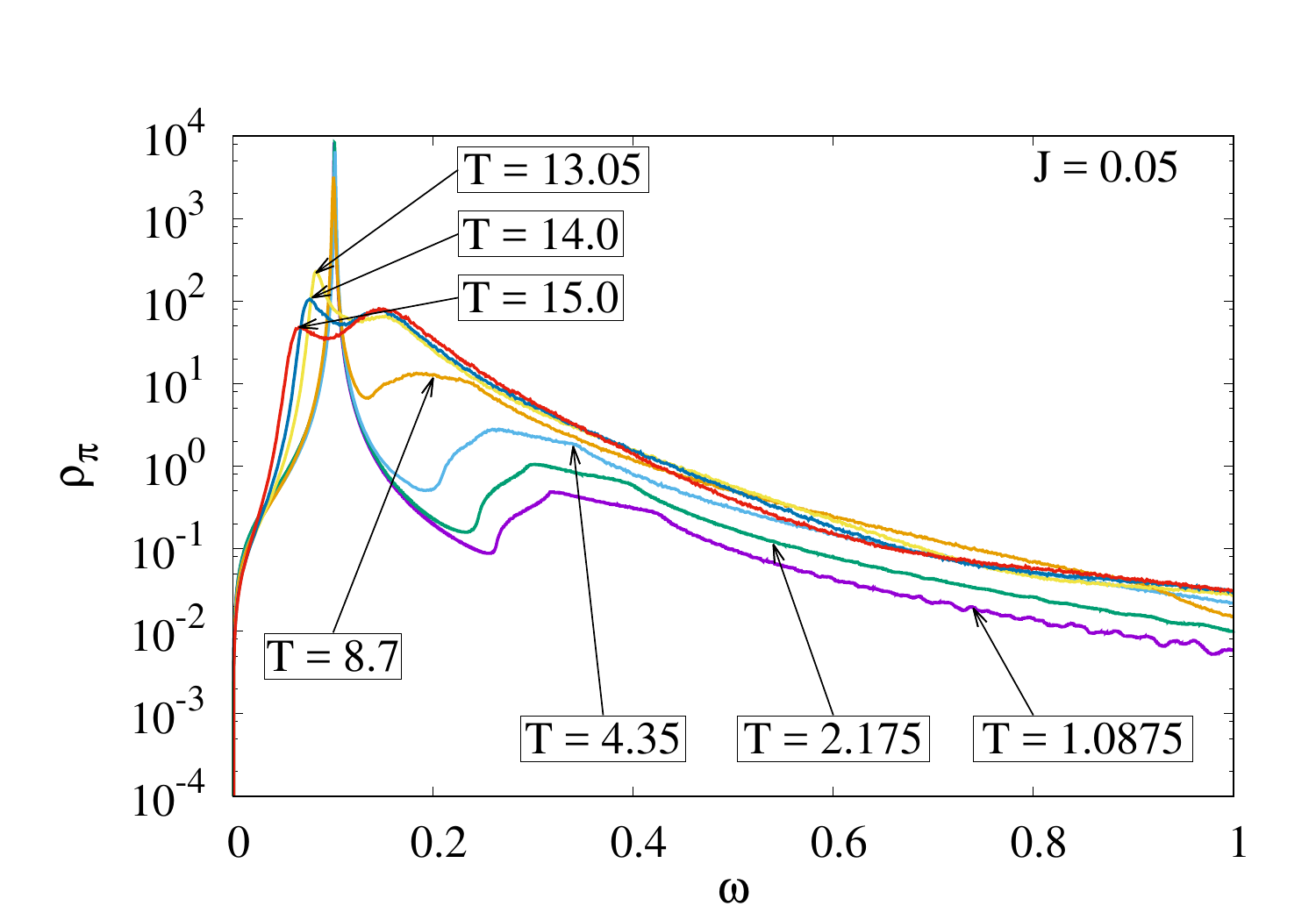}
\end{minipage}
\begin{minipage}{0.33\textwidth}
\includegraphics[width=\textwidth]{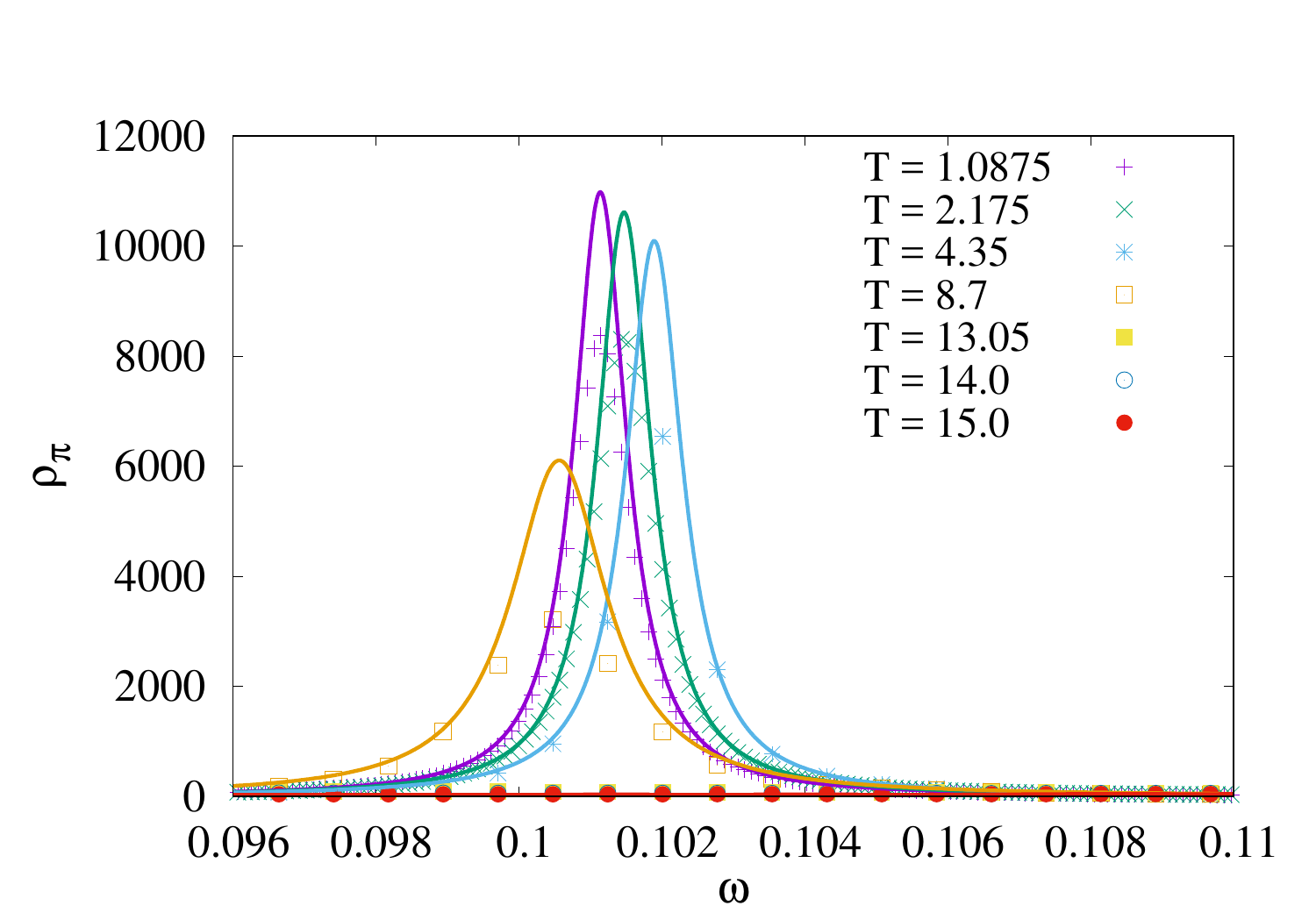}
\end{minipage}
\begin{minipage}{0.33\textwidth}
\includegraphics[width=\textwidth]{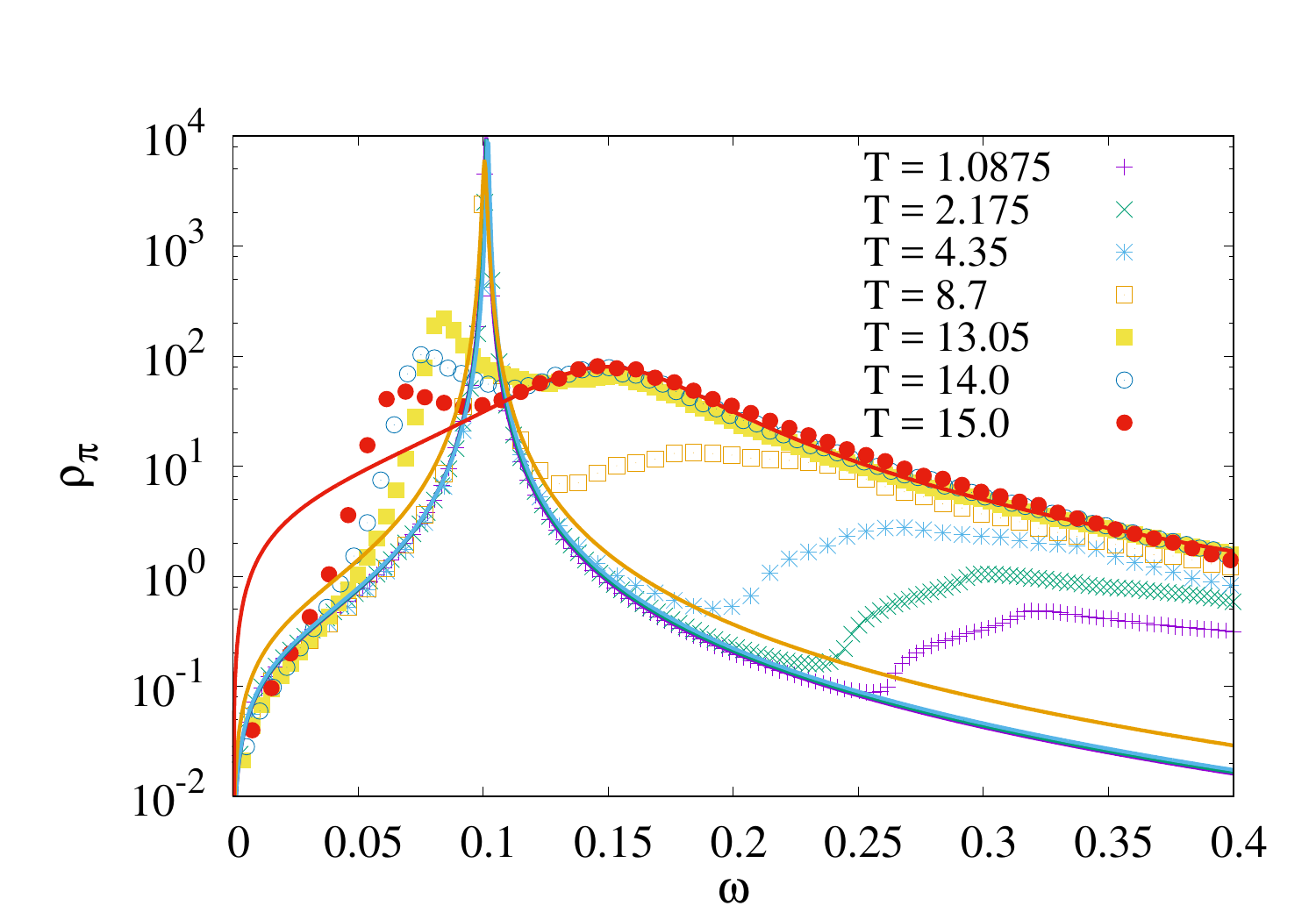}
\end{minipage}
\begin{minipage}{0.33\textwidth}
\includegraphics[width=\textwidth]{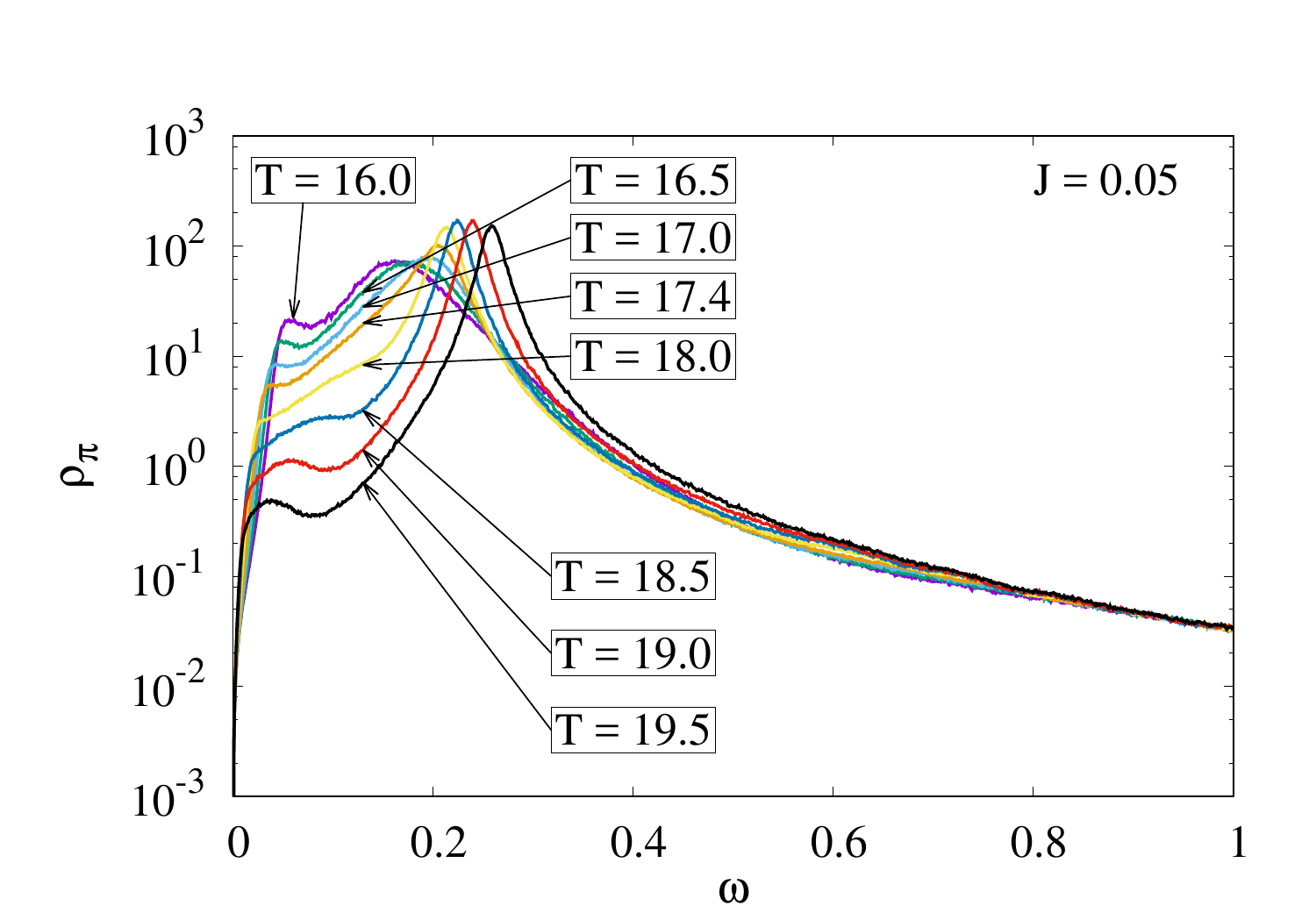}
\end{minipage}
\begin{minipage}{0.33\textwidth}
\includegraphics[width=\textwidth]{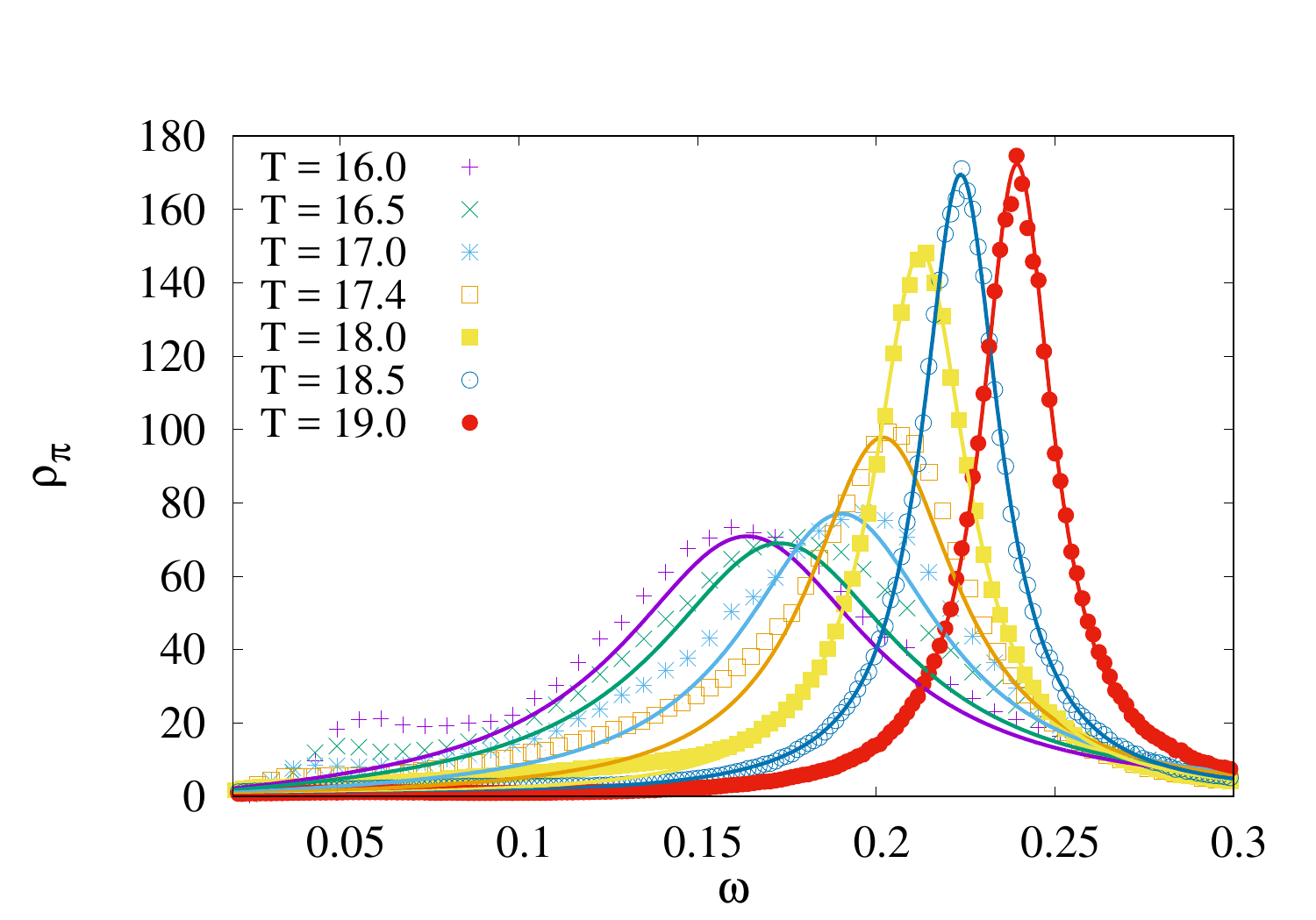}
\end{minipage}
\begin{minipage}{0.33\textwidth}
\includegraphics[width=\textwidth]{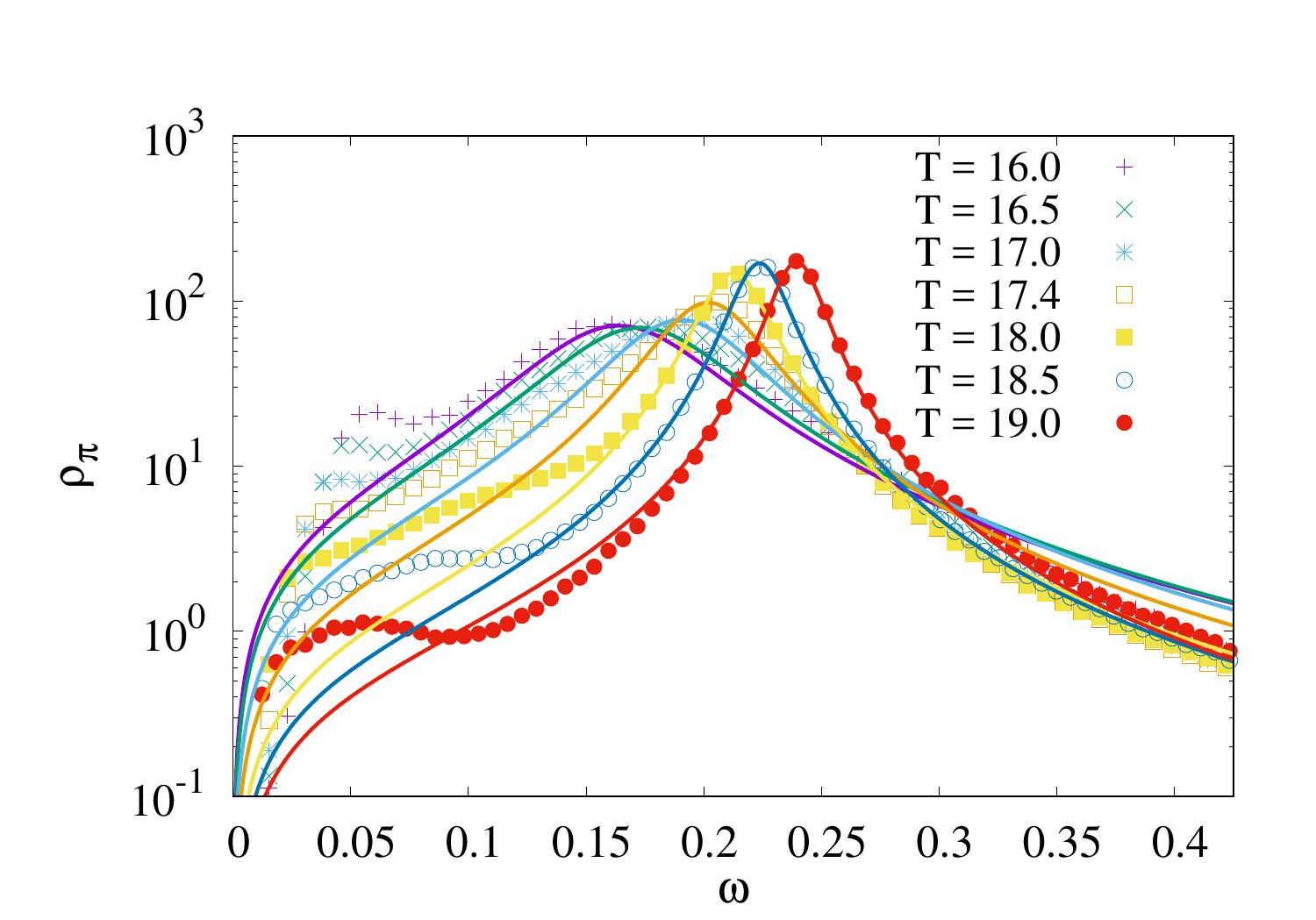}
\end{minipage}
\begin{minipage}{0.33\textwidth}
\includegraphics[width=\textwidth]{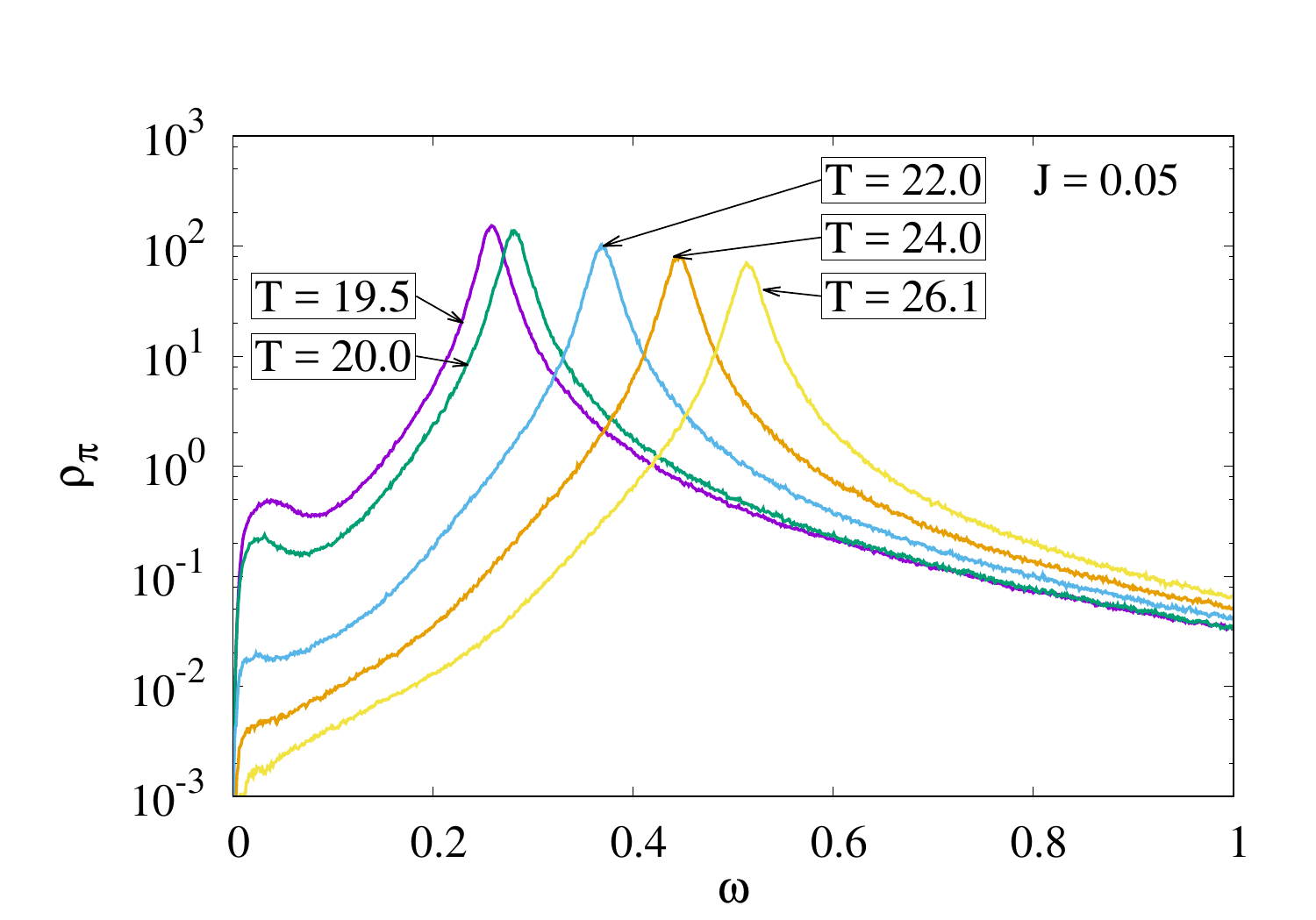}
\end{minipage}
\begin{minipage}{0.33\textwidth}
\includegraphics[width=\textwidth]{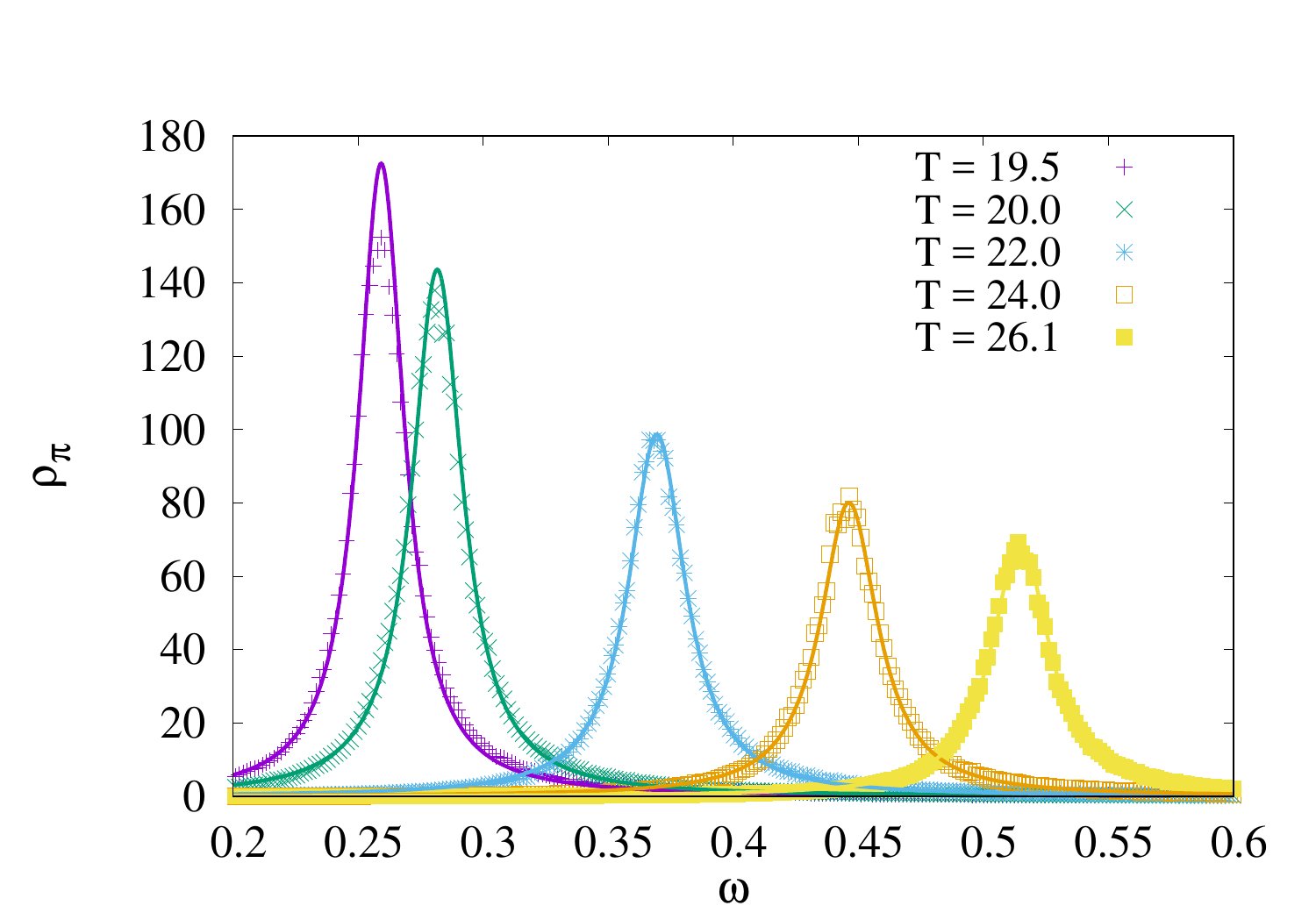}
\end{minipage}
\begin{minipage}{0.33\textwidth}
\includegraphics[width=\textwidth]{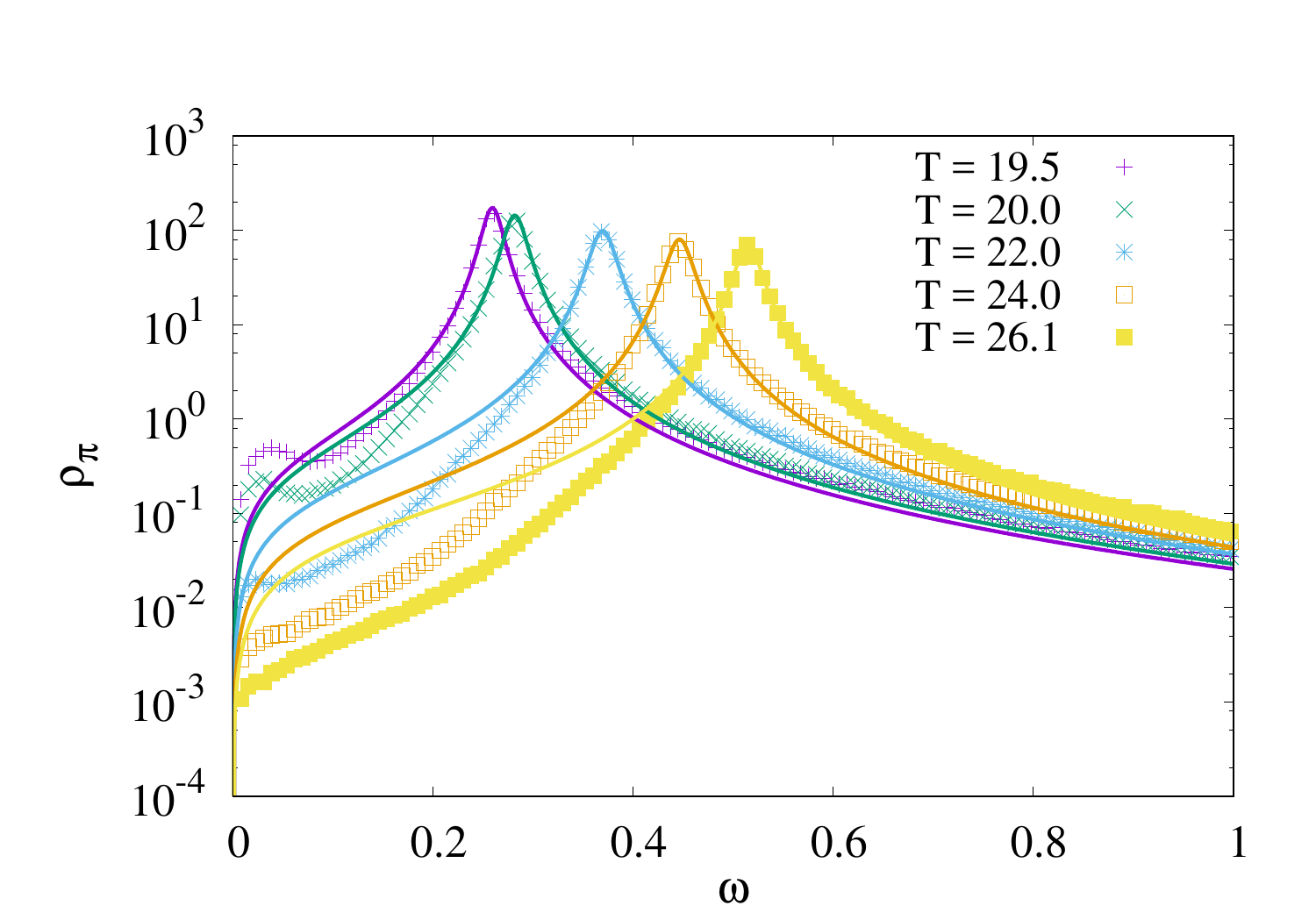}
\end{minipage}
\caption{\label{fig:piJsf} $T$ dependence of $\pi$ spectral functions for $J=0.05$ below (top row), around
(middle row) and above (bottom row) the pseudocritical temperature $T_{pc}\approx19.5$.
Left column: Raw data only. Middle column: Data with Breit-Wigner fits (\ref{eq:breit_wigner}) to
the larger peak, where applicable. Error bars are of the order of the pointsize. Several
intermediate points are not displayed for better visibility but are considered for the fits.
Right column: Same as middle column but with log scale.} 
\end{figure*}

Before we turn our attention to the behavior in the vicinity of the critical point, we first present results for the temperature dependence of the spectral function at a relatively large explicit symmetry breaking $J=0.05$ as indicated by the vertical dashed line in the phase-diagram (cf.~Fig.~\ref{fig:phasediag}). We note that at such large values of $J$ the transition is a relatively smooth cross over, with a pseudo-critical transition temperature $T_{pc}(J=0.05)\approx 19.5$. We have collected all of our results for the $\sigma$- and $\pi$-components of the spectral functions in Figs.~\ref{fig:sigmaJsf} and \ref{fig:piJsf}. Different rows in Figs.~\ref{fig:sigmaJsf} and \ref{fig:piJsf} show the spectral functions in different temperature regimes, starting from very low temperatures in the top row, to temperatures just below the pseudo-critical transition temperature $T_{pc}$ in the middle row, all the way to temperatures above the pseudo-critical transition temperature $T_{pc}$ in the bottom row. Different columns in Figs.~\ref{fig:sigmaJsf} and \ref{fig:piJsf}, all show the same data for the spectral functions but plotted in different ways in order to highlight the various features more clearly. 


\begin{figure*}[t!]
\begin{minipage}{0.5\textwidth}
\includegraphics[width=\textwidth]{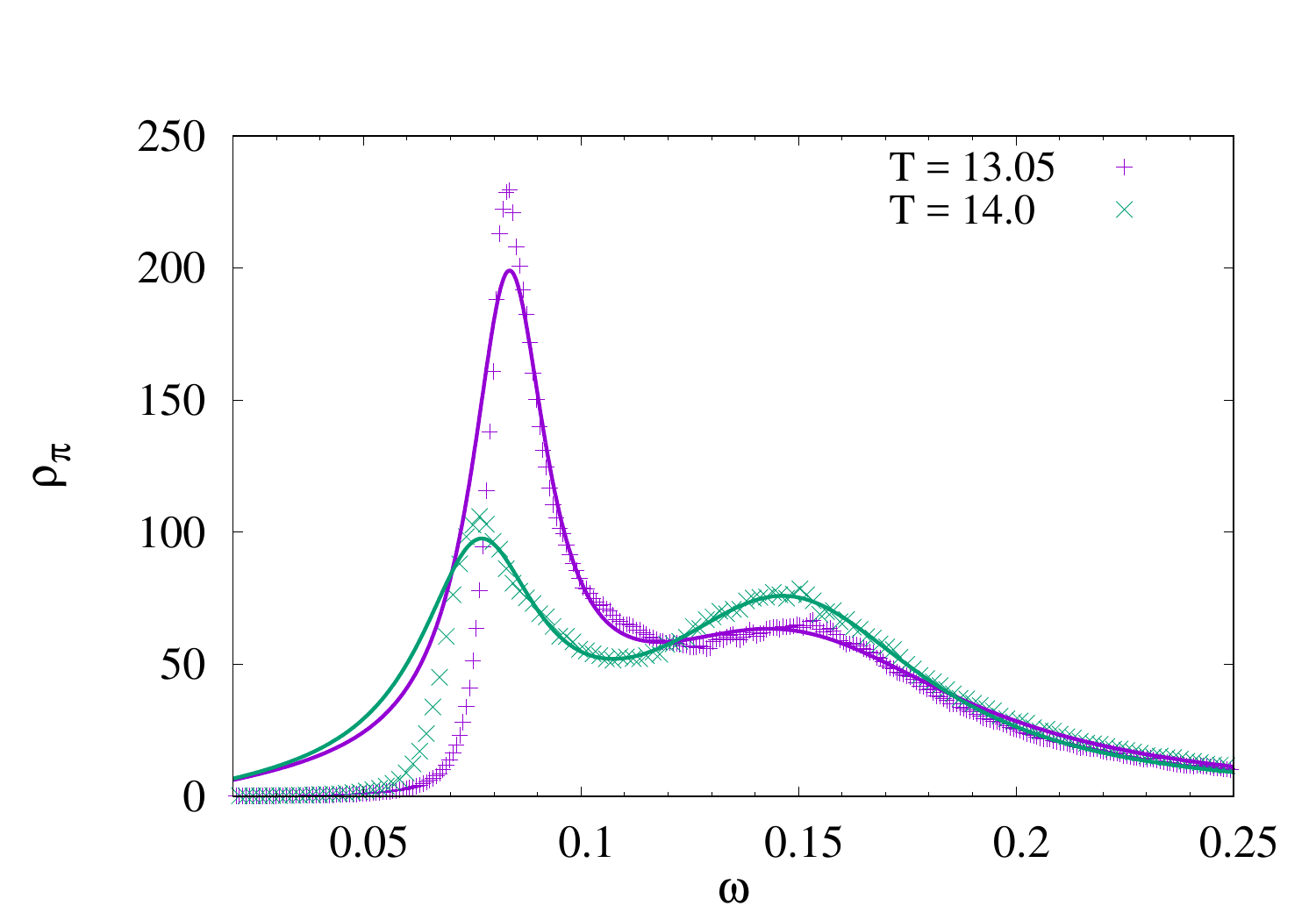}
\end{minipage}
\begin{minipage}{0.5\textwidth}
\includegraphics[width=\textwidth]{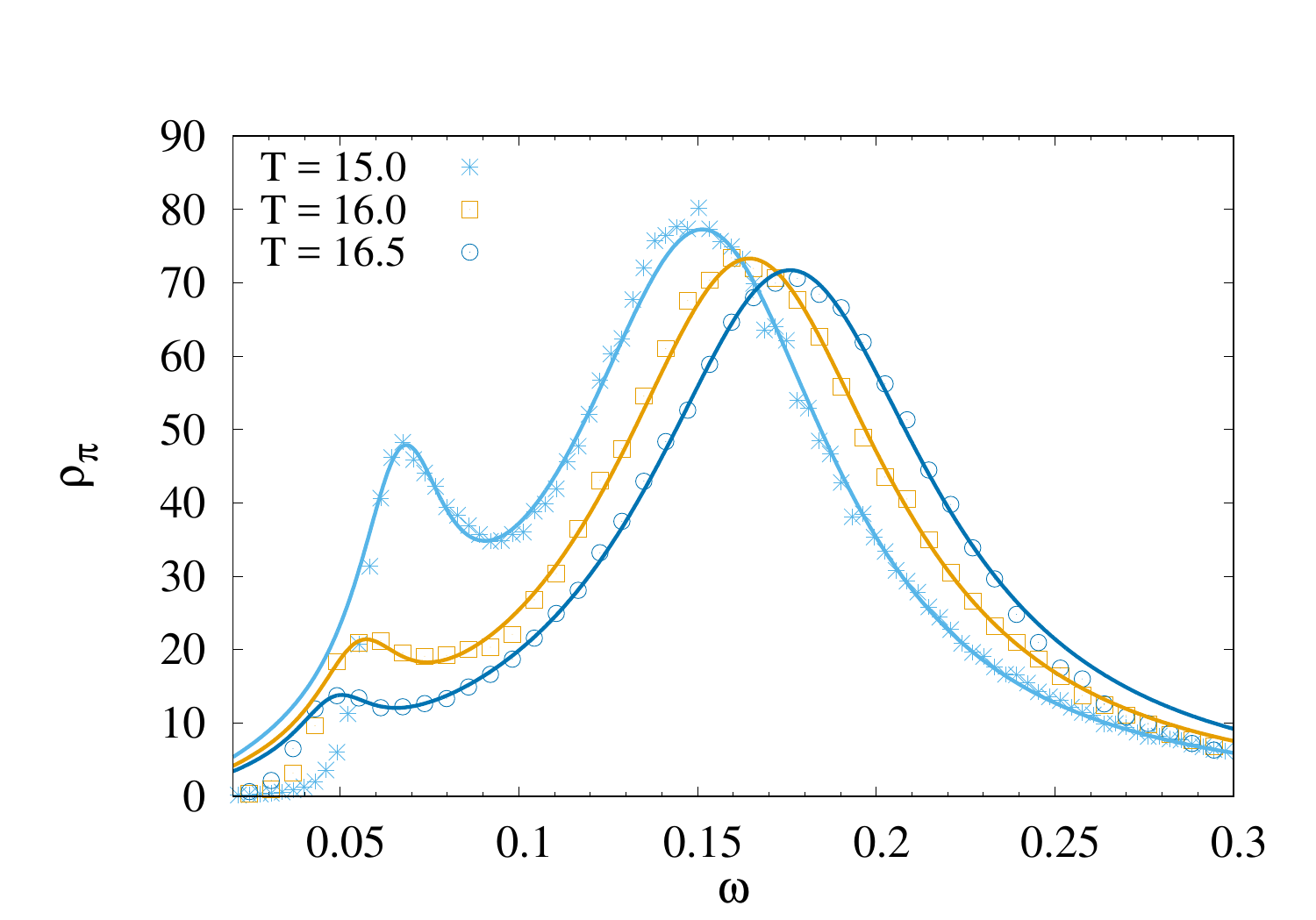}
\end{minipage}
\caption{ \label{fig:DoublePeak} Spectral function $\rho_{\pi}(\omega)$ for $J=0.05$ in the intermediate temperature range where two peaks are visible. Solid lines represent fits to a Breit-Wigner distribution with two resonance (\ref{eq:double_breit_wigner}).} 
\end{figure*}

Before we turn to our simulation results we note that in the limit $T\to0$ thermal fluctuations are suppressed, and the classical-statistical description reduces to the mean-field limit, where
\begin{eqnarray}
\label{eq:RhoMF}
\rho_{\pi/\sigma}(\omega,T=0,p=0)= 2\pi~\text{sign}(\omega)~\delta(\omega^2 -m_{\pi / \sigma}^2)\;, 
\end{eqnarray}
and for $\lambda J^2 \ll -m^6$ (with $m^2<0$) one has 
\[
m_{\sigma}^2 \simeq -2 m^2 + \sqrt{\frac{3}{2 N} \frac{\lambda J^2}{-m^2}}
\, , \;\; \mbox{and} \;\;  m_{\pi}^2\simeq \sqrt{\frac{1}{6 N} \frac{\lambda J^2}{-m^2}}\, .
\]
With our parameters, for comparison with the corresponding peaks at the lowest temperature ($T= 1.0875$) in Figs.~\ref{fig:sigmaJsf} and \ref{fig:piJsf}, this then amounts to $m_\sigma \simeq 1.425$ and $m_\pi \simeq 0.101$.

Clearly, however, this trivial behavior is a result of the classical-statistical approximation, which misses all quantum effects that would otherwise become important in the low-temperature regime $T \ll T_{pc}$, such as for instance the decay process $\sigma \to 2 \pi$ which would result in strong modifications of the vacuum spectral functions. Nevertheless, even though an entirely classical description is thus not particularly well suited for the description of the low-temperature physics, it is still interesting to investigate the behavior of the classical-statistical spectral functions in this regime.

Based on our simulation results in the low-temperature regime $T<15$ shown in the top row of Figs.~\ref{fig:sigmaJsf} and \ref{fig:piJsf}, we find that at low temperatures the classical-statistical spectral functions, exhibit the expected quasi-particle behavior where towards the lowest temperature $T=1.0875$ the masses of $\sigma$ and $\pi$ are already nicely seen to approach the mean-field estimates below Eq.~(\ref{eq:RhoMF}), up to minute shifts and some collisional broadening due to the small but finite residual temperature. Besides these quasi-particle peaks, the $\pi$ spectral function also shows an additional cusp at higher frequencies which at very low temperatures occurs approximately for frequencies $\omega \sim 3m_{\pi}$ and should be attributed to a multi-pion excitation. By further increasing the temperature, the quasi-particle peaks remain, however the mass of $\sigma$ becomes lighter as the vacuum expectation value of the $\sigma$ field decreases. Even though the width of $\sigma$ spectral function also increases, it turns out that except for a small enhancement at low frequency, the spectral function of the $\sigma$ mode can still be well described in terms of a single Breit-Wigner resonance
\begin{equation}
\rho(\omega)=\frac{\omega\Gamma}{(\omega^2-m^2)^2 +\omega^2 \Gamma^2 } \label{eq:breit_wigner}\\
\end{equation}
as indicated by the solid lines, representing Breit-Wigner fits of the spectral function. 

Conversely, the spectral function for the $\pi$ mode exhibits a much more non-trivial behavior as the frequency threshold for the scattering states lowers and the resonance becomes more pronounced as temperature increases. Beyond $T=13.05$ the spectral function $\rho_{\pi}$ features an interesting double peak structure, where the effective mass and spectral weight of the lower frequency peak decrease as a function of temperature, while the upper frequency peak becomes increasingly dominant when further increasing the temperature. We further illustrate this double peak structure in Fig.~\ref{fig:DoublePeak}, which shows a close up of the $\pi$ spectral function in the same temperature regime. In order to track the widths and positions of the individual peaks, we also present fits to a double Breit-Wigner distribution, featuring two distinct resonances, of the form
\begin{equation}
\rho(\omega)=c_1\frac{\omega\Gamma_1}{(\omega^2-m_1^2)^2 +\omega^2 \Gamma_1^2 }
+c_2\frac{\omega\Gamma_2}{(\omega^2-m_2^2)^2 +\omega^2 \Gamma_2^2 } \, .
 \label{eq:double_breit_wigner}\\
\end{equation}
While Eq.~(\ref{eq:double_breit_wigner}) provides a good description of the peaks, it tends to overestimate the spectral weight in the low frequency tails of the spectral function.

Spectral functions just below the crossover transitions at $T_{pc}=19.5$, are presented in the middle rows of Figs.~\ref{fig:sigmaJsf} and \ref{fig:piJsf} and show a smooth continuation of the temperature dependence observed at the lower temperatures. The spectral function of the $\sigma$ mode continues to show a quasi-particle peak, where the effective mass continues to becomes lighter but the
width decreases again as expected when the pseudo-critical transition temperature is approached. While the double peak structure in the $\pi$ spectral function is most pronounced below the crossover temperature (i.e.~around $T\simeq 15$) some remnants of the low frequency peak clearly persists not only up to the pseudo-critical temperature $T_{pc}=19.5$ but also further into the symmetry restored phase. Despite the clear presence of a second peak, we find that for temperatures $T>16$ the dominant peak of the $\pi$ spectral function can again be described to reasonable accuracy by the Breit-Wigner distribution in Eq. (\ref{eq:breit_wigner}), as indicated by the solid lines in Fig.~ \ref{fig:piJsf}.

Beyond the pseudo-critical temperature $T>19.5$, the dominant features of the spectral functions for $\pi$ and $\sigma$ begin to coincide, as can be seen from comparing the results in the bottom rows of Figs.~\ref{fig:sigmaJsf} and \ref{fig:piJsf}. One finds that as the temperature is increased further beyond $T_{pc}$, both $\sigma$ and $\pi$ spectral functions are increasingly well described by the Breit-Wigner ansatz in Eq. (\ref{eq:breit_wigner}), with increasing mass and decay width as a function of temperature. Between $T=19.5$ and $T=22$ the additional low frequency peak in the $\pi$ spectral function slowly disappears, such that at the highest temperature $T=26.1$ the spectral functions for $\sigma$ and $\pi$ become almost degenerate up to small differences at very low frequencies $\omega \lesssim 0.1$, signaling the approximate restoration of the full $O(4)$ symmetry on the level of the spectral functions.

\begin{figure}[t!]
\includegraphics[width=0.5\textwidth]{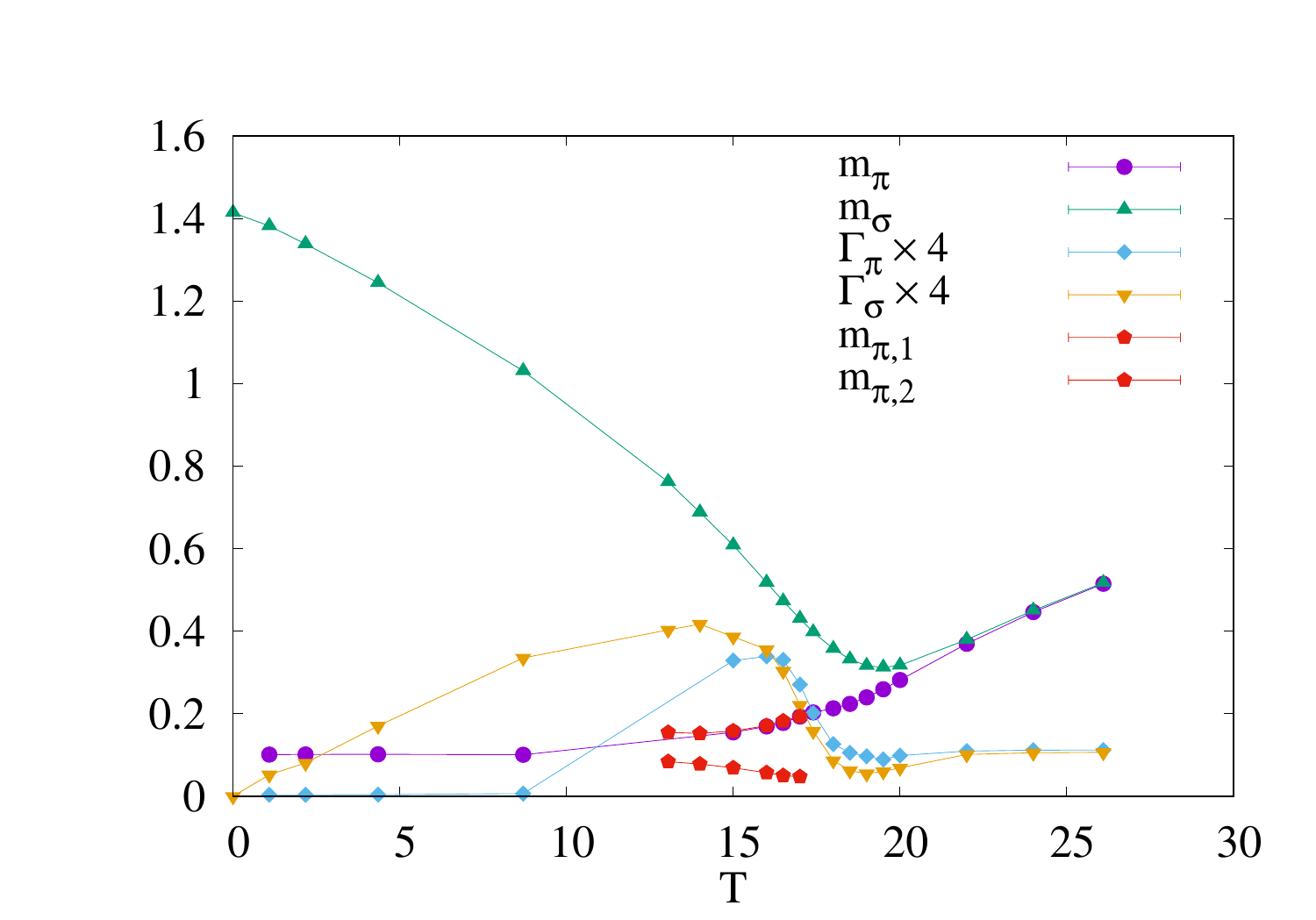}
\caption{\label{fig:BWfitsT} Dependence of (single-peak and double-peak) Breit-Wigner fit parameters for $\pi$ and $\sigma$ spectral functions on the temperature $T$ around the crossover transition at finite explicit symmetry breaking $J=0.05$.} 
\end{figure}

Our results for the temperature dependence of the $\pi$ and $\sigma$ spectral functions in the crossover regime, are compactly summarized  in Fig.~\ref{fig:BWfitsT}, where we show the temperature dependence of the effective masses $m_{\pi/\sigma}$ and decay width $\Gamma_{\pi/\sigma}$ obtained from the (single and double peak) Breit-Wigner fits. While at low temperatures $\pi$ and $\sigma$ spectral functions in the classical-statistical approximation show well defined quasi-particle peaks, the $\sigma$ mass rapidly decreases with increasing temperature, and the in-medium decay widths of $\pi$ and $\sigma$ increase significantly below the pseudo-critical transition temperature $T_{pc}$. The scattering states in the $\pi$ spectral function lead to the development of an additional resonance peak as $T_{pc}$ is approached. Interestingly, it is this emerging second peak which appears to develop further into the resonance peak that eventually becomes degenerate with the $\sigma$ mode as temperature is increased further beyond $T_{pc}$. The original low-temperature quasi-particle peak on the other hand slowly melts and disappears around the pseudo-critical temperature. The two distinct peaks at $m_{\pi,1} $ and $m_{\pi,2}$ around $T\simeq 15$ in fact show signs of an interesting avoided-crossing behavior which has not been observed in the corresponding solutions of analytically continued FRG flow equations, for example, so far.

\subsection{Spectral functions in the crossover regime -- $J$ dependence}
So far we have investigated the temperature dependence of the $\sigma$ and $\pi$ spectral functions in the vicinity of the crossover transition, at a fixed relatively large explicit symmetry breaking $J=0.05$. Since we always have to keep a non-vanishinig explicit symmetry breaking in order to distinguish between $\pi$ and $\sigma$ components, we will now fix the temperature close to $T_c$ at $T=17.4$ and decrease the explicit symmetry breaking $J$ by successive factors of two to approach the critical point, as indicated by the horizontal line in the phase diagram in Fig.~\ref{fig:phasediag}.

\begin{figure*}
\begin{minipage}{0.5\textwidth}
\includegraphics[width=\textwidth]{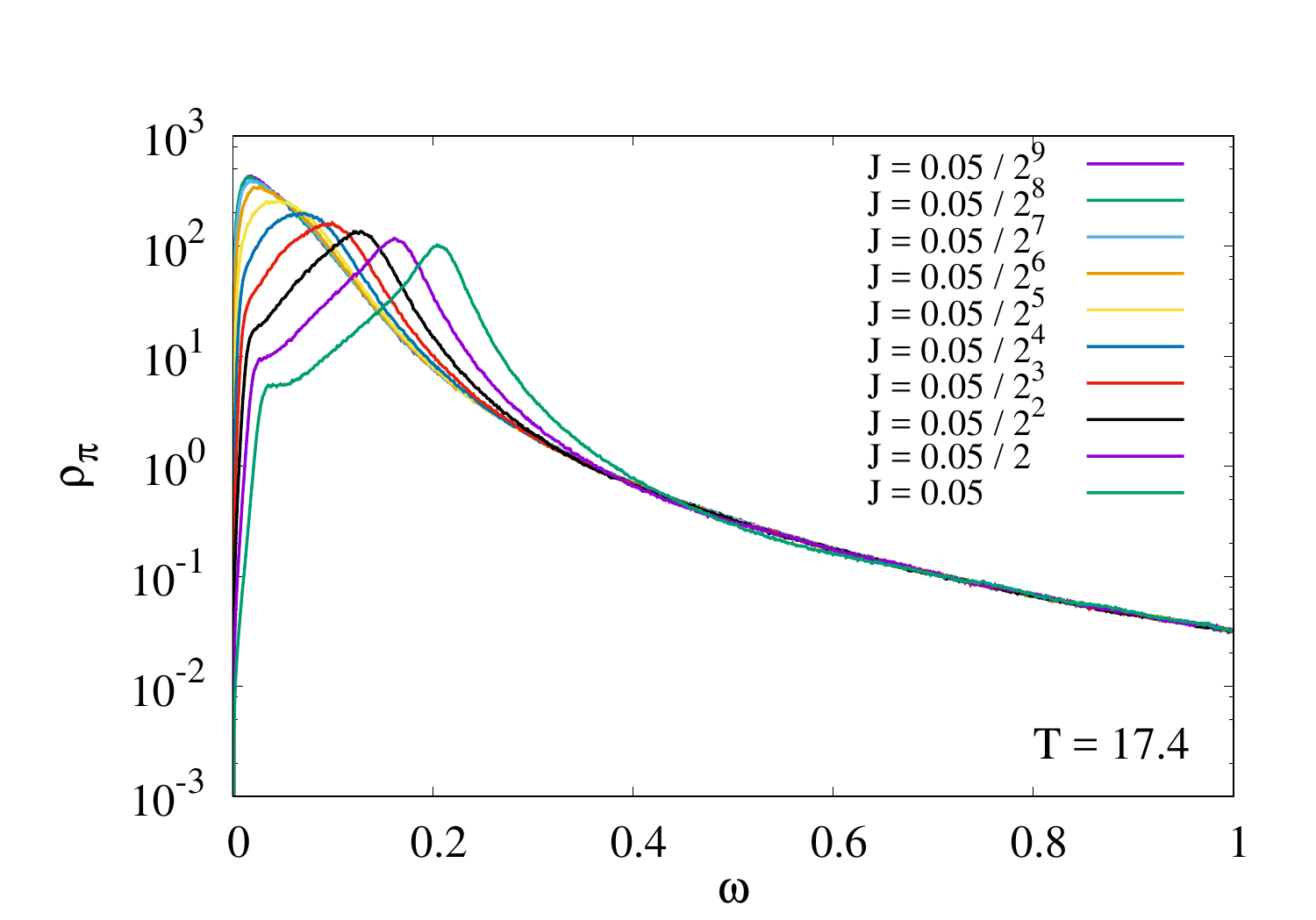}
\end{minipage}
\begin{minipage}{0.5\textwidth}
\includegraphics[width=\textwidth]{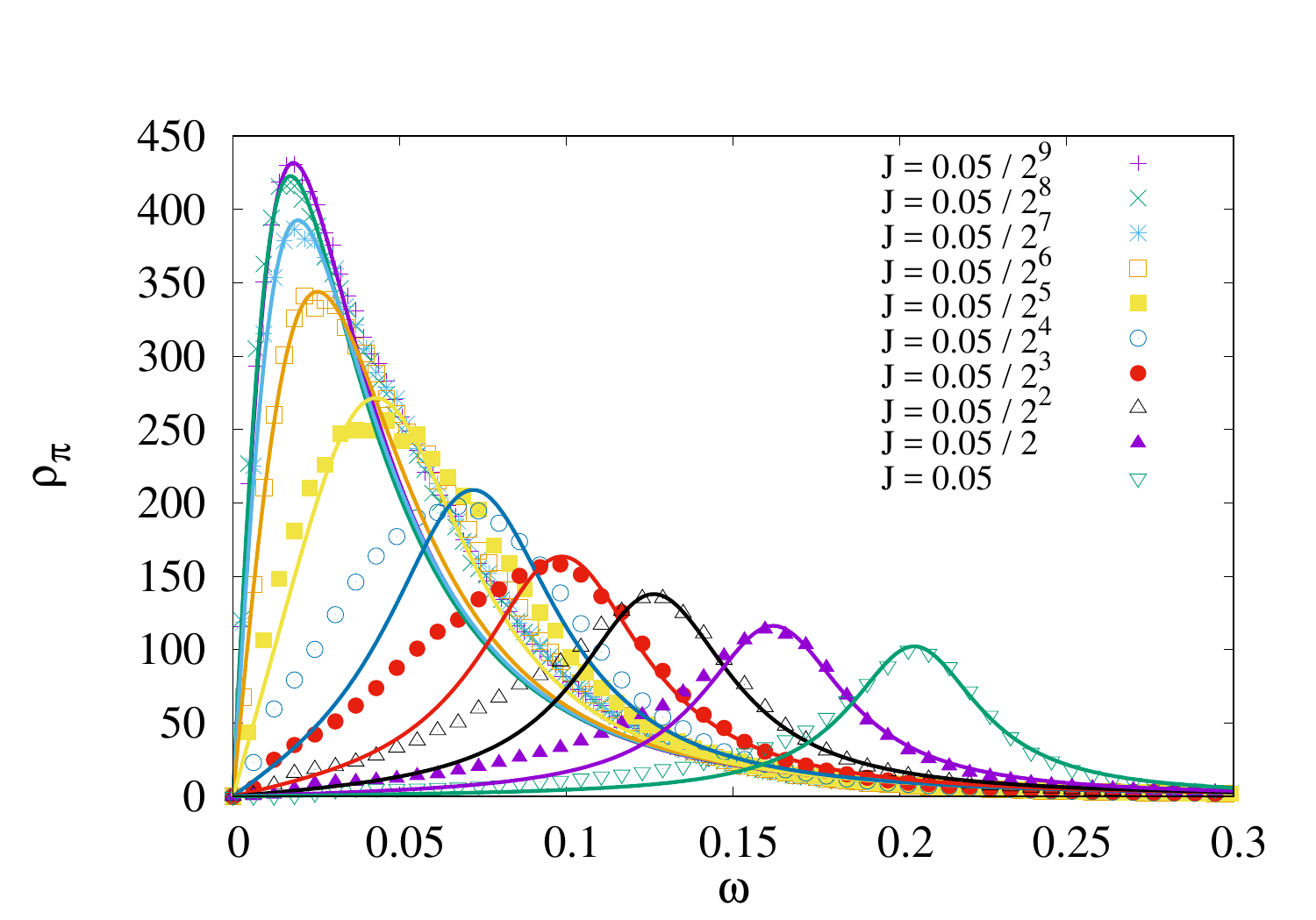}
\end{minipage}
\begin{minipage}{0.5\textwidth}
\includegraphics[width=\textwidth]{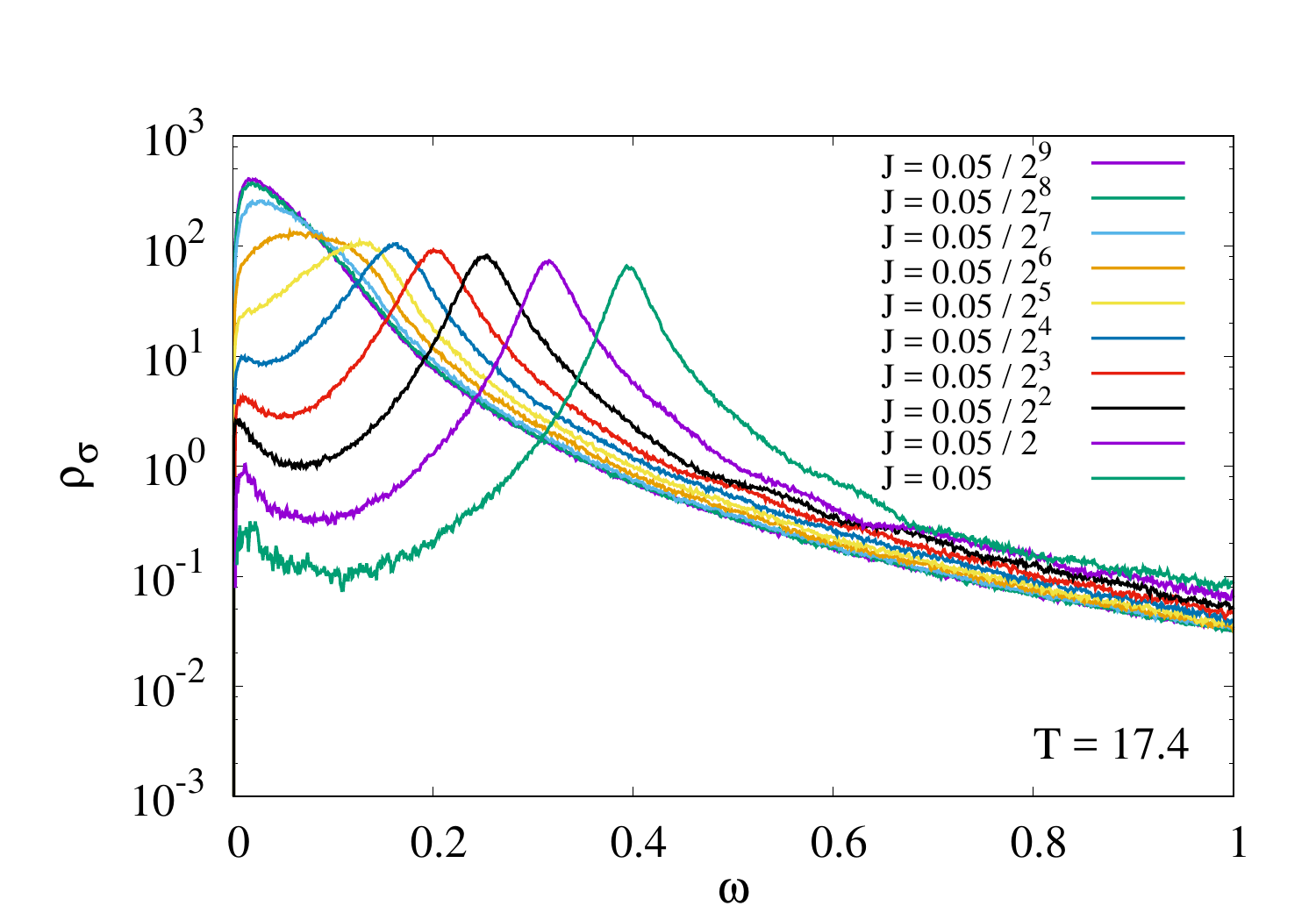}
\end{minipage}
\begin{minipage}{0.5\textwidth}
\includegraphics[width=\textwidth]{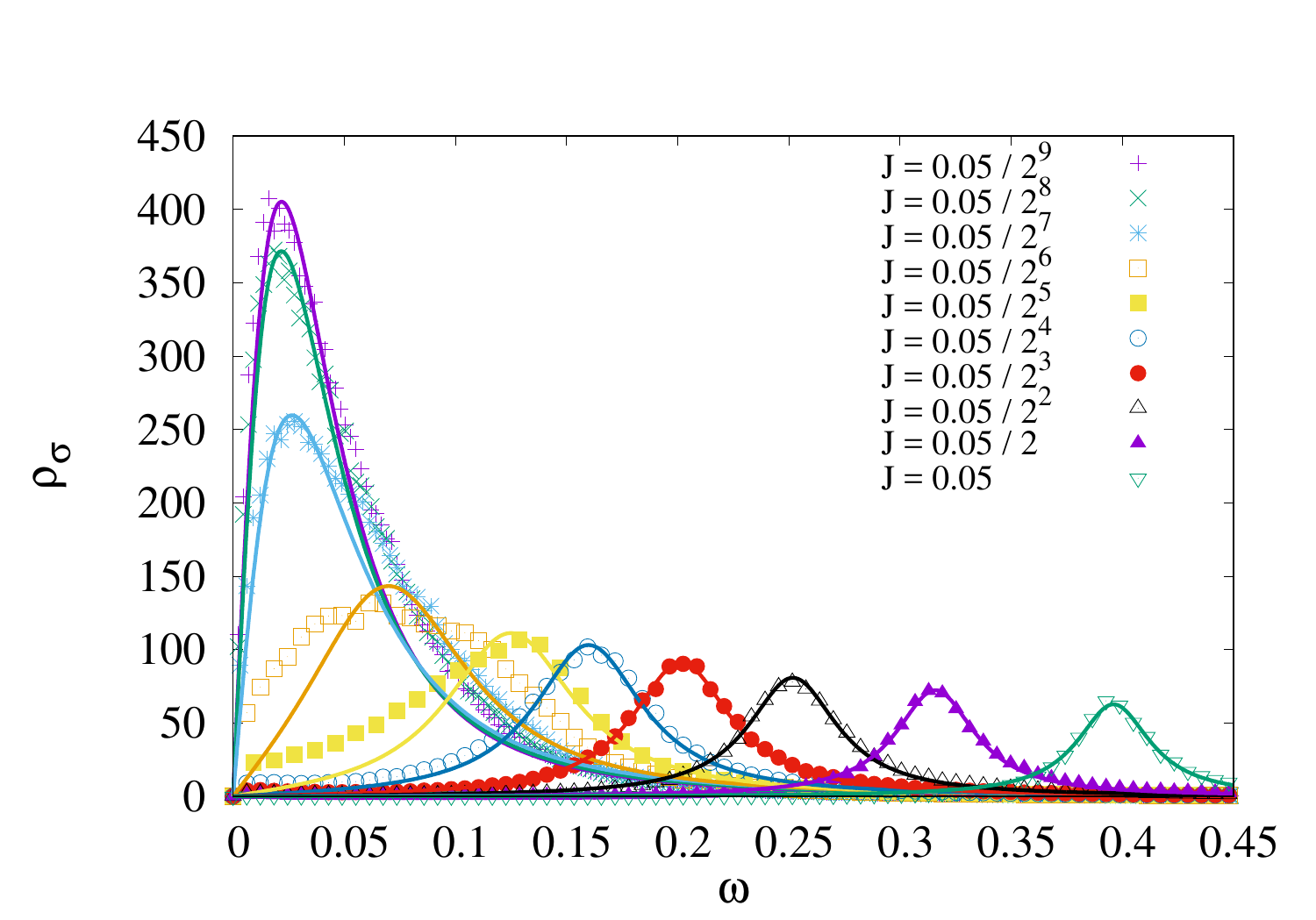}
\end{minipage}
\caption{\label{fig:SFJScan} Spectral functions $\rho_\pi$ and $\rho_\sigma$ for different amounts of explicit symmetry breaking $J$ at a nearly critical temperature $T=17.4$. Data is shown in log-lin representation in the left panel and lin-lin representation in the right panel. Curves in the right panel show fits to a Breit-Wigner distribution (\ref{eq:breit_wigner}).} 
\end{figure*}

\begin{figure}[t!]
\begin{minipage}{0.5\textwidth}
\includegraphics[width=\textwidth]{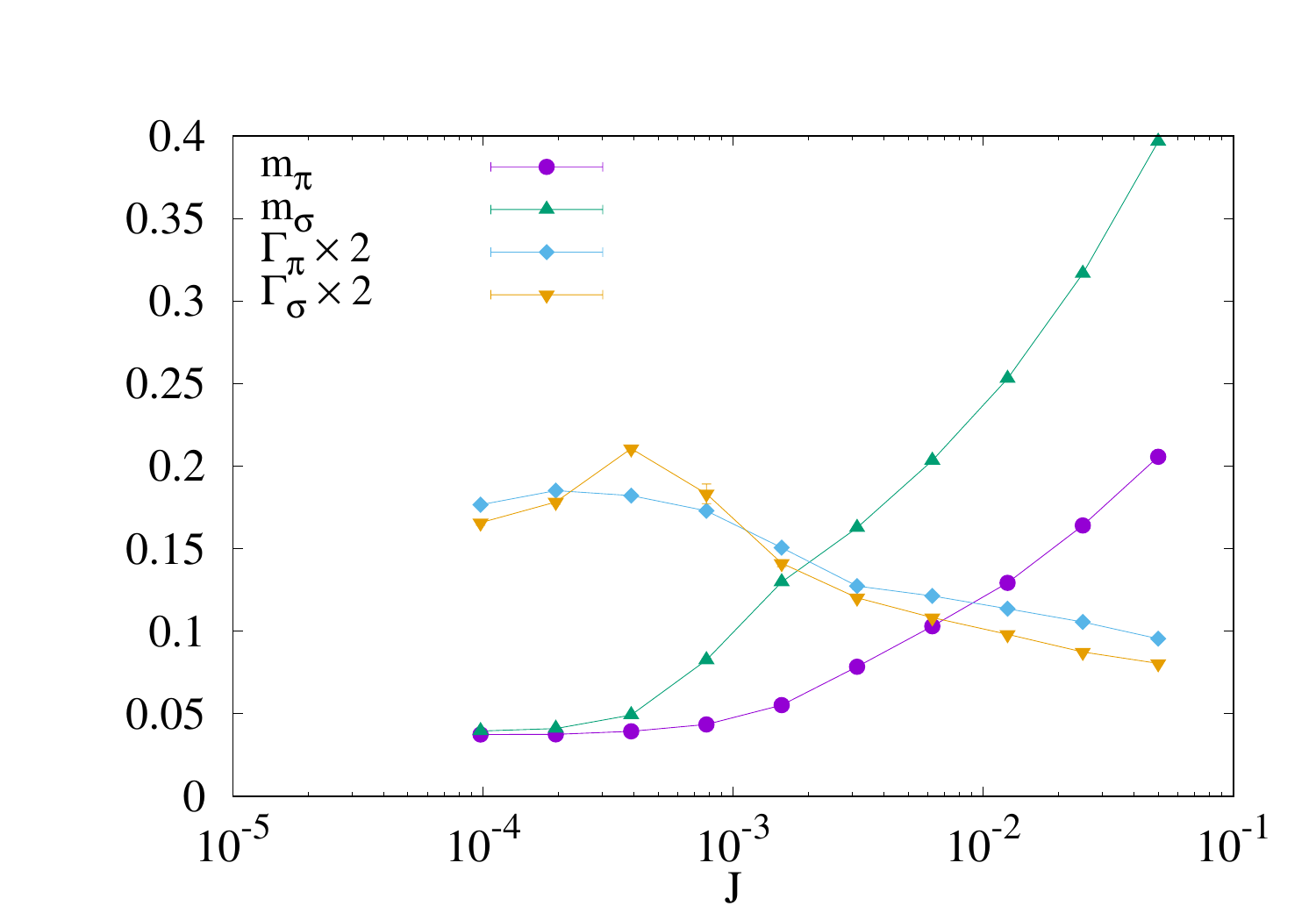}
\end{minipage}
\caption{\label{fig:BWfitsJ} Dependence of Breit-Wigner fit parameters for $\pi$ and $\sigma$ spectral functions on the explicit symmetry breaking $J$ at a nearly critical temperature $T=17.4$.} 
\end{figure}

Our results for the $J$ dependence of spectral functions close to $T_c$ are summarized in Fig.~\ref{fig:SFJScan}, where top and bottom rows show the $\pi$ and $\sigma$ spectral functions at different explicit symmetry breaking. Starting from $J=0.05$ employed in our temperature scan of the crossover transition, we find that lowering the explicit symmetry breaking $J$ results in a rapid decrease of the effective mass of $\sigma$ and $\pi$ along with a simultaneous increase of the decay width. Effectively the combination of these two phenomena leads to a melting of the quasi-particle peaks, in both $\sigma $ and $\pi$ spectral functions, as can be seen from Fig.~\ref{fig:BWfitsJ}, where we present the $J$ dependence of the Breit-Wigner resonance parameters. However, one should caution, that already at $J=0.05 \times 2^{-1}$ the $\pi$ spectral function develops an additional enhancement at low frequencies, which is no longer fully captured by the Breit-Wigner fits. Even though initially the $\sigma$ spectral function can still be reasonably well described in terms of a single resonance, we find that below $J=0.05 \times 2^{-5}$ the description in terms of Breit-Wigner distribution becomes increasingly inaccurate also for the $\sigma$ spectral function, as both spectral functions start to feature a strong enhancement at low frequency, which is no longer captured by a simple quasi-particle peak.

\begin{figure*}
  \vspace*{-.5cm}
\begin{minipage}{0.5\textwidth}
\includegraphics[width=\textwidth]{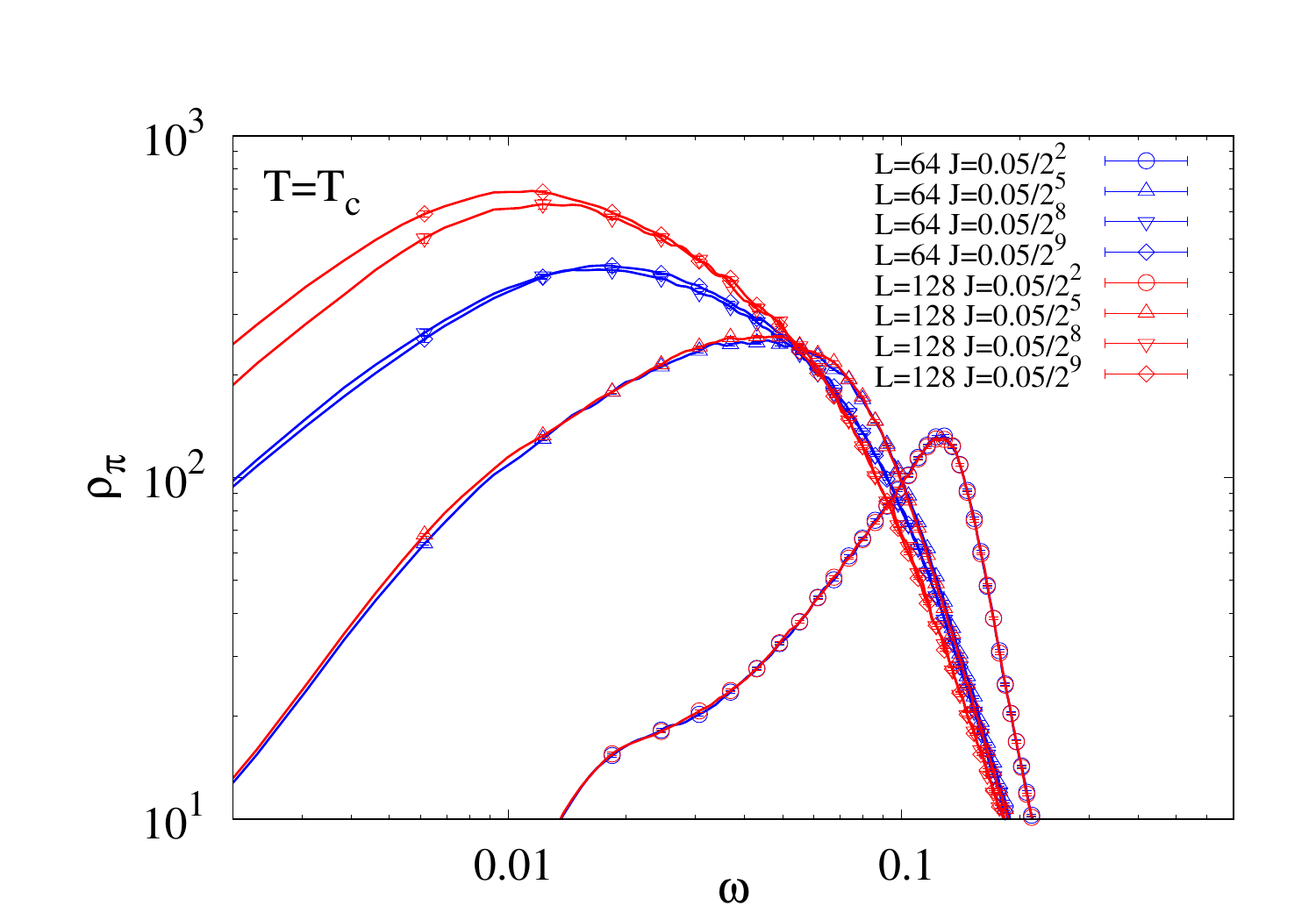}
\end{minipage}
\begin{minipage}{0.5\textwidth}
\includegraphics[width=\textwidth]{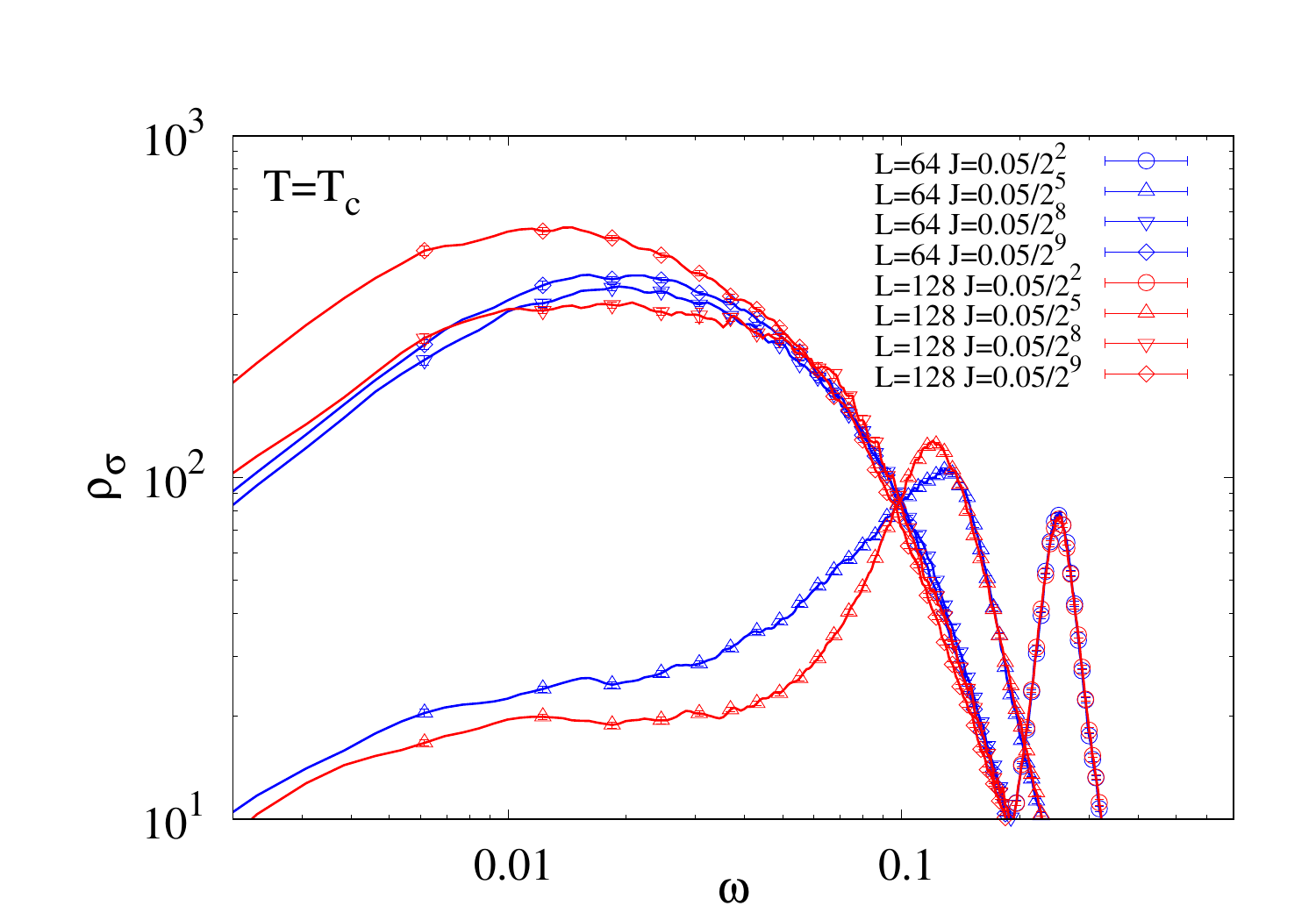}
\end{minipage}
\caption{\label{fig:SFNearCritical} Spectral functions $\rho_\pi$ (left) and $\rho_\sigma$ (right) at $T\approx T_c$ for different amounts of explicit symmetry breaking $J$ and different lattice sizes $L=64$ and $L=128$. In the vicinity of the critical point ($J \to 0$), fluctuations of the order parameter prohibit the distinction between $\pi$ and $\sigma$ modes. Lines represent raw data, points shown to guide the eyes and illustrate errorbars.} 
\end{figure*}

Eventually, the amount of explicit symmetry breaking is no longer large enough to guarantee the alignment of the order parameter in our finite volume system, such that for very small values of $J$ the spectral functions of $\pi$ and $\sigma$ effectively become degenerate. 
Even though this is a finite volume artifact, it is also clear that extending the study to larger and larger lattices will only shift the problem towards smaller and smaller values of $J$, as in any finite system the alignment of the order parameter with the symmetry breaking axis will only be guaranteed above a certain amount of explicit symmetry breaking. 
We illustrate this problem in Fig.~\ref{fig:SFNearCritical}, where we compare the results for the spectral functions $\rho_\pi$ and $\rho_\sigma$ obtained on $L=64$ and $L=128$ lattices. 

One finds that for $J=0.05\times 2^{-2}=0.0125$ the results obtained on $L=64$ and $L=128$ lattices are in good agreement with each other, indicating the absence of finite-volume effects. Decreasing the amount of explicit symmetry breaking $J$ further, to approach the critical point, the distinction between $\pi$ and $\sigma$  becomes less and less prominent, as the simulations develop a significant volume dependence in the vicinity of the critical point. 
In particular, for the $L=64$ lattice, one clearly observes that at some point the relevant infrared cut-off is no longer set by $J$ but rather by the finite system size, leading to $J$ independent results for the spectral function for $J \lesssim 0.05\times 2^{-8}=0.000195$. 
While the results obtained on $L=128$ lattices, continue to show an increase of the low frequency enhancement with decreasing $J$, a clear distinction between $\pi$ and $\sigma$ modes ceases to exist for $J \lesssim 0.05\times 2^{-9}=0.000098$ due to finite volume effects. 

While previous studies of the critical dynamics of a $Z_2$ symmetric $\phi^4$ theory in $2+1$ dimensions dealt with this problem by performing simulations on extremely large lattices \cite{Berges:2009jz}, it is worth pointing out that the problem is substantially more severe for the breaking of a continuous symmetry, where the orientation of order parameter field can rotate continuously over the course of the simulation, and we will therefore have to explore different strategies to study the dynamical critical behavior in the limit $J\to0$ and $T\to T_c$.

\subsection{A glance at critical dynamics}
We now focus on the behavior of the spectral function in the vicinity of the critical point, realized by setting $T\approx T_c$ and $J \to 0$ in our simulations. Since in the vicinity of the critical point $J \to0$ our finite volume simulations do not allow us to distinguish between $\pi$ and $\sigma$ modes, we set $J=0$ directly and investigate the behavior of the combined spectral function
\begin{equation}
\rho\equiv\frac{1}{N}\, \text{tr}~\rho = \frac{1}{4} \rho_{\sigma} + \frac{3}{4} \rho_{\pi}\;.
\end{equation}
In the limit $T \to T_c$, for $\omega \to 0$ and $p \to 0$ the spectral function is expected to exhibit a scaling behavior of the form \cite{Berges:2009jz}
\begin{equation}
\rho(s^z \omega, sp, s^\frac{1}{\nu}\, T_r) \, = \, s^{-(2-\eta)}\, \rho(\omega,p,T_r) \,,
\label{eq:scalingrho}
\end{equation}
where $\eta$ is the anomalous dimension, $\nu$ is the correlation-length exponent and $z$ denotes the dynamic scaling exponent which can be used to classify the time-dependent critical behavior of a system. Taking $\nu=0.7377(41)$ and $\gamma=1.4531(104)$ from \cite{Engels:2014bra}
and using $\gamma/\nu=2-\eta$ we obtain $\eta=0.03022$. Based on the analyis of Halperin and Hohenberg~\cite{RevModPhys.49.435}, the dynamic universality class is determined by the conserved quantities of the system along with their coupling to the order parameter. 

We first note that for the Hamiltonian dynamics of the relativistic $O(4)$ model considered in our study, we have a conserved energy 
\begin{eqnarray}
\epsilon&=& \int d^3x \left[ \frac{1}{2} \pi_{a}(x)\pi_{a}(x) +\frac{1}{2} (\nabla \phi_{a}(x))(\nabla \phi_{a}(x)) \right. \\
 && \left. \qquad +\frac{m^2}{2}  \phi_{a}(x)  \phi_{a}(x) +\frac{\lambda}{4!N}  (\phi_{a}(x)  \phi_{a}(x))^2  \right]\;, \nonumber
\end{eqnarray}
the conversed momentum
\begin{eqnarray}
\Pi_{i}=  \int d^3x~\pi_{a}(x) \nabla_{i}\phi_{a}(x)
\end{eqnarray} 
and a conserved $O(4)$ current
\begin{eqnarray}
j^{ab}= \epsilon^{abcd} \int d^3x~\phi_{c}(x) \pi_{d}(x)
 \end{eqnarray}
Since the Poisson brackets of the $N$-component order parameter $\phi_{a}=\frac{1}{V} \int d^3x \,\phi_{a}(x)$ with the conserved quantities $\epsilon$ and $j^{ab}$ are non-vanishing, the dynamics of these modes can affect the critical dynamics of the order parameter, and the relativistic $O(4)$ model with Hamiltonian dynamics does not belong to one of the standard dynamic universality classes according to the classification scheme of Halperin and Hohenberg~\cite{RevModPhys.49.435}. However, it has been argued \cite{Halperin:1974zz} that for negative values of the specific heat exponent $\alpha<0$  (which is the case for the $O(4)$) the coupling to the conserved energy is irrelevant.  Since for $J=0$ the order parameter becomes an $N =4$ component field (i.e.~carrying information about the orientation as well as the magnitude), the analysis of Wilczek and Rajagopal \cite{Rajagopal:1992qz} suggests that the critical dynamics of the relativistic $O(4)$ scalar theory follows an extension of model $G$, where the dynamical critical exponent $z=\frac{d}{2}$ can be determined from a renormalization group analysis.

\begin{figure*}[t!]
 \begin{minipage}{0.5\textwidth}
\includegraphics[width=\textwidth]{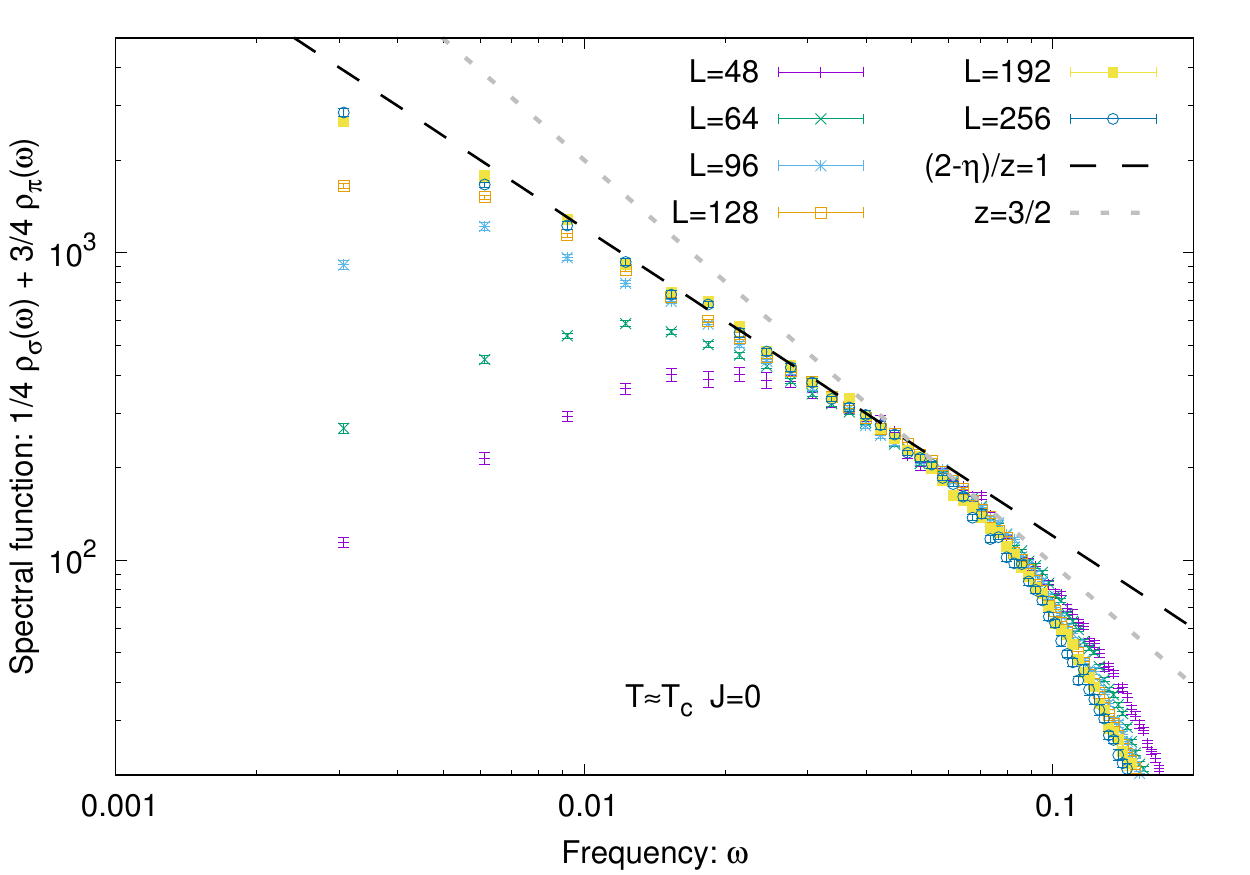}
\end{minipage}
\begin{minipage}{0.5\textwidth}
\includegraphics[width=\textwidth]{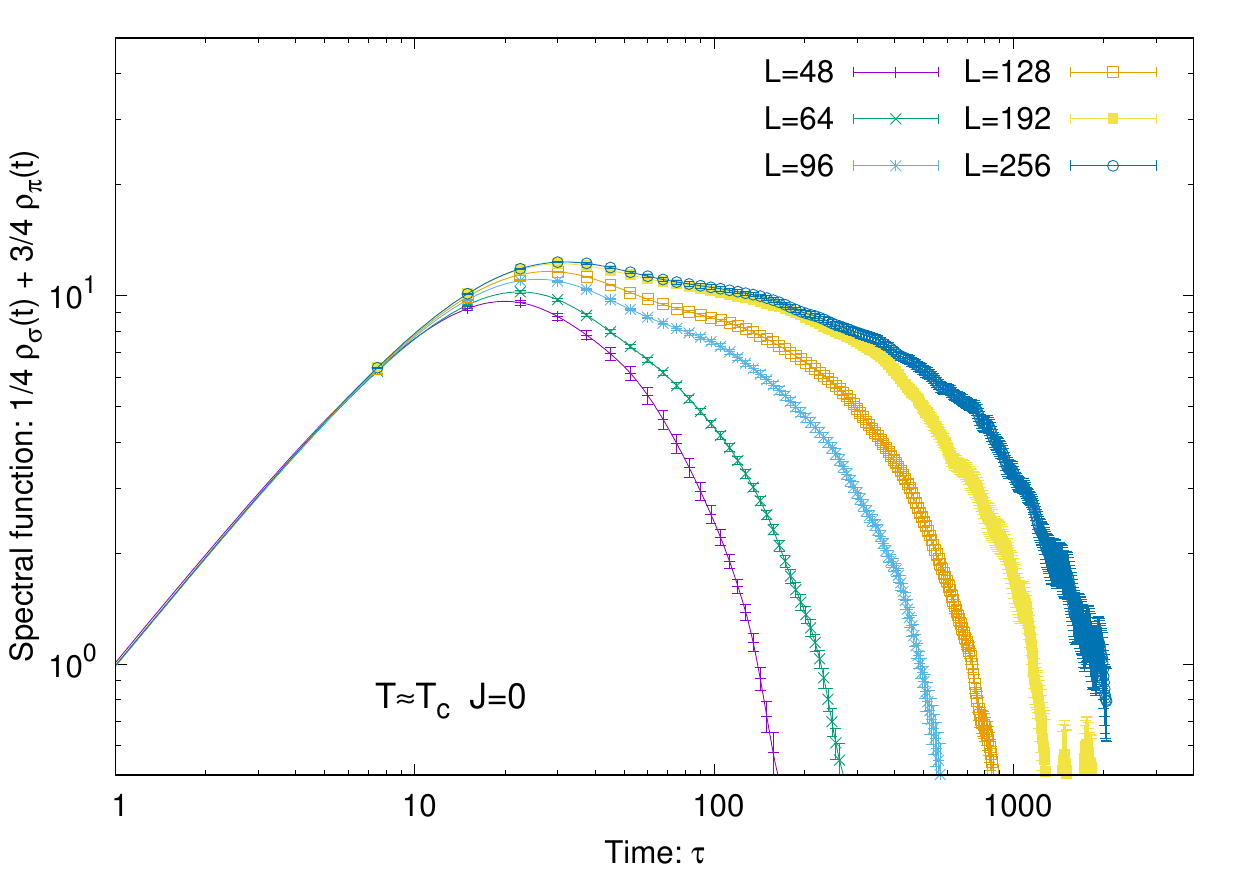}
\end{minipage}
\caption{\label{fig:RhoCrit} Behavior of the spectral function close to the critical point ($T\approx T_c$,$J=0$) for different lattice sizes $L$ in the frequency (left) and time domain (right).
Lines represent raw data and points illustrate error estimates and guide the eyes in right figure.} 
\end{figure*}

Simulation results for $\frac{1}{N} \text{tr}~\rho$ are shown in Fig.~\ref{fig:RhoCrit} where we present results in the frequency and time domain obtained for various different lattice sizes between $L=48$ and $L=256$. We find that in the vicinity of the critical point, a fine time step of the numerical integrator is needed to correctly reproduce the late-time behavior of the spectral function; we have therefore decreased the time step in our numerical integration by a factor of four to $\Delta t=0.00125$ and checked explicitly for our $L=96$ data that reducing the time step by an additional factor of four does not affect the results.

Despite the strong finite-size dependence in our simulations, we also observe first indications of the emergence of an infrared power law dependence of the spectral function in the low frequency domain. Based on the scaling relation in Eq.~(\ref{eq:scalingrho}) the critical spectral function is expected approach the following scaling behavior
\begin{equation}
\rho(\omega,p=0) \propto \omega^{\frac{2-\eta}{z}}
\end{equation}
for an infinite system. Different curves in the left panel of Fig.~\ref{fig:RhoCrit} indicate power law fits, employing the values $z=2-\eta$ and $z=3/2$ for the dynamical critical exponent. We find that our results for the spectral function at zero spatial momentum favor the value $z=2-\eta$ of the so called ``conventional theory" of dynamic critical phenomena, which emerges when the critical divergencies of kinetic coefficients is not taken into account in the scaling analysis~\cite{RevModPhys.49.435}. In fact this observation may be reinforced further by looking at the behavior of the spectral function in the time domain, where values of $z<2-\eta$ would lead to an increase of the spectral function function $\rho(t) \propto t^{1-\frac{2-\eta}{z}}$ as a function at late times, which is clearly not observed in our simulations.  Conversely, a value of $z=2-\eta$ (or $z>2-\eta$) would lead to a logarithmic time dependence (or power law decay) of the spectral function $\rho(t)$, which appears to be more consistent with our results.

\begin{figure}[t!]
\begin{minipage}{0.45\textwidth}
\includegraphics[width=\textwidth]{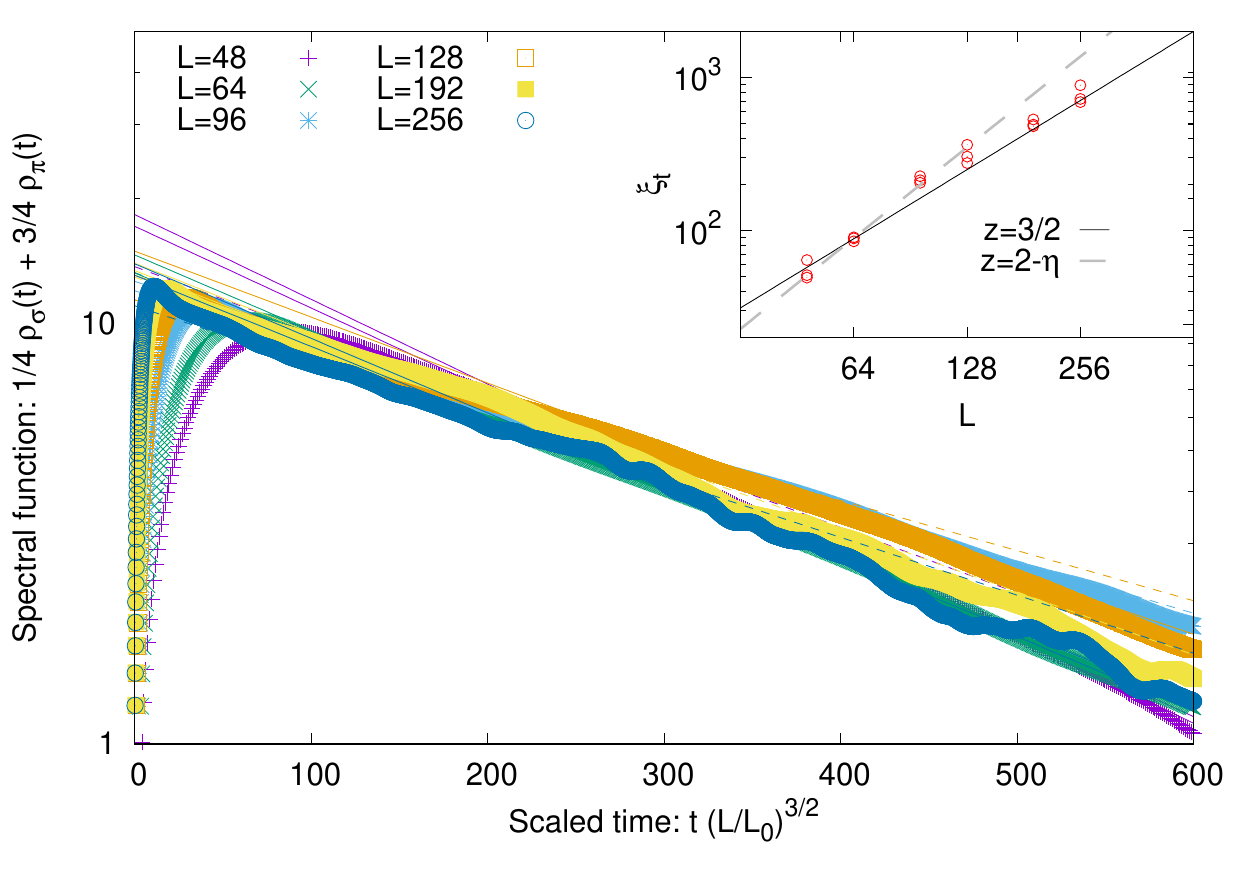}
\end{minipage}
\caption{\label{fig:XiTScaling}Extraction of the auto-correlation time $\xi_{t}(L)$ from the spectral function close to the critical point ($T\approx T_c$,$J=0$). Different curves show results for the spectral function $\rho(t,p=0)$ for different lattice sizes $L$ as a function of the scaled time $t (L/L_0)^{3/2}$ with $L_0=128$.}
\end{figure}

It is also evident from Fig.~\ref{fig:RhoCrit}, that for any finite-size system the critical behavior of the spectral function $\rho(t \to \infty)$ is suppressed by the exponential decay with the auto-correlation time $\propto \exp(-t/\xi_t(L))$.  While for any finite system the auto-correlation time $\xi_{t}(L)$ is finite, it diverges with increasing system size as $\xi_{t}(L) \propto L^{z}$, corresponding to the well known phenomenon of critical slowing down, and the typical way to extract the dynamical critical exponent $z$ in Monte-Carlo simulations. Our results for the analysis of the auto correlation time are compactly summarized in Fig.~\ref{fig:XiTScaling}, where we we present fits to the late time exponential behavior of the spectral function $\rho(t) \propto \exp(-t/\xi_t(L))$ along with the results for $\xi_t(L)$ shown in the inset. 
While for small volumes $L<128$ the scaling of the auto-correlation time $\xi_t(L)$ appears to be consistent with the $z=2-\eta$ predicted by the conventional theory, the behavior of $\xi_t(L)$ for large volumes hints at a weaker divergence of the auto-correlation time consistent with $z=3/2$ on the larger lattices, as indicated by the solid and dashed curves in the inset of Fig.~\ref{fig:XiTScaling}.

Since the frequency dependence of the critical spectral function $\rho(\omega,p=0) \propto \omega^{\frac{2-\eta}{z}}$, and the finite-size scaling of the auto-correlation time $\xi(t) \propto L^{z}$ lead to different extractions of the dynamic critical exponent $z$, we are unable to determine the dynamical critical behavior precisely from our current simulations. One possible explanation of the observed discrepancies could be due to the fact that we have set the spatial momentum $p=0$ prior to taking the limit $\omega \to 0$ (or $t \to \infty$), which may or may not affect the critical scaling of the spectral function. In any case, it would be interesting to investigate the critical dynamics in more detail as a function of $p$ and $\omega$ at non-vanishing $T_r$ and $J$, to further elucidate on the structure of excitations in the vicinity of the critical point.  However, this will require significant computational resources and is well beyond the scope of our present work.

\section{Conclusions}
\label{sec:conclusion}
We have performed a detailed study of classical-statistical spectral functions in the relativistic $O(4)$ model. While the static critical behavior is naturally reproduced correctly within the classical-statistical lattice approach, the focus of our study has been on the behavior of the real-time spectral functions $\rho_{\pi}$ and $\rho_{\sigma}$. While at very low temperatures, the classical-statistical approximation is inadequate and effectively reduces to a mean-field approximation, we argued that a classical-statistical description becomes accurate in the vicinity of a second order phase transition and demonstrated some intriguing features of the spectral functions close to the crossover transition and in the vicinity of the $O(4)$ critical point.

In the broader context of non-perturbative calculations of real-time spectral functions, the results from classical-statistical simulations reported in this paper may provide additional guidance to alternative theoretical approaches, based, e.g.~on functional methods or analytic continuation of Euclidean correlation functions, where prior information on the structure of excitations is required to devise suitable ans\"{a}tze or efficient truncation schemes. Specifically, there is an interesting possibility to benchmark the quality of results obtained within functional approaches, based on a direct comparison of the results obtained in the classical-statistical limit. This is work in progress and will be reported elsewhere.

\section*{Acknowledgements}
This research was supported by the Helmholtz International Center (HIC) for FAIR within the LOEWE initiative of the State of Hesse, and by the Deutsche Forschungsgemeinschaft (DFG) through the grant CRC-TR 211 ``Strong-interaction matter under extreme conditions.''


\bibliographystyle{elsarticle-num}
\bibliography{references}

\end{document}